\documentstyle[12pt,psfig]{article}

\newcommand{\ipar}{\hspace*{6mm}}
\newcommand{\ilskp}{\hspace*{27mm}}
\newcommand{\hlf}{\mbox{$\frac{1}{2}$}}

\newcommand{\case}[2]{\mbox{\small $\displaystyle \frac{#1}{#2}$}}

\newcommand{\etal}{{\protect\it et al.}}
\newcommand{\bcn}{\begin{center}}
\newcommand{\beq}{\begin{equation}}
\newcommand{\beqar}{\begin{eqnarray}}
\newcommand{\Aslash}{\mbox{$\not \!\! A$}}

\newcommand{\dslash}{\mbox{$\not \! \partial$}}
\newcommand{\dpsm}{\frac{d^{4}p}{(2\pi)^4}}
\newcommand{\dP}{\frac{d^{4}P}{(2\pi)^4}}
\newcommand{\dK}{\frac{d^{4}K}{(2\pi)^4}}
\newcommand{\dbq}{\frac{d^{4}Q}{(2\pi)^4}}
\newcommand{\dq}{\frac{d^{4}q}{(2\pi)^4}}
\newcommand{\dPqq}{\frac{d^{4}Pq'q}{(2\pi)^{12}}}
\newcommand{\dqq}{\frac{d^{4}q'q}{(2\pi)^{8}}}

\newcommand{\qslash}{\mbox{$\not \! q$}}

\newcommand{\ecn}{\end{center}}
\newcommand{\eeq}{\end{equation}}
\newcommand{\eeqar}{\end{eqnarray}}

\newcommand{\Eq}[1]{Eq.~(\protect\ref{#1})}
\newcommand{\Eqs}[1]{Eqs.~(\protect\ref{#1})}
\newcommand{\Fig}[1]{Fig.~\protect\ref{#1}}
\newcommand{\Sec}[1]{Sec.~\protect\ref{#1}}
\newcommand{\Table}[1]{Table~\protect\ref{#1}}
\newcommand{\sect}[1]{\section{{\protect\sc #1}}
       \setcounter{table}{0}\setcounter{figure}{0}
       \setcounter{equation}{0}\vspace*{-\parskip}}
\newcommand{\subsect}[1]{\subsection{{\protect\underline{\protect\rm #1}}}
       \vspace*{-\parskip}} 

\begin{document}      
\parbox{185mm}{ \hspace*{\fill}
Preprint Numbers: \parbox[t]{45mm}{KSUCNR-101-97 \\  nucl-th/9705018 \\ }  }

\bcn
{\bf HADRON PHYSICS FROM THE GLOBAL COLOR MODEL OF QCD}\footnotemark[2]
\vspace*{10mm}

PETER C. TANDY

Center for Nuclear Research, Department of Physics, Kent State University,\\
Kent, Ohio, 44242 USA \\
\vspace*{1cm}
May 5, 1997

\vspace*{30mm}

ABSTRACT
\ecn

\hspace*{-\parindent}
We review recent progress in modeling the quark-gluon content of mesons 
and their low-energy  interactions through the Global Color Model field theory.
An emphasis is placed on techniques that are shared with the approach based on
truncations of the Dyson-Schwinger equations of QCD.
In contrast to most other  field theory 
models for QCD degrees of freedom in hadron physics, this approach directly
deals with the derived intrinsic space-time extent of the meson modes  in their
role as  field variables and can accommodate confinement as well as dynamical 
breaking of chiral symmetry.   Various theoretical techniques and 
approximations found useful in this approach are described. 
Selected applications reviewed here include the properties and interactions of 
the Goldstone bosons, interaction vertex functions, low energy chiral observables, 
electromagnetic interactions and form factors, and transition form factors.  Some
initial considerations of Vector Meson Dominance and pion loop processes are 
discussed. 
\bigskip

\ilskp KEYWORDS

hadrons from quarks and gluons; non-perturbative QCD modeling; confinement; 
dynamical chiral symmetry breaking; composite, finite size mesons; meson
interaction and transition form factors; electromagnetic coupling; vector meson
and pion loop processes.

\footnotetext[2]{To be published in {\it Prog. Part. Nucl. Phys.} {\bf 39}, 1997.}

\newpage
\tableofcontents
\newpage
\sect{Introduction}

To investigate hadron physics from the perspective  of nonperturbative
quantum chromodynamics
(QCD) and QCD-based models, a number of approaches are currently employed.
These include  the intensive computational method of lattice gauge field
theory, the method of light cone quantization, QCD sum rules and chiral
perturbation theory (ChPT).  In one sense, the first two constitute a completely 
fresh start in that much of the empirical 
information on hadron dynamics that has been built up over many of the past 
decades through the refinement of model hadronic lagrangians is not easily
joined.   This becomes irrelevant when  S-matrix elements for many-body hadronic
processes can be routinely obtained from  the relativistic gauge field theory 
of quarks and gluons.  However it is apparent that for some time to come, 
model hadronic theories remain an important intermediate form into which
new findings about non-perturbative QCD can be packaged for rapid deployment
in hadron  physics.   One way to link QCD with such an intermediate form is to
subject the quark-gluon action to a restructuring or change of field
variables to expose  hadronic or collective degrees of freedom and integrate
out the quarks and gluons.   At present, most work of this type that is able
to confront a variety of hadronic observables deals with QCD-based model 
field theories.   We review meson physics obtained from the Global Color Model
(GCM) of QCD (Cahill and Roberts, 1985; Praschifka \etal, 1987a).   Other
overviews of this model that emphasize different aspects can be found in Cahill 
(1992) and in Cahill and Gunner (1997).  

What we refer to as a model hadronic field theory was in the past usually called 
an effective field theory.  In either case  this is in the old-fashioned sense that
implies a semi-phenomenological model field theory meant for a limited low energy
domain; but not implying an equivalent representation of an underlying theory
through a systematic expansion and a counting scheme to order corrections.  To 
distinguish the former from ChPT and other recent developments in hadron physics that 
are referred to as effective field theories in the technical sense, we shall use the 
terms model or approximate field theories.  

The GCM is a model field theory of quarks in which the interaction is between
quark color currents mediated by an effective $2$-point gluon function.  The
infrared form of the latter is the phenomenological element of the model,
while it is constrained to match the known perturbative QCD behavior in the 
ultraviolet.  The point coupling limit gives the closely related and widely used 
Nambu--Jona-Lasinio (NJL) model (Nambu and Jona-Lasinio, 1961) that is simpler 
to work with.   Special interest is taken here in the distinguishing features 
of the GCM that include a  naturally accommodated confinement mechanism and 
the finite size of the $\bar q q$ meson modes.   The derived interactions
between these modes produce vertex form factors that are accountable to 
underlying quark-gluon dynamics.  It is this aspect that may provide an 
interesting perspective to the thrust
in nuclear theory that centers around hadronic field theory models.
A key question is whether a consistent quark-gluon basis can be developed for
important hadron interaction vertices that can support the present empirical
description (especially of the form factors) thus correlating many hadronic
parameters with far fewer sub-hadronic parameters. 

The Dyson-Schwinger equation (DSE) approach to non-perturbative QCD 
(Roberts and Williams, 1994) is to 
truncate the infinite system of coupled equations that link the $n$-point 
functions of QCD at low order, use an Ansatz to replace the discarded 
information,  and to seek a self-consistent solution.  There is a close 
correspondance in form  between the content of the GCM action at tree-level 
in the composite hadron fields and a particular low-order implementation of 
the DSE approach.  For this reason, a number of present calculations in each
approach share common elements.  A clearer distinction would become evident  
at the next  level of treatment where the organization of the physics in each 
approach is different.   However, they remain complementary; each providing 
insights into the role of nonperturbative QCD effects in hadron physics.
We shall refer extensively to the mixing of elements of these approaches.

In \Sec{sect_gcm} we introduce the Global Color Model and review the 
auxiliary field method for bosonization to an action for  the meson modes
of the theory.  Also considered there is the decomposition to 
local field variables and the identification of the corresponding meson-quark
vertex functions.  That section ends with a discussion of issues related to
extraction of physics from the formalism and calculations defined in Euclidean
metric, and the relationship of the GCM and DSE approaches.    Important 
elements for producing hadron observables from the GCM are the dressed
quark propagators and forms used to facilitate calculations are summarized
in \Sec{sect_prop}.  We examine in \Sec{sect_mesons}  the GCM description
of the Goldstone boson sector, selected low energy chiral observables and
methods used for other mesons.  Electromagnetic coupling is developed in 
\Sec{sect_EM} where recent calculations of the pion and kaon charge form
factors are reviewed.  Meson interaction and transition form factors are
discussed in \Sec{sect_interact}.  In \Sec{sect_loops} selected issues beyond 
meson tree level are considered and these include vector meson dominance processes
and pion loop contributions.  Concluding remarks are presented in \Sec{sect_conclude}.

\sect{Global Color Model}
\label{sect_gcm}

The action for the GCM is defined in Euclidean metric as (Cahill and Roberts,
1985; Praschifka \etal, 1987a) 
\beq
S_{GCM}[\bar{q},q]=\int d^{4}x \; \bar{q}(x)
\bigl( \gamma \cdot \partial _{x} +m  \bigr) q(x)
 + \case{1}{2} \int d^{4}xd^{4}y \; j_{\mu
}^{a}(x) \, g^2 D_{\mu \nu}(x-y) \, j_{\nu }^{a}(y) ,\label{sgcm}
\eeq
where $m$ is a diagonal matrix of bare quark masses, 
and $j_{\nu }^{a}(x)=\bar{q}(x)\frac{\lambda ^{a}}{2}\gamma _{\nu}q(x)$ 
is the quark colored current.  The employed Euclidean metric is such that $a 
\cdot b= a_\mu b_\mu$, spacelike vectors satisfy \mbox{$a^2 > 0$}, and
\mbox{$\{\gamma_\mu ,\gamma_\nu \}=$}\mbox{$2\delta_{\mu\nu}$}, with
\mbox{$\gamma_\mu= \gamma_\mu^\dagger$}. 
A model field theory of this form can summarize a number of diverse 
investigations in the modeling of hadron physics such as Cahill and Roberts 
(1985), McKay and Munczek (1985), Barducci \etal~(1988), and McKay 
\etal~(1989).  
This model may be viewed as a truncation of QCD and some consideration of that
aspect can indicate the content and role that should be attributed to the 
phenomenological gluon two point function $D_{\mu\nu}$.  The connections 
between QCD and
\Eq{sgcm} have been considered by Cahill and Roberts (1985) and by McKay \etal
~(1989) and a brief summary follows.  Functional integral methods are
convenient for that purpose.  In QCD (see, e.g. Marciano and Pagels, 1978)
the generating functional for quark propagators  is  
\beq
Z[\bar{\eta},\eta] =N \int D\bar{q}DqDA \:{\rm exp}
\left(-S[ \bar{q},q,A] + \int d^4x \;(\bar{\eta}q + \bar{q}\eta)   \right) , 
\label{zqcd}
\eeq
where 
\beq
S[ \bar{q},q,A] = \int d^4x \left(
\bar{q}( \dslash +m  -ig\frac{\lambda^a}{2} \Aslash^a )q 
+\case{1}{4} F^a_{\mu\nu}F^a_{\mu\nu}  \right)  ,
\label{sqcd}
\eeq
and \mbox{$F^a_{\mu\nu} = \partial_{\mu}A^a_{\nu} - \partial_{\nu}A^a_{\mu}
+ g f^{abc}A^b_{\mu}A^c_{\nu}$}.  We leave the gauge fixing term,
the ghost field term and its integration measure to be understood.  
If we introduce
\beq
{\rm exp}\left(W[J]\right)= \int DA \: {\rm exp} \left( - \int d^4x \: 
( \case{1}{4} F^a_{\mu\nu}F^a_{\mu\nu} - J^a_{\mu} A^a_{\mu} ) \right)   ,
\eeq
the generating functional becomes
\beq
Z[\bar{\eta},\eta] =N \int D\bar{q}Dq \:
{\rm exp}\left( - \bar{q}( \dslash +m)q  + \bar{\eta}q + \bar{q}\eta \right) \;
{\rm exp} \left( W \left[ ig\bar{q}\frac{\lambda^a}{2}\gamma_{\mu}q \right] 
\right)  ~,\label{zqcd2}
\eeq
where the abbreviated notation implies the usual spacetime integration.
The functional $W[J]$ is the generator of connected gluon $n$-point functions
without quark-loop contributions.  It may be written as
\beq
W[J] = \case{1}{2} \int d^4x d^4y \: J^a_{\mu}(x) D_{\mu \nu}(x-y) 
J^a_{\nu}(y) + W_R[J]  
\label{w}
\eeq
allowing the formal factorization
\beq
Z[\bar{\eta},\eta] = {\rm exp} \left( W_R \left[ ig \frac{\delta}{\delta\eta}
\frac{\lambda^a}{2} \gamma_{\mu} \frac{\delta}{\delta\bar{\eta}} \right] \right)
Z_{GCM}[\bar{\eta},\eta]  .
\label{zqcdf}
\eeq
Here the quark field variables in the argument of $W_R$ have been replaced by
the corresponding source derivatives.  
The truncated model identified this way is generated from
\beq
Z_{GCM}[\bar{\eta},\eta]= N\int D\bar{q} Dq \:
{\rm exp}(-S_{GCM}[\bar{q},q] + \bar{\eta}q + \bar{q}\eta )  ,
\label{zgcm}
\eeq
where $S_{GCM}[\bar{q}q]$, is given by \Eq{sgcm} in which the dependence upon
the coupling constant $g$ has been made explicit.
The GCM corresponds to the retention of just the first term 
of \Eq{w}, involving the gluon $2$-point function
excluding $\bar q q$ vacuum polarization contributions.   The neglected term
$W_R$ involves the $n$-point functions, \mbox{$n \geq 3$},  from the pure 
gauge sector.  Whether this can successfully model soft hadron physics can 
only be judged by  consideration of the results that follow from a consistent 
phenomenological form for $ g^2 D_{\mu\nu}(x-y)$.   Certainly, any element of
hadron physics that absolutely requires the gluonic coupling of three or more
quark currents can be expected to resist description via the GCM; that finding
in itself would constitute a valuable contribution.  On the other hand, if
a significant body of information can be correlated through a phenomenological
gluon $2$-point function via accessible continuum calculations,  this might
highlight those aspects that deserve more concentrated attention through
lattice-QCD methods (Rothe, 1992).  Even though the model in \Eq{sgcm} 
looks Abelian in
form (QED may be recast that way),  the non-Abelian nature of the
gauge field sector has not been totally ignored.  Those non-Abelian aspects
that contribute to the dressed pure-glue $2$-point function can be implemented
through the phenomenological element of the GCM.  

\subsect{Bosonization to Mesons}

The auxiliary field method consists of a transformation of a field theory 
that introduces additional field variables through which important dynamical
features can be revealed more readily than with the original fields.  
The method
has found wide use in statistical mechanics as the Hubbard-Stratonovich
transformation (Stratonovich, 1957; Hubbard, 1959).   Early applications 
of the auxiliary field method to quark dynamics 
(Eguchi and Sugawara, 1974; Kikkawa, 1976; Kleinert, 1976a) were applied to 
four-fermion interaction models of the NJL type.       
Here the auxiliary fields are introduced so as to assume the role of 
hadronic collective modes generated by combinations of $\bar{q}(x)$ and 
$q(x)$ at the same space-time point. 
The four-fermion interaction is exactly transformed into
boson mass terms plus a local Yukawa coupling of boson fields to the quarks.
The locality of the auxiliary boson fields is a consequence
of the contact nature of the original four-fermion interaction.  
Since there remains only a quadratic dependence upon the quark fields, they
could then be integrated out leaving a bosonized action.
Expansion of the action in quantum
fluctuations about the classical values of these composite fields was used 
to produce an effective action from which a number of important issues, e.g. 
renormalizability of four-fermion coupling (Eguchi, 1976) could
be addressed.  

It was subsequently observed (Kleinert, 1976b; Shrauner, 1977) that the 
auxiliary field method can be generalized to account for the intrinsic
nonlocality in the quark couplings generated by models that have
an Abelian vector gauge field coupled to quarks.  The vector gauge field  
can be integrated out to produce a finite range four-fermion interaction.
The auxiliary boson 
fields needed to reduce this interaction to quadratic form are bilocal 
combinations of $\bar{q}(x)$ and $q(y)$ whose free-field solutions 
turn out to be the ladder Bethe-Salpeter (BS) bound states.
Expansion of the bilocal field in terms of these BS free-field eigenmodes
produces (Kleinert, 1976b; Kugo, 1978) an effective action for the 
finite-size $\bar{q}q$ mesons of the model. 

In these works color was omitted and baryons were not considered.   
The achieved reformulations 
in terms of $\bar{q}q$ objects constitute a true hadronization for the 
case of a single Abelian vector gauge field coupled to any number of flavors 
of quarks.  Such a field theory shares many properties in common with the
strong interaction such as the conserved vector currents, 
global $SU(N_f) \otimes SU(N_f)$ chiral symmetry, and the consequent PCAC.  
Nevertheless, to obtain insight into approximate methods for QCD in the
nonperturbative domain of hadron physics, a study was begun 
(Cahill and Roberts, 1985) on the bilocal field hadronization method for 
colored quarks that interact through finite range coupling of their colored
vector currents.  

We shall consider only the Feynman-like gauge where 
\mbox{$ g^2 D_{\mu \nu}(x-y) = \delta_{\mu \nu} D(x-y)$} and an Ansatz for $D$ 
defines the model.  This is done to permit the 
simplest form of the reordering identities that are to follow.  It should
not be taken to imply that the intent is to match with information from 
a corresponding gauge choice in QCD.  Such a choice is not possible as the 
above Ansatz will depart from the perturbative form in both transverse and
longitudinal parts while the latter will have no interaction or dressing in QCD.   
A reformulation of $S_{GCM}$ in terms of fields that represent $\bar{q}q$ 
objects is achieved by the Fierz reordering identities 
\beq
\gamma^{\mu}_{ij} \gamma^{\mu}_{kl} = K^a_{il} K^a_{kj} \; , \; \; \;
K^a = ( {\bf 1},i\gamma_{5},\case{i}{\sqrt{2}}\gamma _{\nu },
\case{i}{\sqrt{2}}\gamma _{\nu}\gamma _{5} )  
\label{mesonK}
\eeq
for spin,
\beq
\delta_{ij} \delta _{kl} = F^b_{il} F^b_{kj} \; , \; \; \;
F^b_{N_f=2} = ( \case{{\bf 1} }{\sqrt{2}},\case{\vec{\tau } }{\sqrt{2}} ) \; , 
\; \; \; F^b_{N_f=3} =( \case{{\bf 1} }{\sqrt{3}},\case{\lambda^a  }{\sqrt{2}} )
\label{mesonF}
\eeq
for flavor $SU(2)$ singlet plus triplet  (${\bf 1}_f + {\bf 3}_f$), or $SU(3)$ 
singlet plus octet (${\bf 1}_f + {\bf 8}_f$), and 
\beq
\lambda^a_{ij} \lambda^a_{kl} = C^c_{il} C^c_{kj} \; , \; \; \;
C^c = ( \case{4}{3}{\bf 1},\case{i}{\sqrt{3}}\lambda ^{a} )
\label{mesonC}
\eeq
for color $SU(3)$ singlet plus octet (${\bf 1}_c + {\bf 8}_c$).  
With the direct product set of matrices \mbox{$\Lambda^{\theta} 
= \frac{1}{2} K^{a}\otimes F^{b}\otimes C^{c}$} where 
\mbox{$\theta = (a,b,c)$},  the complete reordering needed to express the
the quartic quark current-current coupling in terms of a quadratic coupling of 
$\bar{q}q$ objects is
\beq
\left(\case{\lambda^a}{2}\gamma_{\mu} {\bf 1}_F\right)_{ij}
\left(\case{\lambda^a}{2}\gamma_{\mu} {\bf 1}_F\right)_{kl} =
\Lambda^{\theta}_{il} \; \Lambda^{\theta}_{kj} ,
\label{fierzm1}
\eeq
and we use the summation convention for repeated indices. 
Hence in \Eq{sgcm}, this reordering causes $q(x)$ to form a scalar product with 
$\bar q (y)$ and $\bar q (x)$ with $q(y)$.   
An equivalent form that is useful later is
\beq
\case{\lambda^a}{2}\gamma_{\mu} {\bf 1}_F \: M \:
\case{\lambda^a}{2}\gamma_{\mu} {\bf 1}_F =
\Lambda^{\theta} \: {\rm tr} \: (\Lambda^{\theta} M)
\label{fierzm2}
\eeq
where $M$ is any matrix.   
The Dirac matrix $\sigma_{\mu \nu}$ does not appear in \Eq{mesonK} due to 
the use of the Feynman-like gauge.   If one chooses a family of gauges 
parameterized by $\xi$, a more complicated Fierz reordering involving
$\sigma_{\mu \nu}$ and $\xi$-dependent weights of the spin matrices is
obtained. Because of the phenomenological nature of the effective gluon
propagator here, the simplifications of the Feynman-like gauge are
appealing.

With \Eq{fierzm1} the GCM action in \Eq{sgcm} can be written as
\beq
S[\bar{q},q]=\int d^{4}x \: \bar{q}(x)
\bigl( \gamma \cdot \partial _{x} +m  \bigr) q(x)
 - \case{1}{2} \int d^{4}xd^{4}y \: {\cal J}^{\theta}(x,y)^*
D(x-y) {\cal J}^{\theta}(x,y) ,\label{sgcm2}
\eeq
where the negative sign arises from the anticommuting property of the 
Grassmann variables $q$ and $\bar{q}$.  There appear now   
bilocal currents \mbox{${\cal J}^{\theta}(x,y) = $} \mbox{$ \bar{q}(x)\Lambda^{\theta}q(y)$}
that are hermitian, i.e. 
\mbox{$[{\cal J}^{\theta}(x,y)]^* = $} \mbox{$ {\cal J}^{\theta}(y,x)$}, and carry the 
Lorentz and flavor indices of the candidate $\bar{q}q$ states of this model.
However the color structure of the $\bar q q$ combinations consists of the 
singlet ${\bf 1}_c$ as well as
the octet ${\bf 8}_c$.  The latter may be eliminated in favor of diquark
correlations as discussed later.

As a specific illustration of the content of the interaction in \Eq{sgcm2},
consider, for $N_f=2$, the  point coupling limit where 
\mbox{$D(x-y) \rightarrow \tilde G \delta(x-y)$}.   The
$\Lambda^\theta$ in the color singlet channel are 
\beq
\Lambda^\theta = \case{\sqrt{2}} {3} \tau^\alpha \left\{ {\bf 1}_{D},i\gamma
_{5},\case{i}{\sqrt{2}} \gamma _{\nu }, \case{i} {\sqrt{2}} 
\gamma _{\nu } \gamma _{5} \right\}  ,\label{lam2}
\eeq
where \mbox{$\alpha =0,1,2,3$} and \mbox{$\tau^0 = 1$}. 
It is convenient to separate out numerical factors $n_\theta$ and deal with
the canonical set of matrices $\kappa_\theta$ defined by
\beq
\kappa_\theta = \frac{\Lambda^\theta}{n_\theta} = \tau^\alpha \left\{ 
{\bf 1}_{D},i\gamma_{5},i \gamma _\nu, 
i \gamma _{\nu } \gamma _{5} \right\}  .\label{kappa}
\eeq
Then the interaction term from \Eq{sgcm2} in the color singlet sector becomes
\beq
S_{int}^{NJL} = - \case{1}{2} \int d^4x \: \left\{ 
 G_S \left[ (\bar{q}\tau^\alpha q)^2 + (\bar{q} i\gamma_5 \tau^\alpha 
                                                                  q)^2\right]
+G_V \left[ (\bar{q}i\gamma_\nu \tau^\alpha q)^2 + 
   (\bar{q} i\gamma_\nu \gamma_5 \tau^\alpha q)^2\right] \right\}  .
\label{snjl}
\eeq
Here \mbox{$G_S = \frac{2}{9} \tilde G$}, and 
\mbox{$G_V = \frac{1}{9} \tilde G$}.  This is
one of the simplest forms of the  Nambu--Jona-Lasinio model (NJL) (Nambu and
Jona-Lasinio, 1961) as applied to quarks in the color singlet channel.  
For recent reviews of the NJL model as applied to hadron physics, see
Vogl and Weise (1991) and Klevansky (1992).   The
coupling constant relation \mbox{$G_S= 2 G_V$} evident here has its 
origin in the
simple coupling of quark color currents.  Applications are usually made 
from the point of view of an effective model that implements chiral symmetry 
and its dynamical breaking;
the constants $G_S$ and $G_V$ are allowed to be independent and are fitted
to data.   The GCM retains the distributed coupling of the $\bar{q}q$ 
correlations via an effective gluon $2$-point function $D(x-y)$ and also 
maintains the relative Fierz weighting that accompanies that function to each 
meson channel.  For large momentum $q^2$, the magnitude of $D(q^2)$ 
in the GCM is fixed by the requirement that 
perturbative QCD should be obeyed; this is not possible in the NJL since
the constants $G_S$ and $G_V$ force the same coupling strength in the UV as
in the IR.  The role asked of phenomenology is quite different in the two
models and the outcome can also be quite different. 

We return now to \Eq{sgcm2} and  before restructuring it, we note that
the interaction term in \Eq{sgcm2} is explicitly chirally symmetric as is its
original form.  However that is not true for a fixed value of the summation
index $\theta$.  If one is to truncate that sum, it is necessary to retain
a subset terms that mix under the chiral transformation to preserve the 
symmetry.  Similarly for any symmetry e.g. an electromagnetic gauge 
transformation. 
The four-fermion term of the GCM action in \Eq{sgcm2}, when viewed as the 
product of two
bilocal currents, may now be recast through integration over auxiliary
bilocal field variables that couple in a linear fashion to those bilocal
currents.  The functional path integral formulation is convenient for
illustrating this.  
The generating functional for the quark propagators  or $n$-point functions
can be written as
\beq
Z[\bar{\eta},\eta] =N \int D\bar{q}Dq \:
{\rm exp} \left\{ -(\bar{q} ( \gamma \cdot \partial + m )  q) 
+ \case{1}{2} ( {\cal J}^* D {\cal J} )
+ (\bar{\eta}q +\bar{q}\eta) \right\} , 
\label{z2}
\eeq
where the bracket notation denotes the usual spacetime integrations and
summation over discrete indices as is appropriate to the context.  
For example
\beq
( {\cal J}^* D {\cal J} )
=  \int d^{4}xd^{4}y \: {\cal J}^{\theta }(y,x)  D(x-y) {\cal J}^{\theta }
(x,y) .
\eeq 
Since the normalization factor $N$ of the generating functional is irrelevant
for the quantities of interest derivable from it, we consider now the 
restructuring that can be produced by multiplication of the generating 
functional in \Eq{z2} by the constant
\beq
[det (D^{-1})]^{-\hlf} =
\int D{\cal B} \: {\rm exp} \left\{ - \int d^{4}xd^{4}y \: \frac{{\cal
B}^{\theta }(x,y){\cal B}^{\theta }(y,x)}{2D(x-y)}\right\} 
= \int D{\cal B} \: {\rm exp} 
\left( -\case{1}{2}({\cal B}^* D^{-1} {\cal B} ) \right) .
\label{det}
\eeq
Here one chooses auxiliary field variables ${\cal B}^{\theta }(x,y)$ that have 
the same transformation properties as the associated bilocal currents 
${\cal J}^{\theta}(y,x)$.
In particular, if the flavor basis is hermitian, these fields 
are hermitian, that is, 
\mbox{$[{\cal B}^{\theta}(x,y)]^{*} = {\cal B}^{\theta}(y,x)$}.  
The integration measure is
\mbox{$\int D{\cal B} \equiv \int D{\cal B}^a D{\cal B}^b \cdots$}.   The
result \Eq{det} is a direct generalization of the gaussian integral over 
a real, local field.  After \Eq{z2} is multiplied by \Eq{det}, the 
shift of integration variables 
${\cal B}^{\theta }(x,y)\rightarrow {\cal B}^{\theta}(x,y)
+D(x-y){\cal J}^{\theta }(y,x)$,
will eliminate the current-current (four-fermion) term 
$\hlf ( {\cal J}^* D {\cal J} )$ in favor of a Yukawa coupling term
\mbox{$({\cal J} {\cal B}) \equiv 
(\bar{q}\Lambda^{\theta}{\cal B}^{\theta}q)$} 
and the original quadratic term in ${\cal B}$.

The generating functional is now given by
\beq
Z[\bar{\eta},\eta] =N' \int D\bar{q}DqD{\cal B} \: {\rm exp} 
\left( -(\bar{q} G^{-1}[{\cal B}] q) - \case{1}{2} ( {\cal B}^* D^{-1} 
{\cal B} )   + (\bar{\eta}q +\bar{q}\eta) \right) , 
\label{z3}
\eeq
where the object similar to an  inverse quark propagator  that occurs here is
\beq
G^{-1}(x,y)=\left( \gamma \cdot \partial _{x}+m \right) \delta(x-y)
+\Lambda ^{\theta }{\cal B}^{\theta }(x,y). 
\label{2a18}
\eeq
We thus have
\mbox{$Z[\bar{\eta},\eta]=N'\int D\bar{q}DqD{\cal B} \; {\rm exp} 
\left( -S[ \bar{q},q,{\cal B}] + (\bar{\eta}q +\bar{q}\eta) \right)  $}
where the new action is
\beq
S[ \bar{q},q,{\cal B}] =\int
d^{4}xd^{4}y \: \Biggl\{ \bar{q}(x)\; G^{-1}(x,y) \; q(y) +
\frac{{\cal B}^{\theta }(x,y) \; {\cal B}^{\theta }(y,x)}{2D(x-y)}\Biggr\} .
\label{sqqB}
\eeq
The introduction of auxiliary boson fields thus allows the quark fields to be
integrated out since they appear now  only in a quadratic form.  
This integration is
\beq
\int D\bar{q}Dq \: {\rm exp} \left( -(\bar{q} G^{-1} q) 
+ (\bar{\eta}q +\bar{q}\eta) \right) =   
{\rm exp} \left( {\rm Tr Ln} G^{-1}[{\cal B}] + (\bar{\eta}G\eta) \right)~,
\label{qint}
\eeq
where the fermion determinant has been expressed as 
\mbox{$ {\rm Det}( G^{-1}) =  {\rm exp} \; {\rm Tr Ln} G^{-1}$}.   This yields
\beq
Z[\bar{\eta}, \eta] =  N\int D{\cal B} \: {\rm exp} \left( -{\cal S}[ {\cal B}] 
+ (\bar{\eta}G\eta) \right) ,
\label{z4}
\eeq
where the completely bosonized action is
\beq
{\cal S}[ {\cal B}] = -{\rm TrLn} \, G^{-1}[{\cal B}] 
+  \int d^{4}xd^{4}y \: \frac{{\cal
B}^{\theta }(x,y){\cal B}^{\theta }(y,x)}{2D(x-y)} . 
\label{s4}
\eeq
This result is an exact functional change of field variables from
quarks to $\bar{q}q$ bosons that preserves the physical content of the 
original action.

The point coupling or NJL limit of this action is more familiar and can help
illustrate the content of the more general distributed case.  With
\mbox{$D(x-y) \rightarrow \tilde G \delta(x-y)$}, the necessary  auxiliary
fields take the special form 
\beq
{\cal B}^{\theta }(x,y) \rightarrow \frac{\phi^\theta (x)} {n_\theta} 
\delta(x-y) ,
\eeq
where $n_\theta$ are the purely numerical factors introduced by the Fierz 
reordering such that $\Lambda^\theta = n_\theta \kappa_\theta$ (no summation),
where $\kappa_\theta$ are Dirac matrix covariants (see \Eq{kappa}).  
Thus for $N_f = 2$, \Eq{sqqB} reduces to 
\beqar
S^{NJL} [\bar q, q ,\phi] = \int d^4 x \; \Biggl\{ \bar q \; ( \gamma \cdot 
\partial &+& m +\sigma + i\gamma_5 \vec{\tau} \cdot \vec{\pi} + i\gamma 
\cdot \omega + \cdots ) \; q  \Biggr. \nonumber \\
&+& \Biggl.  \frac{\sigma^2 + \pi^2}{2G_S} + \frac{\omega^2 + \rho^2}{2G_V} 
+ \cdots \Biggr\}   , \label{snjl2}
\eeqar
where $G_\theta = n_\theta^2 \tilde G$.  After integration over the quark 
fields, the point coupling limit of \Eq{s4} is
\beq
{\cal S}^{NJL}[\sigma, \pi, \omega, \cdots] = -{\rm TrLn} \,
(\dslash + m +\sigma + i\gamma_5 \vec{\tau} \cdot \vec{\pi} + \cdots) +
\int d^4 x \; \frac{\sigma^2 + \pi^2}{2G_S} + \cdots ~. \label{snjl3}
\eeq
This is the familar bosonized form of the NJL model.  

With either the GCM or its point coupling limit, it is necessary to identify 
the classical ground state fields.  From \Eq{s4} which covers both cases, 
the classical field configurations 
${\cal B}^{\theta }_{0}(x,y)$ are defined by the saddle point condition 
$\delta {\cal S} / \delta{\cal B}^{\theta }_{0}(x,y) = 0$ and   
translation invariant solutions represent the classical vacuum about which 
quantum fluctuations may be developed.  After functional differentiation of 
\Eq{s4} this equation of motion is
\beq
{\cal B}^{\theta }_{0}(x-y)=D(x-y)\; {\rm tr} \left[ \Lambda ^{\theta }
S(x-y)\right] ,
\label{saddle}
\eeq
where the associated propagator $S$ depends
self-consistently on ${\cal B}^{\theta }_{0}$. In particular,
\begin{equation}
S^{-1}(x-y)= \left( \gamma \cdot \partial _{x}+m \right) \delta (x-y)
+\Sigma (x-y) ,
\label{2b3}
\end{equation}
where
$\Sigma (x-y) = \Lambda ^{\theta }{\cal B}^{\theta}_{0}(x-y)$ 
represents the quark self-energy to the present level of
treatment.  From \Eq{saddle} this self-energy  satisfies
\begin{equation}
\Sigma (x-y) = D(x-y) \Lambda^{\theta}\; {\rm tr}\; \left[ \Lambda ^{\theta }
S(x-y)\right] = \case{4}{3}D(x-y)\gamma _{\nu }S(x-y)\gamma_{\nu }.
\label{sde}
\end{equation}   
The second form is obtained by reversing the Fierz reordering using 
\Eq{fierzm2} and carrying out the color sum to produce  
\mbox{$\frac{\lambda^a}{2} \frac{\lambda^a}{2} = \frac{4}{3}$}.
\Eq{sde} is the Dyson-Schwinger equation (DSE) with a bare
vertex, i.e. the ladder or rainbow approximation.  This illustrates one of  
the efficiencies of the bilocal field method.  Treatment of the bilocal fields
${\cal B}^{\theta}(x-y)$ at the classical level generates a quark self-energy
that sums an infinite subset of quantum loop Feynman diagrams in terms of the
original quark and gluon fields.  

The momentum representation defined by
\mbox{$f(r)=\int \dq e^{iq \cdot r} f(q)$}
allows the inverse propagator to be written as
\mbox{$S^{-1}(q)=$} \mbox{$i\qslash A(q^2)+$} \mbox{$B(q^2)+m$}.  
The rainbow DSE for the Dirac scalar and vector amplitudes of $\Sigma$ 
obtained from \Eq{sde} are
\begin{equation}
\left[ A(p^{2})-1\right] p^{2} = \case{8}{3} \int \dq
 D(p-q)\frac{ A(q^{2})q\cdot p }{ q^2A^2(q^2)+(B(q^2)+m)^2 },
\label{a}
\end{equation}
and
\begin{equation}
B(p^{2})=\case{16}{3} \int \dq
 D(p-q) \frac{ B(q^2)+ m }{ q^2 A^2(q^2)+(B(q^2)+m)^2 }  .
\label{b}
\end{equation}
It is useful to note that the only non-zero classical auxiliary fields 
${\cal B}_0^\theta$ are those for which the index $\theta$ corresponds to 
${\bf 1}_c$ for color, ${\bf 1}_f$ for flavor;  for Dirac spin, there is only
the scalar $1_D$ (which gives $B(p^2)$) and vector $\gamma_\mu$ (which 
gives $A(p^2)-1$). 

In the point coupling or NJL limit, \mbox{$D(k) = \tilde G$}, and the rainbow 
DSE gives \mbox{$A=1$} and \mbox{$ B(p^2) \rightarrow \tilde M$} where
\beq
\tilde M = \case{16}{3} \tilde G \int^\Lambda  \dq
  \frac{ \tilde M + m }{ q^2 +(\tilde M +m)^2 } . 
\label{Mnjl}
\eeq
We have indicated the cutoff that is obviously needed here.  Note that the (point 
coupling) NJL model leads to a free dressed quark propagator with mass 
\mbox{$M(\Lambda)=\tilde M (\Lambda) + m$}.  The cutoff dependence cannot be 
removed and $\Lambda$ becomes a parameter of the non-renormalizable NJL model. 
In contrast, the finite range property of the 
quark DSE, either \Eq{a} and \Eq{b} in the rainbow approximation needed at this 
stage in the GCM, or the exact DSE discussed later for the 
DSE approach, allows renormalization after regularization so that dependence upon
a regulator parameter $\Lambda$ is removed in favor of the renormalization scale
$\mu$ where boundary conditions are imposed.   The quark self-energy is necessarily
momentum dependent and constraints from perturbative QCD (pQCD) can be implemented.
There are a number of limitations that follow from the point coupling NJL limit 
and we will return to this topic in \Sec{sect_njl} after the contrasting features
of the GCM approach have been developed. 

The bosonized action in \Eq{s4} may be expanded about the classical 
configuration
in terms of the quantum fluctuation variables $\hat{{\cal B}}^{\theta }(x,y)=
{\cal B}^{\theta }(x,y)-{\cal B}^{\theta }_{0}(x-y)$ to define 
\mbox{$ \hat{\cal S}[\hat{ {\cal B} }] = {\cal S} [{\cal B}] 
- {\cal S} [{\cal B}_0] $} as the action for the quantum modes.   
With functional generalization of the expansion 
\mbox{$ {\rm ln}(1+x)=$} \mbox{$ \sum_n (-x)^{n+1}/n$}, and noting the fact that terms 
of first-order in fluctuations about an extremum cancel, one obtains
\begin{equation}
\hat{\cal S}\left[ \hat{{\cal B}}\right] = {\rm Tr} \sum
_{n=2}^{\infty }\frac{(-1)^{n}}{n}\left( S \Lambda^{\theta}
\hat{{\cal B}}^{\theta}\right) ^{n}
+ \int d^{4}xd^{4}y \: \frac{\hat{{\cal B}}^{\theta }(x,y)
\hat{{\cal B}}^{\theta }(y,x)}{2D(x-y)}.\label{shat}
\end{equation}
This form of action, truncated to selected low mass mesons, was used to
begin investigations nucleons as mean field solitons (Cahill and Roberts, 
1985), and to initiate 
estimates of meson couplings (Praschifka \etal, 1987a; 1987b).
Also studies and estimates of the possible properties of diquark correlations
in this framework were started (Praschifka \etal, 1989).  

Although only color singlet components are present in the classical field
configurations, the fluctuation fields $\hat{\cal B}^{\theta}$ in the present 
arrangement have, in general, both color octet and singlet components.  
The presence of ${\bf 8}_c$ as well as  ${\bf 1}_c$  $\bar{q}q$
fields in the meson bosonization  is not satisfying for several
reasons.  Color octet bosons are not physically realizable states and therefore
may only have a role as virtual constituents of hadrons.  However, as the
gluon exchange interaction is repulsive in that channel, such fields 
may not be efficient as effective dynamical variables.  Furthermore in order
to treat color singlet baryons, diquark correlations in color anti-triplets 
are needed.   A new change of field variables that leads 
to mesons and baryons was subsequently developed (Cahill, Praschifka and 
Burden, 1989; Reinhardt, 1990).  The method is based on the other possible
Fierz reordering of the original current-current interaction.   That is, 
in \Eq{sgcm}, this reordering causes $q(x)$ to form a scalar product with 
$q(y)$ and $\bar q (x)$ with $\bar q (y)$.   The color part of this Fierz
identity is
\beq
\lambda^a_{ij} \lambda^a_{kl} = \tilde C^0_{il} \tilde C^0_{kj} +
\tilde C^t_{ik} \tilde C^t_{lj}
\; , \; \; \; \tilde C^0 =  \sqrt{\case{4}{3}} {\bf 1} 
\; , \; \; \; \tilde C^t_{ik} =  \sqrt{\case{2}{3}} \epsilon_{tik}  ~,   
\label{mesdiqC}
\eeq
where $\epsilon_{tik}$ is the antisymmetric Levi-Civita tensor.   The first
term introduces  ${\bf 1}_c$ $\bar{q}q$ combinations and  the second term 
introduces $\bar{{\bf 3}}_c$ $qq$ diquark as well as the  corresponding 
${\bf 3}_c$ $\bar{q}\bar{q}$ combinations.  This leads to a bosonization 
in terms of meson and diquark bilocal fields and has been used to obtain   
a set of covariant Faddeev equations for baryons truncated to a quark-diquark format
(Cahill, 1989; Buck, Alkofer and Reinhardt, 1992).
For the purpose of discussing meson physics, the  ${\bf 1}_c$ sector obtained from   
the simpler $\bar{q}q$ meson bosonization is sufficient. 

The quadratic term of the action \Eq{shat} defines the propagator and the
corresponding free field solutions indicate the physical interpretation
required of the bilocal meson fields.  For this purpose, it is convenient
to use a momentum representation in which total meson momentum $P$  is
the Fourier transform conjugate of $R=(x+y)/2$, and the relative $\bar{q}-q$ 
momentum is $q$.  For the equal mass case (e.g. pions with $m_u=m_d$), $q$ 
is the conjugate of $r=x-y$.   Then the hermitian property 
\mbox{$[{\cal B}^\theta(r;R)]^{*} = {\cal B}^\theta(-r;R)$} is equivalent to
\mbox{$[{\cal B}^\theta(q;P)]^{*}={\cal B}^\theta(q;-P)$}. 
In the cases with unequal masses for $\bar{q}$ and $q$, such as the kaon, the
choice of relative momentum variable will be discussed as the need arises.
The choice of non-hermitian flavor basis needed to form charge eigenstates
leads to \mbox{$[{\cal B}^\theta(q;P)]^{*}=\bar{{\cal B}}^\theta(q;-P)$} 
where $\bar{{\cal B}}$ is the antiparticle field corresponding to ${\cal B}$.

The quadratic term of \Eq{shat} can be written as 
\beq
\hat{\cal S}_{2}[\hat{{\cal B}}] = \case{1}{2} \sum_{\theta'\theta} \int 
\dPqq \:
\hat{{\cal B}}^{\theta'}(q';P)^* \; {\cal D}^{-1}_{\theta'\theta}(q',q;P) \; 
\hat{{\cal B}}^\theta(q;P), 
\label{S2}
\eeq
where the inverse propagator is
\beq
{\cal D}^{-1}_{\theta'\theta}(q',q;P) =  \delta_{\theta'\theta} D^{-1}(q'-q) 
+(2\pi)^4 \delta(q'-q)\;
{\rm tr}\;\left[ \Lambda^{\theta'} S(q_+)\Lambda^\theta S(q_-)\right] 
,\label{dinv}
\eeq
with tr denoting a trace over spin, flavor and color and $q_\pm = q\pm P/2$.  
Note that $S(q)$ is a diagonal matrix in flavor space and that the
flavor component of the $\Lambda^{\theta}$ can be used to project onto 
the appropriate propagators for each flavor.
The first term of the inverse propagator in \Eq{dinv} is the momentum
representation of $1/D(r)$, while the second term is a quark loop with the 
momentum integral not yet carried out.  
The equation of motion for free mesons is
\mbox{$\delta \hat{\cal S}_2[\hat{\cal B}] / \delta 
\hat{ {\cal B} }^{\theta'} = 0$}  which yields
\beq
\sum_{\theta} \int \dq \: {\cal D}^{-1}_{\theta'\theta}(q',q;P_n) 
\hat\Gamma^{\theta}_n(q;P_n) = 0 .
\label{eqm}
\eeq
Here $P_n$ indicates that solutions (labeled by $\hat\Gamma_n$) are possible 
only for special (on-mass-shell) values of the total momentum.  
A more recognizable form is produced by multiplication in the operator 
sense by $D$ giving
\beq
\sum_{\theta} \int \dq D(q'-q) 
{\rm tr} \left[ \Lambda^{\theta'} S(q_+)\Lambda^\theta S(q_-)\right] 
\hat\Gamma_n^{\theta}(q;P_n) = - \hat\Gamma_n^{\theta'}(q';P_n) ,
\label{omsbs}
\eeq
which is the ladder BS equation separated into components via a trace 
technique.  To make this clear, use of the definition
\mbox{$\hat\Gamma_n(q;P)= \Lambda^{\theta} \hat\Gamma_n^{\theta}(q;P)$}, and 
the reversal of the Fierz reordering via \Eq{fierzm2}, converts 
\Eq{omsbs} to the standard ladder BS form 
\beq
\hat\Gamma_n(q';P_n)  =
-\int \dq D(q'-q) 
\case{\lambda^a}{2} \gamma_\nu S(q_+)\hat\Gamma_n(q;P_n) S(q_-) 
\case{\lambda^a}{2} \gamma_\nu   .
\label{omsbs2}
\eeq
The ladder BS amplitude is $\hat\Gamma_n(q;P)$ and it is a matrix of functions 
in the direct
product space of color, flavor and spin, corresponding to mesons with masses 
given by $P_n^2=-m_n^2$.  The solution label $n$ denotes the mass as well as 
the quantum numbers  $J^{\pi C},\: G$ etc.  In a given meson channel, e.g.  
$\pi, \: \rho, \: \omega \: $ etc, a number of Dirac spin matrices are 
involved through the various $\Lambda^{\theta}$ that are coupled in 
\Eq{omsbs} to form 
\mbox{$\hat\Gamma_n(q;P)= \Lambda^{\theta} \hat\Gamma_n^{\theta}(q;P)$}.  The 
$\hat\Gamma_n^{\theta}(q;P)$ are not all invariants  since in general they also
contribute factors of momentum that are contracted with the Dirac matrices 
of the $\Lambda^{\theta}$ to form  true covariants whose coefficients are
invariant amplitudes.  With the $\omega$-meson for example, 
the set of true covariants for 
$\Gamma_\nu (q;P)$ includes not only the canonical Dirac matrix $\gamma_\nu$,
but also  the covariant $q_\nu 1_D$ involving the Dirac unit matrix, and
\mbox{$q_\nu \gamma \cdot q$}, etc.

\subsect{Effective Local Meson Fields}
\label{sect_eigen}

The free bilocal meson field solutions (\Eq{eqm}) are on-mass-shell 
ladder BS bound states.  An expansion of the general interacting bilocal meson 
field variables in terms of the set of eigenmodes of that ladder BSE kernel 
serves to define a set of effective local meson field 
variables.  Essentially  the continuous internal $\bar{q}q$ degree of freedom
is decomposed into this basis.  The local fields 
will couple to quarks in a well-defined distributed fashion described by
vertex amplitudes that are determined by the ladder BSE kernel.  This
technique is an important step to both clarify the physical content of the
bilocal meson fields and to provide a link to familiar local field models and
methods.   This seems
to have been first exploited by Kleinert (1976b) and Kugo (1978).  Here we 
will follow a variation on the eigenmode expansion method as formulated by
Cahill (1989).

It is helpful to introduce an abbreviated notation so that the inverse 
meson propagator \Eq{dinv} can be expressed as 
\beq
{\cal D}^{-1}(P) = D^{-1} + {\cal G}(P) ,
\label{dinvop}
\eeq
in terms of quantities that are  operators in the space with the completeness 
relation \mbox{${\bf 1}=\sum_{\theta} \int \dq$} 
\mbox{$ |\theta, q \rangle \langle \theta, q|$} and   
normalization \mbox{$\langle\theta',q'| \theta, q\rangle =$}
\mbox{$(2\pi)^4\;\delta_{\theta'\theta} 
\delta(q'-q)$}.  
In this operator form, $D^{-1}$ is diagonal in $\theta$, while ${\cal G}(P)$
is diagonal in $q$.  Then, the extension of the ladder BS \Eq{omsbs} into an
eigenvalue problem defined for any $P^2$, namely
\beq
\sum_{\theta} \int \dq D(q'-q) {\cal G}_{\theta' \theta}(q;P) 
\hat\Gamma_n^{\theta}(q;P) = (\alpha_n(P^2)-1) \hat\Gamma_n^{\theta'}(q';P) ,
\label{eigen}
\eeq
can be summarized as
\beq
D{\cal G}(P) | \hat\Gamma_n(P)\rangle 
= (\alpha_n(P^2)-1) | \hat\Gamma_n(P)\rangle . 
\label{eigenr}
\eeq
Here \mbox{$\hat\Gamma_n^{\theta}(q;P)\equiv \langle \theta, q | 
\hat\Gamma_n(P)\rangle$}.   We shall refer to $\alpha_n$ as the eigenvalues
and they  are dimensionless and   
physical on-mass-shell solutions correspond to
\mbox{$\alpha_n(P_n^2=-M_n^2)=0$}.  That relation will be seen to provide
the free equation of motion for the effective local meson fields that will
emerge.
The corresponding anti-meson solution 
\mbox{$ \bar{\hat\Gamma}(q;P) = [C^{-1}\hat\Gamma(-q;P) C]^T$} has the same  
eigenvalue.  In the limit of exact flavor symmetry, 
$\bar{\hat\Gamma}=\hat\Gamma$.
Due to the above-mentioned properties of the bilocal fields, the conjugate 
solution can be expressed as   
$\bar{\hat\Gamma}_n^{\theta}(q;-P) \equiv \langle \hat\Gamma_n(P)
|\theta,q\rangle$.   

The symmetry property \mbox{${\cal G}_{\theta' \theta}(q;P)=
{\cal G}_{\theta \theta'}(q;-P)$}
allows the conjugate eigenproblem to be written
\beq
\langle \hat\Gamma_n(P) | {\cal G}(P) D
= \langle \hat\Gamma_n(P) | (\alpha_n(P^2)-1) . 
\label{eigenl}
\eeq
With the convenient preliminary normalization,
\beq
\langle \hat\Gamma_m(P)|D^{-1} | \hat\Gamma_n(P)\rangle = \delta_{m,n}  , 
\label{prenorm}
\eeq
expansion in eigenmodes gives
\beq
D{\cal G}(P) = \sum_n |\hat\Gamma_n(P)\rangle (\alpha_n(P^2)-1)
\langle \hat\Gamma_n(P)|D^{-1} . 
\eeq
Thus the mode expansion of the bilocal inverse propagator is
\beq
{\cal D}^{-1}(P) = \sum_n D^{-1}| \hat\Gamma_n(P)\rangle \alpha_n(P^2)
\langle \hat\Gamma_n(P)| D^{-1}  .
\label{dinvexp}
\eeq
Also of use is the operator for the free propagator of $\bar q q$ correlations
that carry meson bound state quantum numbers.  In this formulation it is the
inverse of \Eq{dinvop} and this gives
\beq
{\cal D}(P) = D - D \; {\cal G}(P) \; {\cal D}(P) .
\label{dop}
\eeq
The more familiar form for this object can be recovered from this abbreviated
or operator notation as follows.  Form \mbox{$ \Lambda^{\theta^\prime} \langle 
\theta',q'| {\cal D}(P) |\theta, q\rangle \Lambda^\theta $} summed over discrete
indices, and use \Eq{fierzm1} to reverse the Fierz reordering.  Then the
first term of \Eq{dop} yields
\beq
{\cal D}_{\theta^\prime\theta}(q',q;P) = D(q^\prime - q) \;
\frac{\lambda^a}{2}\gamma_{\mu}\; \otimes \; \frac{\lambda^a}{2}\gamma_{\mu}
 + \cdots  ~, \label{oge}
\eeq
which is the one-gluon exchange term.  The obtained index structure of the
kernel  term produces the ladder summation. 
Thus it is the ladder $\bar q q$ scattering operator or T-matrix operator 
that plays the role of the propagator of the bilocal meson fields as we have
chosen to define them.  Other choices are possible but all should agree near
the mass-shell poles.  

\begin{figure}[ht]

\psrotatefirst
\centering{\
\psfig{figure=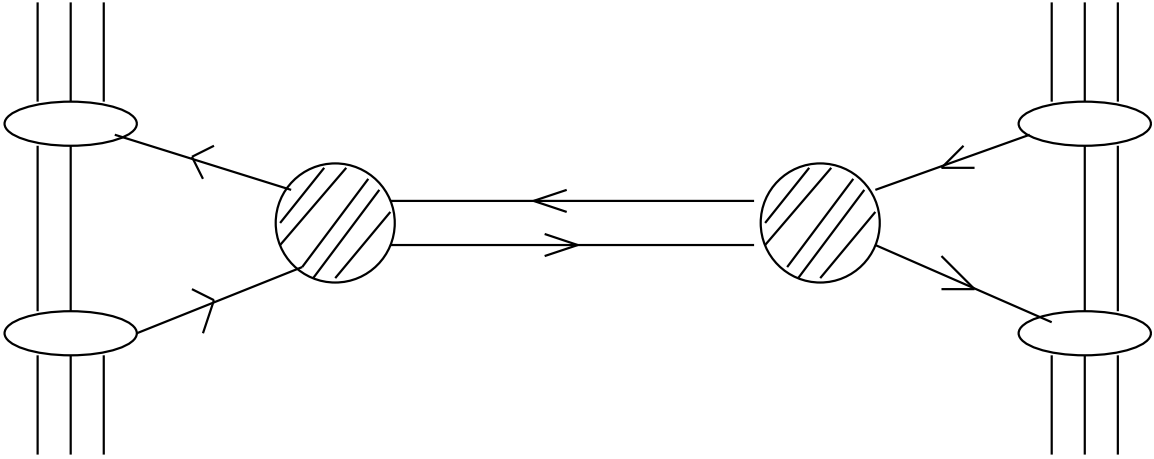,rheight=7.0cm,height=6.0cm,angle=-90} }  

\parbox{130mm}{ \caption{Exchange of a $\protect\bar q q$ correlation mode or meson
between sources which might be considered nucleons. \label{mexch} } }
\end{figure}

The eigenmode expansion, which is easily seen to be
\beq
{\cal D}(P) = \sum_n 
\frac{| \hat\Gamma_n(P)\rangle \; \langle \hat\Gamma_n(P)| }
{\alpha_n(P^2)} ~,  \label{dexp}
\eeq
can be useful for analysis of how the composite meson fields propagate from,
and couple to, given sources.   Some of the issues that emerge can be 
pointed out with reference to Fig.~\ref{mexch} and \Eq{dexp}.   
The dependence upon the total $t$-channel momentum $P$ is carried by the 
denominator $\alpha_n(P^2)$ (which can be associated with an effective
local propagator represented by the parallel $\bar q q$ lines in 
Fig.~\ref{mexch}), and also by the meson-$\bar q q$ vertex amplitudes $\hat 
\Gamma_n(q;P)$ (which are represented by the cross-hatched circles in 
the Figure).  These vertex amplitudes enter an integration over the quark 
amplitudes of the sources to produce effective meson sources which are 
momentum-dependent and constitute form factors.  For a given mode, the 
distribution of momentum dependence and normalization between numerator and 
denominator of the representation in \Eq{dexp} is quite an arbitrary choice.
This  can be  changed by a field redefinition which will also change the 
interactions accordingly if they are  consistently derived.  
Only at the pole is there no ambiguity.  Away from a pole the
question of what is the propagator of a $\bar q q$ correlation with the
quantum numbers of a given meson bound state cannot be answered uniquely by
any separation of numerator from denominator in \Eq{dexp} or by any separation
of one eigenmode in the sum from another.   The above eigenmode 
representation is probably not a feasible
route to calculations of the $\bar q q$ T-matrix. 
We view \Eq{dexp} as simply a formal
device to make contact with  effective local meson actions and models.
The point coupling counterpart of this T-matrix in the NJL model has been
used to investigate meson exchange mechanisms for the nucleon-nucleon 
interaction (Celenza \etal, 1996).  As we will see shortly, in that limit
there is usually only one mode per meson in \Eq{dexp} and the numerator in
\Eq{dexp} becomes a constant.  Thus the issues raised above do not arise 
because the dynamics is simpler.

Substitution of the mode expansion for ${\cal D}^{-1}(P)$ from \Eq{dinvexp}  
into the quadratic or free term \Eq{S2} of the bilocal field action yields
\beq
\hat{S}_{2}[\hat b] = \case{1}{2} \sum_n \int \dP \:
\hat b_n^\ast(P) \alpha_n(P^2) \hat b_n(P) , \label{S2b}  
\eeq
where the $\hat b_n(P)$ are effective local field variables defined by the 
projection  $\langle \hat\Gamma_n(P)|D^{-1}|\hat{{\cal B}}\rangle$, that is
\beq
\hat b_n(P) = \sum_{\theta} \int \dqq \:
\bar{\hat\Gamma}_n^{\theta}(q';-P)D^{-1}(q'-q) \hat{{\cal B}}^{\theta}(q;P)  .
\label{bp}
\eeq
That is, the bilocal fields $\hat{{\cal B}}^{\theta}(q;P)$ have been expanded in
terms of the complete set of internal BS eigenstates according to
\beq
\hat{{\cal B}}^{\theta}(q;P) = \sum_n \hat\Gamma_n^{\theta}(q;P) \hat b_n(P)  .
\label{mexp}
\eeq
The label $n$ characterizes the ladder Bethe-Salpeter spectrum obtained from
\Eq{eigen}.  The original set of bilocal fields ${\cal B}^\theta$ were
finite in number and labelled by the spin-flavor-color index $\theta$. The
eigenmode expansion has converted the continuous degree of freedom labelled
by $q$ into an infinite discrete classification included in the label $n$.
Also included in $n$ are the labels for the various spin-flavor 
representations that diagonalize the ladder Bethe-Salpeter kernel 
$D{\cal G}$.   

The form achieved in \Eq{S2b} is often loosely called a localization 
procedure.   However it does not ignore meson substructure but rather
produces a dynamically equivalent formulation
in terms of local field variables that experience nonlocal coupling through
the relevant BS amplitude.  According to \Eq{shat} the tree-level 
interactions
of the mesons are generated by a quark loop.  If the loop momentum is
$q$ the vertex describing the coupling of the local meson field $\hat b_n(P)$ 
to quarks is given by \Eq{mexp} as 
\beq
\bar{q} \Lambda^\theta \hat{{\cal B}}^{\theta}q = 
\sum_n \bar{q}(q_+) \hat\Gamma_n(q;P) \hat b_n(P) q(q_-)  . \label{coup}
\eeq
Note that it is the label $n$ that characterizes the meson, and not the label 
$\theta$, which identifies the various matrices making up the basis for the
BS amplitude $\hat\Gamma_n$

Physical normalization at the mass-shell is achieved by writing
\mbox{$\alpha_n(P^2)=$}\mbox{$(P^2+m_n^2(P^2)){\cal Z}_n^{-1}$} or equivalently 
\mbox{$\alpha_n(P^2)=$}\mbox{$(P^2+m_n^2){\cal Z}_n^{-1}(P^2)$}, where 
\mbox{$m_n = m_n(P^2=-m_n^2)$} are the physical masses.
Identification of the field renormalization constant
\beq
{\cal Z}_n = {\cal Z}_n(P^2=-m_n^2) = [ 
\alpha_n^\prime(-m_n^2)]^{-1}  ,
\label{Zn}
\eeq
produces the physical fields \mbox{$b_n(P)= \hat b_n(P)/\sqrt{ {\cal Z}_n}$}.
Then with a dynamical mass function defined by 
\beq
P^2 + m_n(P^2) = \alpha_n(P^2) {\cal Z}_n  ,
\label{massfn}
\eeq
we have
\beq
\hat{S}_{2}[b] = \case{1}{2} \sum_n \int \dP \:
b_n^\ast(P) (P^2 + m_n(P^2)) b_n(P) . 
\label{S2bn}  
\eeq
Note that the dynamical mass function $m_n(P^2)$, which appears because
the meson mode is not elementary,  has a zero first derivative 
at the mass-shell point.  That is, the residue at the propagator pole      
is unity.  This is equivalent to choosing a proper normalization for the
BS amplitude. To see this, we note due to the choice of
preliminary normalization in \Eq{prenorm}, the eigenvalue can be written as
\mbox{$\alpha_n(P^2) = 1 + $} \mbox{$
\langle \hat\Gamma_n(P)| {\cal G}(P) | \hat\Gamma_n(P)\rangle$},  that is
\beq
\alpha_n(P^2) = 1 + {\rm tr} \int \dq \: \bar{\hat\Gamma}_n(q;-P) 
S(q_+)  \hat\Gamma_n(q;P)  S(q_-)   .
\label{eigenv}
\eeq
Application of the derivative $\partial/ \partial P_\nu$, followed by 
multiplication of both sides by ${\cal Z}_n$ produces at the mass-shell  
\beq
2P_{\nu} = \left. \frac{\partial}{\partial P_{\nu}} {\rm tr} \int \dq \: 
\bar{\Gamma}_n(q;-K) S(q_+) \Gamma_n(q;K)    
S(q_-)\right| _{P^2=K^2=-m_n^2}  ,
\label{bsnorm}
\eeq
where we have defined
\beq
\Gamma_n(q;K) = \sqrt{{\cal Z}_n} \hat\Gamma_n(q;K)  .
\label{physbs}
\eeq
This is the physical BS amplitude since the relation in \Eq{bsnorm} satisfied 
by this quantity is the canonical BS normalization condition in the form 
appropriate to the BS kernel being independent of total momentum. (Itzykson 
and Zuber, 1980).  

The minimum information that needs to be carried forward from the above 
analysis is that the bilocal fields may be factorized into ladder 
BS amplitudes and corresponding local field variables as in \Eq{coup}. 
An explicit and simple form for the resulting meson action
can be presented if we make the approximation that only the
canonical Dirac matrix covariant is employed for each meson.  That is,
each meson BS amplitude is described by just one scalar function 
\mbox{$\hat\Gamma_{PS},\;\hat\Gamma_V,\;\cdots$}. 
We can leave the normalization of the internal amplitudes to be revisited later
and, for just the lowest mass modes,  
the meson action for effective local field variables
\mbox{$\vec{\pi}, \vec{\rho}_\mu, \omega, \cdots$} can then be written, by 
substituting \Eq{mexp} into Eq.~(\ref{shat}), as 
\beqar
\hat{\cal S}[\pi,\rho,\omega, \cdots]&=& 
{\rm Tr} \sum_{n=2}^\infty\frac{(-)^n}{n}
[S\;(i\gamma_5\vec{\tau}\cdot\vec{\pi}\hat{\Gamma}_{PS}
+i\gamma_\mu\omega_\mu \hat{\Gamma}_V
+i\gamma_\mu\vec{\tau}\cdot\vec{\rho}_\mu \hat{\Gamma}_V + \cdots)]^n 
\nonumber \\
& &+ 9\int d^4x~d^4y \;
\frac{\frac{1}{2}\vec{\pi}\cdot\vec{\pi} \, \hat{\Gamma}_\pi^2
+\omega^2 \, \hat{\Gamma}_V^2 +\vec{\rho}\cdot\vec{\rho} \, \hat{\Gamma}_V^2
+ \cdots}
{2D(x-y)}.
\label{mesonaction}
\eeqar
This dominant covariant approximation is for illustrative purposes as it
allows a simpler view of the essential physics.  It has been the common approximation
in most work within the GCM until recently.  

In general, a meson BS amplitude involves a number of distinct matrix
covariants.  For example, the physical pion BS amplitude has the form 
\mbox{$\Gamma_\pi = i\gamma_5 E_\pi(q;P) + i\gamma_5 \gamma \cdot P 
F_\pi(q;P) + \cdots$} where \mbox{$E_\pi,\; F_\pi,\;\cdots$} are scalar 
functions.  The canonical Dirac covariant, which does not require momenta for its
construction (e.g., $\gamma_5$ for the pion), is usually the dominant one as is 
confirmed  by studies of the BSE such as that by Jain and Munczek (1993).   The 
consequences of limiting the number of covariants allowed has not been subjected to
systematic investigation, especially for various observables that can be generated.
Some initial results in this line by Burden \etal~(1997) indicate that for the light
$\pi$ and $K$ mesons the sub-dominant pseudovector amplitude can contribute up to 
$15$\% to the mass and up to $35$\% to the weak decay constant.  For the next heaviest
mesons, the contributions are indicated to be about $5$\% and to decrease rapidly with
mass.  The more general case of many covariants is already covered by the BSE 
eigenmode analysis covered earlier in  this Section.  Thus with general flavor and spin
matrix BS amplitudes $\Gamma$, physically normalized, the action for lowest mass modes 
can be written  
\beqar
\hat{\cal S}[\pi,\rho,\omega, \cdots]&=& 
\case{1}{2} \sum_{n,m} \int \dP \:
\phi_n^\ast(P) \;  \Delta_{nm}^{-1}(P) \; \phi_n(P) 
\nonumber \\
& &+ {\rm Tr} \sum_{n=3}^\infty\frac{(-)^n}{n}
[S\; (\vec{\Gamma}_\pi\cdot\vec{\pi}
+\Gamma_\mu^\omega \omega_\mu 
+\vec{\Gamma}_\mu^\rho \cdot \vec{\rho}_\mu  + \cdots)]^n ~,
\label{mesaction}
\eeqar
where the local fields are \mbox{$\phi_n = \pi^i,\; \omega_\mu,\; \rho_\mu^i,
\; \cdots$}.  The generalized inverse propagators are
\beq
\Delta^{-1}_{nm}(P) = \delta_{nm}\, \left[ P^2 + m_n^2(P^2)\right] + \Pi_{nm}(P) ~,
\label{geninvpr}
\eeq
where the off-diagonal elements that exist give the relevant mixed self-energies.
It should be emphasized that the meson fields at this level are bare
or tree-level fields in the sense that quantum dressing effects from
the cubic and higher order couplings among meson  fields
have not been applied.  Nevertheless there is significant dynamical
content in these bare fields as evidenced by the ladder Bethe-Salpeter
structure they carry.  From \Eq{mesaction} one may extract the distributed
vertex function for a meson vertex of interest and its momentum structure
will depend upon the internal structure of the mesons as well as the 
gluon substructure within the dressed quark propagators.  
All interaction terms allowed by symmetries are present in \Eq{mesaction} and
as an illustration, the low order form that is generated from use of a 
derivative expansion, and has been investigated (Praschifka \etal, 1987a), is
\beqar
\hat{\cal S} [\pi,\rho,\omega, \cdots] = &&\int d^4x \left[ \case{1}{2}
[(\partial_{\mu}\pi)^2+m_{\pi}^2\pi^2] + \case{1}{2}[\vec{\rho}_{\mu\nu} \cdot
\vec{\rho}_{\mu\nu} + m_{\rho}^2 \vec{\rho} \cdot \vec{\rho}]
+ \case{1}{2}[\omega_{\mu\nu} \; \omega_{\mu\nu} + m_\omega^2 \omega^2]
 \right.  \\ \nonumber
 &&- \left.
 g_{\rho\pi\pi} \vec{\rho}_\mu \cdot \vec{\pi} \times \partial_\mu \vec{\pi}
-i \epsilon_{\mu\nu\sigma\tau}\omega_\mu\partial_\nu \vec{\pi} \cdot
 \partial_\sigma \vec{\pi} \times \partial_\tau \vec{\pi}
 -  i g_{\omega\rho\pi} \epsilon_{\mu\nu\sigma\tau} \omega_\mu
\partial_\nu \vec{\rho}_\sigma \cdot \partial_\tau \vec{\pi} \right.
\\ \nonumber
 &&+ \left. \case{i\lambda} {80\pi^2}
\epsilon_{\mu\nu\sigma\tau} {\rm tr} ( \vec{\pi}\cdot \vec{\tau} \partial_\mu
\vec{\pi}\cdot \vec{\tau} \partial_\nu \vec{\pi}\cdot \vec{\tau}
\partial_\sigma  \vec{\pi}\cdot \vec{\tau} \partial_\tau \vec{\pi}\cdot
\vec{\tau} )  + \cdots \right]  ~.
\label{gcmlocal}
\eeqar
Due to the quark loop structure, the chiral anomaly terms are properly embedded 
in this approach and will occur with the tensor $\epsilon_{\mu\nu\sigma\tau}$
and a factor of $i$ from the Euclidean metric. 

\subsect{NJL Limit of Local Meson Fields}
\label{sect_njl}

Local meson fields also arise in the bosonized form of the point coupling
NJL model.  To explore the different dynamical content at work within the GCM,
it is useful to summarize the point coupling limit of the analysis in the
previous Section.  With \mbox{$D(k) \rightarrow \tilde G$} and
\mbox{$\hat{{\cal B}}^{\theta}(q;P) \rightarrow \phi^\theta (P)/n_\theta$},
\Eq{S2} for $\hat{\cal S}_{2}[\hat{{\cal B}}]$ the quadratic term of
the action, becomes
\beq
\hat{S}_{2}[\phi] = \case{1}{2} \sum_{\theta^\prime \theta} \int \dP \:
\phi^{\theta^\prime\ast}(P) \; I^{-1}_{\theta^\prime \theta}(P) \; 
\phi^{\theta}(P) ~. \label{s2njl}
\eeq
Here, the point limit has identified
\beq
I^{-1}_{\theta^\prime \theta}(P) = \frac{1} {n_{\theta^\prime} n_\theta } \; 
\int^\Lambda \dqq \: {\cal D}^{-1}_{\theta^\prime \theta}(q',q;P)   ~,
\label{I}
\eeq
in terms of the bilocal inverse propagator  defined in \Eq{dinv}.   We have
indicated that this integral must be regulated.   One finds that
\mbox{$I^{-1}_{\theta^\prime \theta}(P) = \delta_{\theta^\prime \theta}  
G^{-1}_\theta + J_{\theta^\prime \theta}(P)$}
where the standard NJL model integral is
\beq
J_{\theta^\prime \theta}(P) = {\rm tr} \int^\Lambda  \dq \left[ 
\kappa_{\theta^\prime} \;S(q_+) \;\kappa_\theta \;S(q_-)\right]  ~.
\label{J}
\eeq
In general $J_{\theta^\prime \theta}(P)$ is a matrix that is diagonal except
for coupling between the  channels characterized by the Dirac matrices $1_D$
and $\gamma_5$ and also $\gamma_\nu$ and $\gamma_\nu \gamma_5$.  The
off-diagonal elements are often ignored in applications of the NJL model. In
this case, the connection to the previous eigenmode analysis for the distributed
coupling case is most transparent.  Since the point coupling limit of
\Eq{eigen} shows that the eigenvectors $\hat\Gamma_n^\theta(q;P)$ become
independent of $q$, and the diagonal assumption makes them independent of $P$,
we have \mbox{$\hat\Gamma_n^\theta \propto \delta_{n,\theta}$}.
That is, the modes are
characterized only by the label $\theta$ which distinguishes among the flavor
and Dirac matrices.  There is only one mass state for each type of meson, and
only a single covariant (the canonical one, i.e. $i\gamma_5$ for $\pi$,
$i\gamma_\nu$ for $\omega$, etc) appearing in the BS amplitude.

When off-diagonal aspects of $I^{-1}_{\theta^\prime \theta}(P)$ are
included, as they should be for generality, diagonalization is needed for 
physical fields.  This can be
summarized  as the point coupling limit of the distributed case
described in Sec.~\ref{sect_eigen}.  The relevant limit of \Eq{eigen} is
\beq
\sum_\theta  G_{\theta^\prime} J_{\theta^\prime \theta}(P)  
\hat\Gamma_n^\theta(P) =
(\alpha_n(P^2) -1) \hat\Gamma_n^{\theta^\prime} (P)  .\label{eigennjl}
\eeq
Thus  using the preliminary normalization  
\mbox{$\bar \Gamma_n^\theta \; G^{-1}_\theta \;\Gamma_n^\theta = \delta_{nm}$}
instead of the point limit of \Eq{prenorm},  one finds the expansion
\beq
I^{-1}_{\theta^\prime \theta}(P) = \sum_n  G^{-1}_{\theta^\prime}
\hat\Gamma_n^{\theta^\prime}(P) \; \alpha_n(P^2) \; 
\bar{\hat\Gamma}_n^\theta(-P) G^{-1}_\theta  , \label{J3}
\eeq
where 
\beq
\alpha_n(P^2) = 1 + {\rm tr}\int^\Lambda \dq \bar{\hat\Gamma}_n^{\theta^\prime}
(-P)
\left[ \kappa_{\theta^\prime} \;S(q_+) \;\kappa_\theta \;S(q_-)\right]
\hat\Gamma_n^\theta (P) 
\label{alnjl}
\eeq
is the point limit counterpart of \Eq{eigenv} from the GCM.
The quadratic term of the meson action from \Eq{s2njl}  therefore becomes
\beq
\hat{S}_{2}[\hat b] = \case{1}{2} \sum_n \int \dP \:
\hat b_n^\ast(P) \alpha_n(P^2) \hat b_n(P) , \label{S2bpt}
\eeq
which is identical in form to \Eq{S2b} for the distributed case.  However the
local fields here come from the expansion
\beq
\phi^\theta(P) = \sum_n \hat\Gamma^\theta_n(P) \; \hat b_n(P)  ,
\label{locexp}
\eeq
in terms of the small number of basis eigenvectors $\hat\Gamma^\theta_n$.  There
are no more than $2$ coupled terms since at most $2$ Dirac matrix covariants  
are coupled by $J_{\theta^\prime \theta}(P)$ given in \Eq{J}.  When such
off-diagonal coupling is ignored, the fields $\hat b_n(P)$ are just rescaled
$\phi^\theta(P)$ fields.  In either case, the analysis of \Eq{S2bpt} to 
physically
rescale the fields to produce unit residue at the mass pole proceeds in
the same way as previously discussed.   We note that the inverse of $I^{-1}$
satisfies \mbox{$I(P) = G - G\; J(P) \;I(P)$} in a matrix notation and this
is the $\bar q q$ T-matrix in the NJL model.  It is the point limit
counterpart of \Eq{dop} for the finite range ladder T-matrix operator 
${\cal D}(P)$ that arises in the GCM or truncated DSE approaches.   

Aspects in the description of meson physics can now be used to identify
several consequences of the point coupling NJL limit by comparison to the GCM/DSE
approach that retains the finite range nature of the interaction.  
For a successful NJL phenomenology, the cutoff $\Lambda$ is typically rather low: 
$\Lambda \sim 1~{\rm GeV}$.  The lack of an interaction at higher momentum suggests
that the physics will not be adequate for form factors of physical hadron processes
except at low momenta and for coupling constants.  The point coupling limit 
produces meson $\bar q q$ Bethe-Salpeter amplitudes that are momentum constants 
in the NJL.   The quark loop
integrals for meson interaction vertices will require a regulator parameter which 
is not necessarily the same as that for the self-energy.    In contrast to this,
the finite range or momentum dependent structure of the meson BS amplitudes in the
GCM/DSE approaches (and in QCD) provide natural regularization of many of the quark
loop integrals for meson vertex functions.   On a more fundamental level, the 
presence of a momentum cutoff in the various DSEs of the NJL model destroys the
translation invariance required for the variable shifts necessary to derive the
various Ward-Takahashi identities (Ward, 1950; Takahashi, 1957) such as the one 
relating the photon-quark vertex to the quark propagator.   An extension of the 
NJL model that modifies the point interaction by matching to a one-gluon-exchange 
tail for momenta beyond a typical hadronic scale ($\sim \Lambda_{QCD}$) has 
recently been put forward to address these issues (Langfeld, Kettner and 
Reinhardt, 1996).

The lack of a confinement capability in the NJL means that quark loops for 
physical meson 
processes will generate a spurious $\bar q q$ production threshold at 
\mbox{$s=4M^2$}.  For a typical constituent quark mass of about $300$~MeV,
mesons heavier than the kaon would receive a spurious width.  This problem can be
avoided in the GCM/DSE approaches by taking advantage of the  indications 
from studies of both the gluon and quark DSEs.   This is that the gluon $2$-point 
function shows sufficient enhancement in the IR to suggest confinement of both gluons 
and quarks through momentum-dependent dressed self-energies that
do not produce a physical mass-shell.  An extension of the NJL to add a 
phenomenological confinement interaction has been 
made recently by Celenza \etal~ (1995).   There the confinement interaction is used 
selectively to dress the meson-$\bar q q$ vertices but the quark self-energy remains 
a constant.   

\subsect{From Euclidean Metric to Physics}
\label{sect_eucl}

Some important considerations in the type of approach considered here are the 
following.  A typical first step is to use the effective meson action of
\Eq{mesaction} to produce, at meson tree-level, the meson self-energy
amplitudes  as integrals involving the dressed
quark propagator $S(p)$ and the ladder BS amplitudes $\Gamma$.
Equivalently, the inverse propagators for the meson modes are calculated.
An important structure that arises is a quark loop with two $\Gamma$
vertices and one independent external momentum.   A subsequent step is to
calculate an interaction term such as the $\rho\pi\pi$ vertex function or form
factor.  Here the calculation requires a quark loop with three $\Gamma$
vertices and two independent external momenta.  The approach here requires 
Euclidean
variables for the momenta and thus the above-mentioned loop integrals for
space-like values of external hadronic momenta are accommodated naturally.
It will be necessary however to produce physical quantities, such as masses
and coupling constants, defined at time-like values of external hadronic
momenta.  This has usually been done by an analytic continuation in the 
external momentum variables before evaluation of the loop integrals.
The momenta
occurring in the quark propagator parts of loop integrands are linear
combinations of the internal integration momentum and the external momenta.
Since one or more of the components of external momenta must be continued to
complex values, the quark propagators must be evaluated in a domain of the
complex momentum plane. 

This procedure is
well-defined for the loop integrals mentioned above in those works we review
where the dressed quark
propagator  is an entire function in order to implement quark
confinement.  The quark propagators will not contribute spurious production
cuts to the loop integrals.  The vertex functions of the loop integrals, being
ladder $\bar{q}q$ BS amplitudes, have no production cuts either.  
For meson tree-level quantities, the needed continuation in the external
momenta to define mass-shell quantities is not hindered by singularities.
Beyond tree-level, there is, for example, a $\pi\pi$ production cut in the
$\rho$ self-energy  that first
occurs at the one-meson-loop level of treatment.   These  quantities
are explicitly constructed  by a pion loop and the production
cuts are automatically generated by the pion loop integration of functions that
themselves have well-defined continuations previously established.

The form of the GCM field theory model, and indeed that of the more general
phenomenological approach based on the DSEs of QCD (Roberts and Williams,
1994), is such that
use of the Euclidean metric is necessary to draw upon the existing body of
experience and results from previous studies of the dressed quark and gluon
propagators.  Such results with applications to hadron physics, either from 
lattice-QCD or from continuum studies through 
the DSEs,  are invariably obtained in Euclidean metric for practical
reasons.  With the Minkowski metric, non-perturbative treatment of DSEs is extremely
difficult due to the singularity structure and the indefinite norm.   For the 
Bethe-Salpeter equation with scalars, a direct four-dimensional approach in Minkowski 
metric has only recently been developed and tested (Kusaka and Williams, 1995).   
There is no known treatment for fermions in that approach.   

The transcription of non-perturbative equations  (such as the DSEs)  from one metric 
to another is not necessarily equivalent to analytic continuation of the solution
amplitudes in the (external) momenta.  In the former case, establishment of the 
connection between solutions in each metric requires explicit
integration along the closed contour generated by the Wick rotation 
(Wick, 1954).  This is difficult except in very special cases.   
The singularity structure of the solutions in one metric dictates whether they
are related to the solutions of the equations that differ only in form due to
transcription into another metric.  In the case of approaches that use an 
entire function quark propagator, there is an essential singularity at infinity and
the Wick rotation cannot be used to justify change of metric by simple
transcription of the form of equations.   Analytic continuation of just the 
solution amplitudes (and not the total form of the non-perturbative equations) from
the space-like domain near \mbox{$P^2=0$} to the time-like domain near 
\mbox{$P^2=-m^2$}, where $m$
is not too large, requires analyticity information of the solution amplitude
only in that neighborhood.   This is the technique used in the GCM/DSE approaches 
to put external hadronic momenta on the relevant mass shell.  

One can take the point of view that the 
field theory in use is defined in Euclidean metric.  The consistency of this 
point of view and the generation of physical quantities by continuation of the 
Euclidean $n$-point functions of a field theory, and of non-perturbative QCD 
models in particular, has been discussed in some detail by Roberts and 
Williams (1994).  We note that this is the approach within lattice-QCD 
(Rothe, 1992) where
the continuation of Euclidean $n$-point functions is achieved by taking the  limit
of large Euclidean time intervals in correlation functions of hadronic interpolating 
currents.   Through the Cauchy integral analysis, this is equivalent  to a 
continuation of the $n$-point functions in the Euclidean hadronic momentum 
component $P_4$ from the real domain to the imaginary  point corresponding to 
the nearest 
mass-shell.   In the GCM/DSE approaches, specific model truncations are imposed
that imply a limited set of quark-gluon interactions and dressings, but otherwise
the Euclidean aspects and the definition of mass-shell quantities  is equivalent 
to that of lattice-QCD.    In the former case, those $n$-point functions that are 
presently used to provide the phenomenological input, such as the gluon $2$-point 
function, should be viewed as interim parameterizations of dynamical quantities that
are, in principle, well-defined from the Euclidean version of QCD and have clear
counterparts in the lattice formulation.    

Given that the GCM/DSE approach involves analytic continuation of $n$-point functions
(say $f$) in the external hadronic Euclidean momenta (say $P^2$), there is more than 
one way to approach this.   Initial investigations simply fit $f(P^2)$ on $P^2>0$
with known functions and then evaluated the fit at the mass-shell point where 
$P^2<0$.   Most of the work in recent years that we shall review here use the 
procedure that hadronic external momenta are to be continued before internal 
loop integrations are performed.  As discussed at the beginning of this Section, 
this requires knowledge of various propagators and vertices in domains of the 
complex plane and the entire function parameterization of the confining dressed
quark propagator has facilitated such work.  A third method has recently been put
forward and tested by Burkardt, Frank and Mitchell (1997).  Drawing upon the 
above-mentioned connections with lattice-QCD methods for extracting physical limits
from the Euclidean metric,  this method is to Fourier transform the external 
hadronic Euclidean energies $P_4$ to Euclidean time and  then take the large time
interval limit.  This removes the explicit need to determine or postulate the
behavior of  loop propagators and vertices in a complex momentum domain and, 
like lattice methods, requires only that the hadronic spectrum in a Lehmann
representation of external hadronic correlation functions be known.   In a simplified
version of the impulse approximation or quark loop mechanism for the pion charge
form factor,  Burkardt, Frank and Mitchell (1997) have demonstrated that this method
reproduces the results from the direct or second approach mentioned above. 

\subsect{The DSE Approach}
\label{sect_DSEcf}

Instead of considering the underpinnings of the present
investigations to be a result of bosonization of the GCM action to produce
an effective meson action, there is a more general viewpoint (Roberts and 
Williams, 1994).  Selected truncation
of the tower of coupled Dyson-Schwinger equations of QCD, together with use
of a generalized impulse approximation, produces a formulation of meson vertex
functions that are identical in form to what is generated in the effective
meson action from the GCM.  The content can be more general.  In the 
generalized impulse approximation for form factors, the dressed quark 
propagator and the meson BS amplitudes are formally defined exact $n$-point
functions.   The interpretation of the role of phenomenology for those 
functions is then quite different from what it is for the same calculation
taken from the tree-level GCM effective meson action where the dynamical
content of those functions is ladder/rainbow DSE.   Quantum meson loop
dressing from the GCM would have to be applied to begin to recover a
comparable interpretation.  

The distinction only becomes clearer when higher order terms are considered
in each formulation.  Progression through the loop expansion in the effective 
meson action derived from the GCM defines a specific ordering of quantum effects.
The tree-level ladder mesons become dressed with ladder mesons, and then dressed 
again, etc.   The point of view implemented through the DSE approach is that 
the quark-gluon content of important $n$-point functions for hadron physics is
decomposed only in the way necessary for construction from other $n$-point functions 
that are active at the level of truncation being employed.
There is potentially more freedom to develop an efficient ordering
with the DSE approach as one moves beyond the impulse approximation for 
hadronic processes and considers less severe truncations of the DSE tower.
Either route is difficult.  Present efforts, for practical reasons, are focussed 
upon meson tree-level within the GCM, and meson and photon insertions in dressed
quark loops within the DSE approach.   At this level, the distinction between the 
GCM and the DSE is a subtle one of interpretation of dynamical content represented
by the employed parameterization of the dressed quark propagators; this becomes 
important mainly as next order corrections are considered. 

We have mainly taken the effective meson action viewpoint that the GCM naturally 
leads to.   One advantage of this is that it emphasizes the relevance to model 
hadronic field theories which summarize efficiently most of what is known and 
unknown about low
energy hadron physics.   It is becoming quite feasible to generate such 
models with many of the previously phenomenological coupling form factors 
now given a quark basis that, although approximate, captures the dominant
influence on dynamics at the hadron size scale.   For a quark-gluon model field
theory to provide information on form factors for use in model hadronic theories,
it is necessary to make contact with the same  prescriptions traditionally used  
in such models for off-shell propagation of hadrons.   This is really only defined
by the form of interacting hadronic lagrangian or action, and this is the reason for
the emphasis upon the bosonization procedure described earlier in this Section.   At
the very minimum, it provides a consistent identification of the propagators for 
the $\bar q q$ correlations with meson quantum numbers.   The same information
is available in the DSE approach, but consistent DSE truncations that amount to
specification of a model action for mesons have not been developed so far.   
  
\sect{Quark and Gluon Dressed Propagators}
\label{sect_prop}

In QCD the fully-dressed and renormalized quark propagator is given by 
the  Dyson-Schwinger equation (DSE)
\begin{equation}
 S^{-1}(p) = Z_2 [i \gamma\cdot p + m_0(\Lambda)]
     + Z_1 \; \case{4}{3} \int^\Lambda \frac{d^4k}{(2\pi)^4} g^2 D_{\mu \nu}(p-k)
        \gamma_\mu S(k) \Gamma_\nu^g (k,p) ,     \label{fullDSE}
\end{equation}
where $m_0(\Lambda)$ is the bare mass parameter and $\Lambda$ is the regularization
parameter which can be taken to be a cut-off for the integral which otherwise is
divergent. Here the dressed gluon propagator
$D_{\mu\nu}(q)$  and the dressed quark-gluon vertex $\Gamma_\mu^g(k,p)$ 
are the renormalized quantities and they satisfy their own DSEs required higher order
$n$-point functions as input.  The coupling
constant $g$ is also the renormalized one.  The meaning of the renormalization
constants is illustrated by the following relations 
\beq
S^{-1}(p) = Z_2 \; S^{-1}_\Lambda (p) ~, \ipar
D^{-1}_{\mu\nu}(q) = Z_3 \; D^{-1}_{\mu\nu \Lambda}(q) ~, \ipar
g = \case{Z_2 \sqrt{Z_3}}{Z_1} \; g_\Lambda  ~, \ipar
\Gamma^g_\nu = Z_1 \; \Gamma^g_{\nu \Lambda} ~, 
\label{renorm}
\eeq
where $Z_1, Z_2, Z_3$ are the renormalization constants for the 
quark-gluon vertex function, the quark field and the gluon field respectively. 
Further details can be found in Itzykson and Zuber (1980).   
In quantum electrodynamics (QED), the
Ward identity  entails that \mbox{$Z_1 = Z_2$}.   In modeling QCD  via the quark 
DSE, it is common to model  $g^2 D_{\mu \nu}(q)$ in terms of a running coupling
constant and such an Abelian approximation leads to  the simplification
\mbox{$Z_1 = Z_2$}.  

At the chosen renormalization point $\mu$, the  constraint
\begin{equation}
\label{bcrenorm}
S^{-1}(p)|_{p^2=\mu ^2}=i\gamma \cdot p +m 
\end{equation}
allows identification of  $Z_2(\Lambda, \mu)$ and the renormalized current quark
mass $m(\Lambda)$.  In general the cut-off $\Lambda$  is different from 
the renormalization point $\mu$.   For sufficiently large $\mu$ and $\Lambda$, 
renormalized quantities for fixed $\mu$ become independent of $\Lambda$ as 
$\Lambda\to\infty$.  It is common to choose $\mu$=$\Lambda$ since for large enough
$\Lambda$ one can take $Z_2$ as unity.   Discussions of these issues can be 
found in Fomin {\it et al.} (1983) and references therein.  For our purposes here,
it is sufficient to note that \Eq{fullDSE} has been subjected to many detailed
investigations under a variety of approximations and renormalization schemes. 
A review can be found in  Roberts and Williams (1994).   Later works that provide
useful details of numerical solutions under a subtractive renormalization scheme 
are Hawes and Williams (1995) and Frank and Roberts (1996).  

We have already seen that bosonization of the GCM 
produces a tree-level meson action that involves the dressed quark propagator 
from the rainbow approximation (\mbox{$\Gamma_\nu^g (k,p) 
\rightarrow \gamma_\nu$}) DSE.   Although there is a large body of early work
that employed a phenomenologically successful form for $ g^2 D_{\mu\nu}(q)$,   
most of the more recent hadron physics investigations within the GCM 
format that we shall review in the following Sections are based on use of 
parameterized model forms for dressed quark propagators. There are several 
reasons.  

Firstly it is difficult to prepare  numerical solutions of the quark DSE that are 
sufficiently accurate and well-defined in the quark momentum domain
required for the subsequent multi-dimensional integrals needed to produce
hadron physics quantities.  One of the complications is that while 
the solution is more readily, and hence usually, addressed in Euclidean metric,
continuation of the solutions to the complex quark momentum domain is usually 
required for application to hadron physics due to mass-shell constraints
on external hadronic momenta.   When improvements over initial analyticity
assumptions were sought, complex plane singularities generated by the rainbow
form of DSE proved problematical (Stainsby and Cahill, 1992; Maris and Holties,
1992).   The realization that dressing of the gluon-quark vertex and a 
realistically strong infrared-enhanced $2$-point gluon function could produce
a viable implementation of quark confinement (Burden, Roberts and Williams, 
1992; Roberts, Williams and Krein, 1992), and could effectively remove propagator
singularities, led to a focus on this aspect.
Secondly, there is at present insufficient knowledge of the infra-red behavior
of the $n$-point functions of QCD to prefer numerical solutions of the quark
DSE (or other DSEs) over reasonable phenomenological models of the infra-red 
behavior for the purpose of investigating whether a wide selection of hadron 
physics can be 
reproduced.  In fact it is worthwhile to take the point of view that 
well-selected hadronic observables might provide a valuable constraint on the 
allowable infrared behavior thus providing a reference point to link with,
and be confronted by, more direct methods such as lattice QCD. 

To discuss the principal models of the dressed gluon $2$-point function and
the quark propagator that have recently been used within the general 
GCM/DSE format, we first briefly recall the leading asymptotic behavior 
of the gluon and quark propagators in QCD.  A more detailed treatment of this 
topic and the relation to renormalization procedures can be
found in the review by Roberts and Williams (1994).
In Euclidean metric the Landau gauge gluon propagator is
\begin{equation}
g^2 D_{\mu\nu}(q) =
\left(\delta_{\mu\nu} - \frac{q_\mu q_\nu}{q^2}\right)
\frac{g^2}{q^2[1+\Pi(q^2)]} \label{glprp1}
\end{equation}
where $\Pi(q^2)$ is the gluon vacuum polarisation.  If ghost contributions are
unimportant for soft hadron physics,  the effective coupling constant 
in \Eq{glprp1} satisfies the same renormalization group equation as the QCD 
running coupling constant $\alpha(q^2)$ (Bar-Gadda, 1980).
This leads to the approximate form
\begin{equation}
\label{glprp2}
g^2 D_{\mu\nu}(q) =
        \left(\delta_{\mu\nu} - \frac{q_\mu q_\nu}{q^2}\right)
                \frac{4\pi\alpha(q^2)}{q^2}~.
\end{equation}
for the dressed gluon propagator in this gauge. This ``Abelian'' form 
accompanied by the QED-like relation $Z_1 = Z_2$, is a common approximation
from which many studies of the quark DSE have begun.
It is reasonable to begin investigations of modeling hadron physics from QCD
with this Abelian-like approximation.  Ghost contributions in DSE studies 
have been studied in
Landau gauge and shown not to modify the qualitative features of quark and
gluon propagators (Brown and Pennington, 1988a; 1988b; 1989).  Quantitatively, 
ghosts provide a small ($<10$\%) effect.   

In Landau gauge, the two-loop renormalization group expression for the 
running coupling constant only receives small ($\sim 10$\%) corrections 
from higher orders for spacelike-$q^2>1$~GeV$^2$ and hence can be said 
to provide an accurate representation on this domain (Brown and Pennington,
1989). 
For $q^2<1$~GeV$^2$, however, $\alpha(q^2)$ can only be calculated 
nonperturbatively and it is not known.  The recent status of lattice-QCD
studies of the transverse gluon $2$-point function is represented by 
Bernard \etal~(1994).   The current
status of DSE studies is summarised in Roberts and Williams (1994).  
Present phenomenological
quark-DSE studies rely on an Ansatz for $\alpha(q^2<1~{\rm GeV}^2)$ motivated
by these gluon-DSE studies.
For spacelike $q^2\gg\Lambda^2_{QCD}$, the running coupling
constant is given at leading-log or one-loop order by
\beq
\alpha(q^2) = {d \pi \over \ln(q^2/\Lambda^2_{QCD})}
\label{alpha_s}
\eeq
where $d = 12/(33-2N_f)$ is the anomalous dimension of the mass, $N_f$ is 
the number of quark flavors and $\Lambda_{QCD}\approx 0.20~{\rm GeV}$ is the 
scale parameter of QCD.   

The general form of the dressed quark propagator is
\begin{equation}
S(p) = - i \gamma\cdot p\, \sigma_V(p^2) + \sigma_S(p^2)
     = \frac{1}{i\gamma\cdot p\, A(p^2) + B(p^2) + m }~.
\label{svsbab}
\end{equation}
When \mbox{$B \neq 0$}, there is dynamical chiral symmetry breaking (DCSB).
The behavior of the above amplitudes in the perturbative or UV asymptotic
region is well known from the QCD renormalization group and operator product
expansion (OPE) and QCD sum rules ( Politzer, 1976,1982; Gasser 
and Leutwyler, 1982; Reinders \etal,~1985).   The behavior has been
summarized in the course of explicit numerical solutions and model building
(Williams, Krein and Roberts, 1991). 
For $q^2 \gg \Lambda^2_{QCD}$  the leading-log result for the running 
mass (\mbox{$M = (B+m)/A = \sigma_S / \sigma_V$}) is
\beq
M(q^2) = \frac{\hat m} { \left[ \hlf \ln(q^2/\Lambda^2_{QCD}) \right]^d  }  \;,
\label{MQ2}
\eeq
where $\hat m$ is a scale invariant parameter to be eliminated in 
favor of the renormalized current quark mass \mbox{$m =$}\mbox{$M(\mu^2)$} 
that has been chosen for the constraint \Eq{bcrenorm}.  This holds when 
\mbox{$m \neq 0$}, i.e. for explicit chiral symmetry breaking (ECSB).  
When there is no ECSB  ($m=0$), there is exact chiral
symmetry, and conservation of the axial-vector current leads to (in Landau
gauge) 
\beq
M(q^2) \ \mathrel{\mathop=_{q^2\to \infty}}\ 
{\kappa \over q^2 \;\left[\ln(q^2/\Lambda^2_{QCD})\right]^{1-d} }   \;,
\label{chiral_q_mass}
\eeq
where $\kappa$ is a constant independent of $\mu$ and is given by
\beq
\kappa \simeq - {4\pi^2 d\over 3}\;{\langle \bar qq \rangle \over 
\left[\ln(\mu^2/\Lambda^2_{QCD})\right]^d} \;.
\label{c_condensate}
\eeq
This implies the scaling behavior \mbox{$\langle\bar q q\rangle \sim $} 
\mbox{$\left[\ln(\mu^2/\Lambda^2_{QCD})\right]^d$} for the quark condensate;
hence \mbox{$m\;\langle\bar q q\rangle$} is a renormalization scale invariant
quantity.
The quark condensate $\langle\bar qq\rangle$ is a measure of dynamical 
chiral symmetry breaking; in the absence of ECSB, the condensate for each 
flavor  is
\beq
{\langle \bar qq \rangle}_\mu = - \lim_{x\to 0^+}\;{\rm tr}_{sc} \; S_0(x) = -
12 \int^\mu
\dpsm { B_0(p^2) \over p^2 A_0^2(p^2) + B_0^2(p^2) } \;,
\label{q_condensate}
\eeq
where $S_0(x)$ is the chiral limit quark propagator and 
where the trace ${\rm tr}_{sc}$ is over spin and color.  Quoted values for 
quark condensates in this paper will be at the scale 
\mbox{$\mu^2 = 1~{\rm GeV}^2$}.

Early investigations with the GCM (Cahill and Roberts, 1985) employed the
simple confining model quark propagator due to Munczek and Nemirovsky (1983).
This is defined by the use of the $\delta$-function form
\mbox{$D_{\mu\nu}(q) = $} \mbox{$\delta_{\mu\nu} (2\pi)^4 \; \delta^4 (q)
\; 3 \alpha^2 /16$} in the rainbow DSEs, \Eq{a} and \Eq{b}.  Equivalently,
Landau gauge can be used and the same results are obtained by replacement of 
$\delta_{\mu\nu}$ above by $\frac{4}{3}(\delta_{\mu\nu}-q_\mu q_\nu /q^2)$.
This Ansatz for the dressed gluon
propagator models the infrared behavior of the quark-quark interaction in
QCD via an integrable infrared singularity.  The chiral limit DSEs reduce to 
non-linear algebraic equations and the solution that exhibits DCSB, also
minimizes the vacuum energy density. The solution is
\begin{eqnarray}
A(p^{2}) & = & \left\{
\begin{array}{cl}
2, & p^{2}\leq \frac{\alpha ^{2}}{4} \\
\frac{1}{2}\left[ 1+\left(1+\frac{2\alpha ^{2}}{p^{2}}\right)
^{\frac{1}{2}}\right] , & p^{2}\geq \frac{\alpha ^{2}}{4}
\end{array}\right. \nonumber \\ & &  \nonumber \\
B(p^{2}) & = & \left\{
\begin{array}{cl}
(\alpha ^{2}-4p^{2})^{\frac{1}{2}}, &
\; \; \; \; \; \; \; \; \; \; \; \; p^{2}\leq \frac{\alpha ^{2}}{4} \\
0, & \; \; \; \; \; \; \; \; \; \; \; \; p^{2}\geq \frac{\alpha ^{2}}{4}
\end{array}\right. . \label{MN}
\end{eqnarray}
For $p^{2}\leq \frac{\alpha ^{2}}{4}$ this gives the simple mass function 
$M^{2}(p^{2})=\frac{\alpha ^{2}}{4}-p^{2}$.  There is no mass-shell, i.e.
\mbox{$p^2 + M^2(p^2) \neq 0$} for any $p^2$, and there is quark confinement.
In model investigations, the above behavior was approximately corrected for
the influence of a perturbative UV tail for the gluon $2$-point function.
Since recent realistic studies of the  DSE for the gluon $2$-point function 
(Brown and Pennington, 1989) have found strong infrared enhancement in
qualitative agreement with a regularized singularity, the above Ansatz is
more representative for low $p^2$ than it might seem. 

A number of detailed studies of mesons, diquarks and the nucleon within 
the GCM (Praschifka \etal, 1989; Burden \etal, 1989; Cahill \etal, 1989a;
Hollenberg \etal, 1992) were carried out with quark solutions of the 
rainbow DSE generated 
from \mbox{$ g^2 D_{\mu\nu}(q) = \delta_{\mu\nu} D(q)$} with
\beq
D(q) = \frac{3\pi^2 \chi^2}{\Delta^2} \; {\rm exp} \left(-q^2/\Delta \right)
+\frac{16\pi^2}{11\, q^2 \ln(1 + \epsilon +  q^2/\Lambda^2)}  ~,
\label{gcmD}
\eeq
with \mbox{$\Lambda = 0.19~{\rm GeV}$}, \mbox{$ \Delta= 0.002~{\rm GeV}^2$}, 
\mbox{$\chi= 1.14~{\rm GeV}$}, and \mbox{$\epsilon = 2.0$}.   The first
term simulates the infrared enhancement and confinement; and the second term 
matches to the leading log renormalization group result in the UV.  The parameter
$\epsilon$ can be varied in the range $1-250$ without significant change in
results.  
This was found to be phenomenologically successful for selected light mesons (
\mbox{$\pi, \rho/\omega, f_1$}), various diquark correlations and the nucleon. 
Hadron mass-shell information was obtained for $P^2 < 0$ (time-like) from 
extrapolated $P^2 >0$ calculations.  Efforts to improve on this experienced 
difficulties with the numerical continuation of the resulting quark 
propagator into the complex plane  due to complex conjugate pairs of 
logarithmic branch  cuts (Stainsby and Cahill,
1992) that often are generated by rainbow-DSEs.   This considerably 
complicates  the extraction of hadron observables beyond the lightest Goldstone
bosons. 

Important guidance for extending the range of hadronic observables accessible
through modeling the quark propagator in the GCM/DSE approaches is provided 
by the work of Burden, Roberts and Williams (1992). This  provides an analytic 
solution to the DSE in Eq.~(\ref{fullDSE}) with the following simple 
one-parameter Ansatz
\begin{eqnarray}
\label{BRWMod}
g^2 D_{\mu\nu}(k) = \left(\delta_{\mu\nu} - \frac{k_\mu k_\nu}{k^2}\right)
                8 \pi^4 D \delta^4(k) \;\; & \; \mbox{and} \; &
\;\; \Gamma_\mu(p,p) = -i \partial_\mu S^{-1}(p)~,
\end{eqnarray}
where $D$ is a mass-scale parameter.  This Ansatz ensures confinement, in the
sense described below.  The dressed-quark gluon vertex used here, and 
resembling the
Ward identity constraint (Ward, 1950) for an Abelian theory, is a
result of extensive analysis of its general form (Ball and Chiu, (1980); 
Curtis and Pennington, (1992)).  The explicit
solution of Eq.~(\ref{fullDSE}) obtained by Burden, Roberts and Williams 
(1992) is
\begin{eqnarray}
\label{SS}
\bar\sigma_S(x) =
\frac{C}{2 \bar{m} y}\exp(-2x) J_1(4 \bar m y)
+ \frac{\bar{m}^2}{y}
        \int_0^\infty\,d\xi\, \xi\,K_1(\bar m \xi)\,J_1(y \xi)
        \exp\left(-\frac{\xi^2}{8}\right)~,
\end{eqnarray}
with $\bar{m} = m/\sqrt{2D}$, $x= y^2= p^2/(2D)$, and where $J_1$ and 
$K_1$ are Bessel functions, and
\begin{equation}
\label{SV}
\bar\sigma_V(x) =
        \frac{1}{\bar m}\left(
        \bar \sigma_S(x) + \frac{1}{4 y}\frac{d}{dy} \bar \sigma_S(x)\right)~,
\end{equation}
with $\bar\sigma_S(x) = \sqrt{2\,D}\,\sigma_S(p^2)$ and $\bar \sigma_V(x)
= 2\,D\,\sigma_V(p^2)$.  In Eq.~(\ref{SS}),  the first term is finite in
the chiral limit and it is associated with dynamical chiral symmetry breaking.
The parameter $C$ is  not determined by
Eq.~(\ref{fullDSE}) with Eq.~(\ref{BRWMod}). The second term of \Eq{SS} is
zero in the chiral limit and is associated with explicit chiral symmetry 
breaking.

An important feature of this model, and its derivatives described below, 
is that both $\sigma_S$ and $\sigma_V$ are
entire functions in the complex-$p^2$ plane.
As a consequence the quark propagator does not have a Lehmann representation
and can be interpreted as describing a confined particle (Roberts, Williams 
and Krein, 1992).  This becomes evident
when quark loop contributions to hadron couplings are considered.  The pion
electromagnetic vertex considered later in \Eq{piemver} is one example. 
The entire function property ensures the absence
of free-quark production thresholds, under the reasonable assumptions that
the pion BS amplitude is regular for spacelike-$p^2$ and the dressed 
photon-quark vertex is regular for spacelike photon momenta (Roberts, 1996).  
There are other, more complicated, ways to realize the absence
of free-quark production thresholds as
discussed by Roberts and Williams (1994) but the effect is the same.  
Some of the phenomenological
implications of a model with a simple realization of this confinement
mechanism have been discussed by Efimov and Ivanov (1993).
\begin{figure}[ht]

\centering{\
\psfig{figure=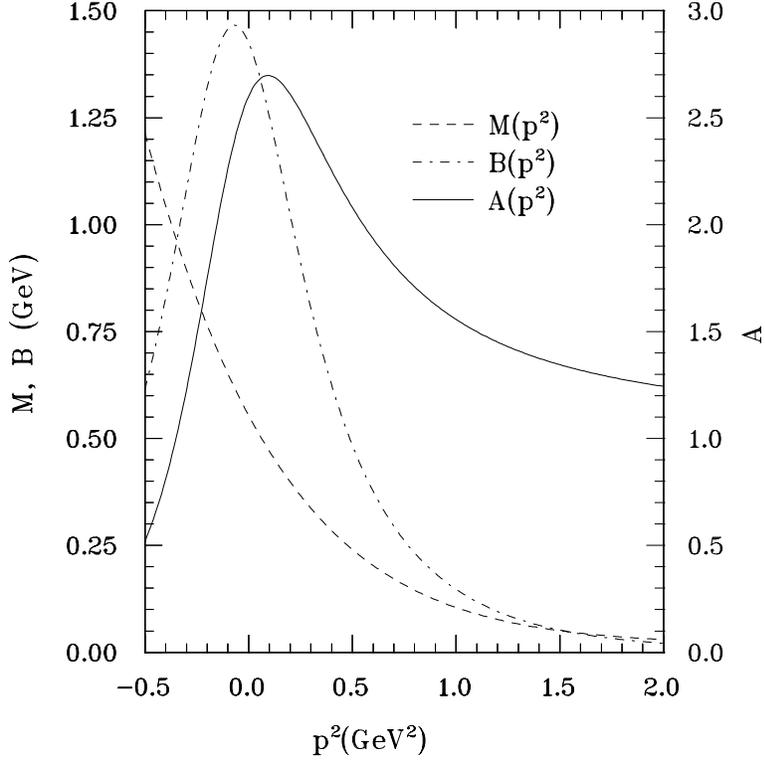,height=10.0cm} }

\parbox{130mm}{\caption{Confined dressed quark amplitudes $A$ (right scale)
and $B$ and the dynamical mass function $M$ (left scale) corresponding to the 
simplified propagator form in  Eqs.~(\protect\ref{sigmas}) and 
(\protect\ref{sigmav}).   Taken from Mitchell and Tandy (1997). \label{MAB} } }
\end{figure}

For the quark loops needed for hadron physics the entire function model
given by Eqs.~(\ref{SS}) and (\ref{SV}) has a well-defined behavior in the
complex $p^2$ plane.  However, the involvement  of Bessel functions and the
integral representation complicate applications.  With the expectation that the
details of that solution are less important than the qualitative features, 
Mitchell \etal ~(1994) and Mitchell and Tandy (1997) have employed a simplified
version in studies of $\rho^0-\omega$ mixing and the $\rho \pi \pi$ vertex.
This was obtained by expansion of Eqs.~(\ref{SS}) and (\ref{SV}) to first
order in $\bar m$.   This produces  
\beqar
\bar{\sigma}_{S}(x)&=&C(\bar{m})~e^{-2x}+\frac{\bar{m}}{x}
\left( 1-e^{-2x} \right) \label{sigmas},       \\
\bar{\sigma}_{V}(x)&=&\frac{ 2x-1 + e^{-2x} } {2x^2}
- C(\bar{m})~\bar{m}~e^{-2 x} \label{sigmav},
\eeqar
where recognition is given to the fact that the previous undetermined 
parameter $C$ should depend upon $\bar m$ if the current mass dependence of
realistic DSE solutions (Williams, Krein and Roberts, 1991) is to be simulated.
This is also necessary to produce a reasonable value for  $m_\pi$.  It was 
verified
by direct comparison that, for typically small current masses, these 
approximate expressions are accurate representations of the original 
expressions  to better than $0.1\%$ in the complex $p^2$ domain needed for
a typical  quark loop such as the self-energy of the $\rho$.   

The parameters $\lambda= \sqrt{2D}=0.889~{\rm GeV}$,
$~C(\bar{m})=0.581$, $~m = 16~{\rm MeV}$,  produce the 
soft chiral physics quantities $f_{\pi} = 90.1$~MeV, $m_{\pi} = 143$~MeV, and
$\langle\bar{q}q\rangle_{\mu^2} =(-173~{\rm MeV})^3$ at 
\mbox{$\mu^2 = 1~{\rm GeV}^2$}.   The formulation of calculations such as 
these will be described in \Sec{sect_GB}.
With the produced pions and vector mesons 
from this model, the pion loop mechanism gives 
\mbox{$m_\rho - m_\omega = -21~{\rm MeV}$} (Mitchell and Tandy, 1997) 
which is in reasonable accord with the  experimental value $-12.0\pm 0.8$~MeV.
The  dressed quark amplitudes
$A(p^2)$, $B(p^2)$ as well as the dynamical mass function
\mbox{$M(p^2)=$} \mbox{$(B(p^2)+m)/A(p^2)$} that follow from \Eqs{sigmas} and 
(\ref{sigmav}) are displayed in \Fig{MAB}.  This model was found to be too
crude to accurately represent typical DSE solutions at 
\mbox{$p^2 \geq 1~{\rm GeV}^2$}.   Apart from the weak strength
at such momenta, and other details, the behavior shown in \Fig{MAB} is
typical of the entire function propagator models discussed here.

The large spacelike-$p^2$ behavior of the original entire function model 
solution, given in Eqs.~(\ref{SS}) and (\ref{SV}), is found to be
\begin{eqnarray}
\label{SUV}
\sigma_S(p^2) \approx \frac{m}{p^2} - \frac{m^3}{p^4} + \ldots \;\;
& \; \mbox{and} \; & \;\;
\sigma_V(p^2) \approx \frac{1}{p^2} - \frac{D+m^2}{p^4} + \ldots ~.
\end{eqnarray}
The leading behavior of $\sigma_V$ in this model agrees with that from QCD, 
but the leading behavior of $\sigma_S$ in this model is missing the QCD 
logarithmic correction due to the anomalous mass dimension.  In particular, 
since \mbox{$\sigma_S(p^2) \rightarrow M(p^2)/p^2$} the leading QCD behavior
deduced from \Eq{MQ2} is   
\beq
\sigma_S(p^2) \approx  {m \over p^2}  \left[{\ln(\mu^2/\Lambda^2_{QCD})\over
\ln(p^2/\Lambda^2_{QCD})}\right]^d \;.
\label{ssqcd}
\eeq
Thus the model defined by Eqs.~(\ref{fullDSE}) and (\ref{BRWMod})
incorporates asymptotic freedom neglecting $\ln(p^2)$ terms.  

The main defect of this model solution can be seen by setting $\bar m = 0$ 
in Eq.~(\ref{SS}), which yields \mbox{$\bar{\sigma}_S(x) = C e^{-2 x}$}.
This is a poor representation for finite momentum since without a bare mass
and when chiral symmetry is dynamically broken, \Eq{chiral_q_mass} implies
that
\begin{equation}
\label{PolSS}
\left.\sigma_S(p^2)\right|_{p^2 \rightarrow \infty}
\rightarrow \, \frac{4 \pi^2 d}{3} \frac{\kappa}
                       {p^4\,\left[\ln(p^2/\Lambda_{QCD}^2)\right]^{1-d} } ~.
\end{equation}
This defect results from
the fact that although the form of $D_{\mu\nu}$ in Eq.~(\ref{BRWMod})
generates confinement, it underestimates the strength of the coupling in QCD
away from $k^2=0$.  In the numerical studies of Eq.~(\ref{fullDSE}) that have
used a better approximation to $D_{\mu\nu}(k)$ (Williams, Krein and Roberts, 
1991) there is no such defect.
\begin{table}[ht]
\begin{center}
\parbox{130mm}{\caption{ Parameters for the model quark propagators 
in Eq.~(\protect\ref{ssb}) and Eq.~(\protect\ref{svb}).  Taken 
from Burden, Roberts and Thomson (1996).   Note 
\protect\mbox{$\protect\Lambda= 10^{-4}$} is not varied.  The scale parameter is
\protect\mbox{$D=0.160~{\protect\rm GeV}^2$}.  At the finite bare mass values, 
\protect\mbox{$C^u_{m_u} = C^s_{m_s} =0$}.  The resulting $\protect\pi$ and $K$ 
observables
are displayed in Table~\protect\ref{tab_pk_res}.  \label{tab_pk} } }

\begin{tabular}{|ccc|}\hline
                  & $u/d$-quark  & $s$-quark    \\ \hline
  $\bar{m} \; $   &  ~0.00897    &  ~0.224      \\ \hline
  $b_0     \; $   &  ~0.131      &  ~0.105      \\ \hline
  $b_1     \; $   &  ~2.90       &  ~2.90       \\ \hline 
  $b_2     \; $   &  ~0.603      &  ~0.740      \\ \hline 
  $b_3     \; $   &  ~0.185      &  ~0.185      \\ \hline
  $C_{m_f=0} \;$  &  ~0.121      &  ~1.69       \\ \hline 
\end{tabular}
\end{center}
\end{table}

To expedite investigations of a variety of hadron physics quantities, a useful
parameterization of quark DSE experience has been developed from modifications
to  Eqs.~(\ref{SS}) and (\ref{SV}) that simply  restore the
missing strength at intermediate-$p^2$. This provides a better
approximation to the numerical solutions, while retaining the confining
characteristics present in Eqs.~(\ref{SS}) and (\ref{SV}).  The following
approximating algebraic forms were introduced by Roberts (1996) in the $u/d$
sector and extended by Burden, Roberts and Thomson, (1996) to include the $s$
sector. The scalar and vector parts of 
the quark propagators are defined in terms of dimensionless functions:
\begin{equation}
\sigma_V^f(s) = \frac{1}{\lambda^2}\bar\sigma_V^f\left(x\right), \hspace{5 mm}
\sigma_S^f(s) = \frac{1}{\lambda}\bar\sigma_S^f\left(x\right)~,
\end{equation}
with $s=p^2$, $x=s/\lambda^2$, \mbox{$\lambda^2 = 2D$}, where $D$ is a 
mass-scale parameter.   The explicit forms are
\begin{eqnarray}
\lefteqn{\bar\sigma_S^f(x)  = C^f_{\bar{m}_f} e^{-2x} +
  \frac{\bar{m}_f}{x + \bar{m}_f^2}
      \left(1 - e^{-2(x + \bar{m}_f^2)}\right) + }\nonumber\\
& &  \frac{1 - e^{-b_1^f x}}{b_1^f x} \frac{1 -
e^{-b_3^fx}}{b_3^fx}
    \left(b_0^f + b_2^f  \frac{1 - e^{-\Lambda x}}{\Lambda x}\right) ,
\label{ssb}
\end{eqnarray}
and
\begin{eqnarray}
\bar\sigma_V^f(x) = \frac{2(x + \bar{m}_f^2)
   - 1 + e^{-2(x + \bar{m}_f^2)}}{2(x + \bar{m}_f^2)^2}
- \bar{m}_f C^f_{\bar{m}_f} e^{-2x} ~.
\label{svb}
\end{eqnarray}
Here \mbox{$\bar{m}_f = m_f/ \lambda$}, and $\Lambda=10^{-4}$ is not
a free parameter.   In this work
the $u$ and $d$ quarks are considered to be identical, except for their
electric charge.
The dressed-quark propagator described by Eqs.~(\ref{ssb}) and (\ref{svb}) is
also an entire function in the finite complex $p^2$-plane and  therefore 
retains the  same confined particle interpretation as the previous
analytic model solution that motivates it.   The exponential function 
form that ensures this is suggested by that
analytic solution of the model DSE given in \Eq{SS} and \Eq{SV}.   The
behavior of Eqs.~(\ref{ssb}) and (\ref{svb}) on the spacelike-$p^2$ axis is
such that, neglecting quantitatively unimportant $\ln[p^2]$ corrections 
associated with the anomalous mass dimension  in QCD, asymptotic freedom is 
manifest.  In Eq.~(\ref{ssb}) the third
term  allows for the representation of dynamical chiral symmetry
breaking, and the second term represents explicit chiral symmetry breaking.
\begin{table}[ht]
\begin{center}
\parbox{130mm}{\caption{ $\protect\pi$ and $K$ observables calculated by Burden,
Roberts and Thomson (1996) using the parameters of
Table~\protect\ref{tab_pk}.  See text for references for the various calculations.
%
Experimental values are extracted from Particle Data Group (1996)
except that  $r_\protect\pi$ is from Amendolia {\protect\it et al.}~(1986a), 
$r_{K^\protect\pm}$ is from Amendolia {\protect\it et al.}~(1986b), 
and $r_{K^0}^2$ is from Molzon {\protect\it et al.}~(1978).
The experimental $\protect\pi$-$\protect\pi$ scattering lengths, $a^I_J$, are 
obtained from
Pocani\protect\'c (1995) and Gasser (1995).   \label{tab_pk_res}  }  }

\begin{tabular}{|cll|}\hline
   & Calculated  & Experiment  \\ \hline
  $f_{\pi} \; $    &  ~0.0924 GeV &   ~0.0924 $\pm$ 0.001     \\ \hline
  $f_{K} \; $    &  ~0.113  &   ~0.113 $\pm$ 0.001     \\ \hline
  $m_{\pi} \; $    & ~0.1385  & ~0.1385  \\ \hline
  $m_{K} \; $    & ~0.4936  & ~0.4937  \\ \hline
  $m^{\rm ave}_{1\,{\rm GeV}^2}$ & ~0.0051 & ~0.0075
                \\ \hline
 $m^s_{1\,{\rm GeV}^2}$ & ~0.128 & ~0.1 $\sim$ 0.3 \\ \hline
 $-\langle \bar u u \rangle^{\frac{1}{3}}_{1\,{\rm GeV}^2}$ & ~0.221 &
        ~0.220\\ \hline
 $-\langle \bar s s \rangle^{\frac{1}{3}}_{1\,{\rm GeV}^2}$ & ~0.205 &
        ~0.175 $-$ 0.205\\ \hline
 $r_{\pi^\pm} \;$ & ~0.56 fm & ~0.663 $\pm$ 0.006  \\  \hline
 $r_{K^\pm} \;$ & ~0.49  & ~0.583 $\pm$ 0.041  \\  \hline
 $r_{K^0}^2 \;$ & -0.020 fm$^2$ & -0.054 $\pm$ 0.026  \\  \hline
 $g_{\pi^0\gamma\gamma}\;$ & ~0.505 (dimensionless) & ~0.504 $\pm$ 0.019\\\hline
 $F^{3\pi}(4m_\pi^2)\;$ & ~$1.04$ & ~$1$ (Anomaly)\\\hline
 $a_0^0 \;  $ & ~0.17  & ~0.26$\pm$ 0.05 \\ \hline
 $a_0^2 \;  $ & -0.048 & -0.028$\pm$ 0.012 \\ \hline
 $a_1^1 \;  $ & ~0.030 & ~0.038 $\pm$ 0.002\\ \hline
 $a_2^0 \; $  & ~0.0015 & ~0.0017 $\pm$ 0.0003\\ \hline
 $a_2^2 \;  $ & -0.00021 & ~0.00013 $\pm$ 0.0003 \\ \hline
  $f_{K}/f_\pi \; $    &  ~1.22 &   ~1.22 $\pm$ 0.01     \\ \hline
 $r_{K^\pm} /r_{\pi^\pm}$ & ~0.87  & ~0.88 $\pm$ 0.06 \\ \hline
\end{tabular}
\end{center}
\end{table}

\begin{figure}[ht]
\psrotatefirst
\centering{\
\psfig{figure=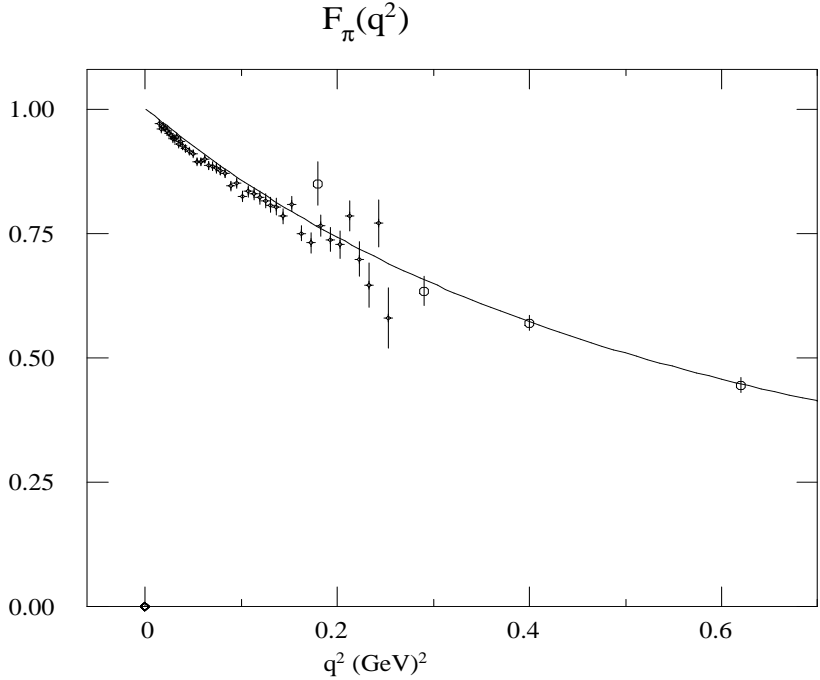,rheight=9.5cm,height=9.0cm,angle=-90} }

\parbox{130mm}{\caption{The pion electromagnetic form factor fit at low 
$q^2$.   Adapted from  Roberts (1996).  The data shown by crosses are from 
Amendolia {\protect\it et al.}~(1986a), the data shown by circles are from Bebek 
{\protect\it et al.}~(1976). 
\label{fig_piff1} } }
\end{figure}

\begin{figure}[ht]
\psrotatefirst
\centering{\
\psfig{figure=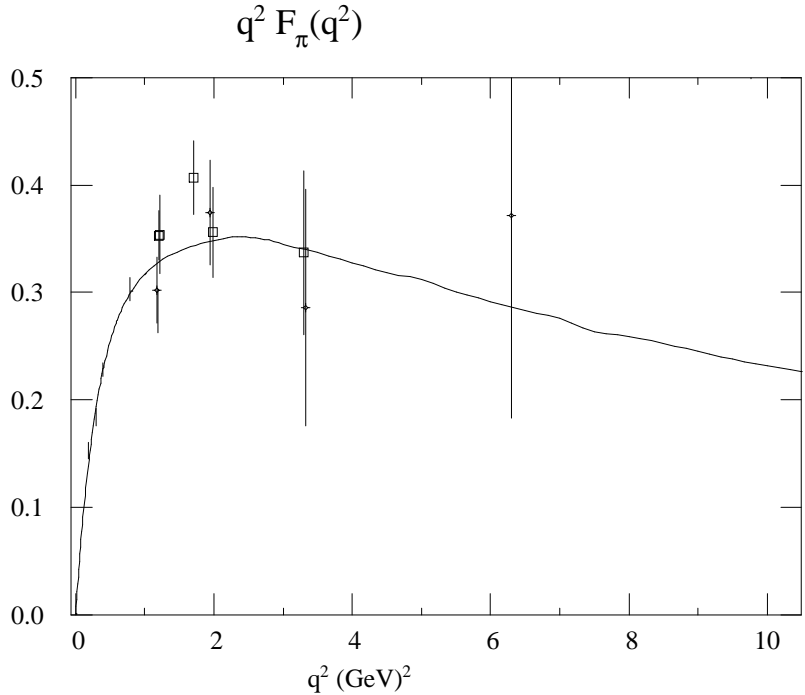,rheight=9.5cm,height=9.0cm,angle=-90} }

\parbox{130mm}{\caption{The pion electromagnetic form factor fit at high 
$q^2$.  Adapted from  Roberts (1996).  The data shown by crosses are from Brown 
{\protect\it et al.}~(1973), the data shown by diamonds are from Bebek 
{\protect\it et al.}~(1976), the data shown by
circles are from Bebek {\protect\it et al.}~(1978).  \label{fig_piff2} } }
\end{figure}

In Burden, Roberts and Thomson, (1996), 
the five parameters $\{\bar m_u,b_0^u,\ldots,b_3^u\}$ in
Eqs.~(\ref{ssb}) and (\ref{svb}) were varied in order to determine whether this
model form could provide a good description of the pion observables:
$f_\pi$; $m_\pi$; $\langle\bar q q\rangle$; $r_\pi$; the $\pi$-$\pi$ scattering
lengths; and the electromagnetic pion form factor.  The procedures followed were
essentially those developed in an earlier work (Roberts, 1996) where more details
can be found.  A very good fit was found with the $u$-quark parameter values listed in
Table~\ref{tab_pk}.   The low energy $\pi-\pi$ scattering  partial wave amplitudes 
were not part of the fitted data, but a reasonable description of the qualitative 
features of the data for \mbox{$l=0,1,2$} and \mbox{$E \leq 4 m_\pi$} was obtained.
Dyson-Schwinger equation studies (Williams, Krein and Roberts, 1991) indicate 
that while it is a
good approximation to represent the $u$- and $d$-quarks by the same
propagator, this is not true for the $s$-quark.  For example; contemporary
theoretical studies suggest that \mbox{$2m_s/(m_u+m_d)\sim
17-25$} (Particle Data Group, 1996) and 
\mbox{$\langle\bar s s\rangle\sim 0.5-0.8 \;\langle
\bar u u\rangle$} (Narison, 1989) which is a nonperturbative difference.  In
Burden, Roberts and Thomson, (1996), with this in mind, the model forms in 
Eqs.~(\ref{ssb}) and
(\ref{svb}) were employed in a study of the kaon observables: $f_K$;
$\langle\bar s s\rangle$; $r_{K^0}$; $r_{K^\pm}$; and the electromagnetic
form factors of the charged and neutral kaon.  The sensitivity of these
observables to $\bar m_s$ and $\langle\bar s s\rangle$ was too weak for an
independent determination and therefore $\bar m_s=25 \bar m_u$ and $b_0^s=0.8
b_0^u$, which ensures $\langle\bar s s\rangle=0.8\langle\bar u u\rangle$,
were chosen for consistency with other theoretical estimates.  The parameter
$b_2^s$ was allowed to vary to provide a minimal residual difference between
the $u/d$- and $s$-quark propagators and a very good fit to the kaon
observables was obtained with the values listed in Table~\ref{tab_pk}.  
The resulting $\pi$ and $K$ charged form factors are displayed in 
\Fig{fig_piff1},  \Fig{fig_piff2} and \Fig{kff}.  The predictions for other 
$\pi$ and $K$ observables are displayed in \Table{tab_pk_res} where $a^I_l$ are 
the \mbox{$\pi-\pi$} scattering lengths.  The notation \mbox{$m^{ave} = $}
\mbox{$(m_u +m_d)/2$} has been used and the scale is \mbox{$\mu^2 = 1~{\rm GeV}^2$}.
The quoted experimental values of
$m^{\protect\rm ave}_{1 \protect\, {\protect\rm GeV}^2}$, 
$m^s_{1 \protect\, {\protect\rm GeV}^2}$, 
$\protect\langle \protect\bar u u \protect\rangle_{1 \, {\protect\rm GeV}^2}$ and 
$\protect\langle \protect\bar s s \protect\rangle_{1 \, {\protect\rm GeV}^2}$ are
representative of current phenomenology including QCD sum rule analysis.
The calculation of $g_{\pi^0\gamma\gamma}$ is from Roberts (1996),
and $F^{3\pi}(4m_\pi^2)$ is from Alkofer and Roberts (1996).   
Predictions for other observables from use of this model, and related
models, are discussed in following Sections.

It is desirable to confirm that the above quark propagator modeling is indeed
representative of what can be generated from the quark DSE with an appropriate
model gluon dressed propagator.  An initial step in that direction was taken
recently by Frank and Roberts (1996) in showing that a one-parameter model
confined-gluon propagator used in the  rainbow-DSE  produces a quark
propagator with no pole on the real $p^2$ axis.  The qualitative behavior
for real $p^2$ is similar to that of the above entire function 
parameterizations  except the there is a broad resonance-like peak centered
at \mbox{$p^2 \approx - 0.55~{\rm GeV}^2$}.  This suggests an interpretation
of a "constituent-quark mass".   The employed gluon propagator model in
Landau gauge \mbox{$g^2 D_{\mu\nu}(q) =$} \mbox{$ (\delta_{\mu\nu} -q_\mu q_\nu 
/q^2)$} \mbox{$ \Delta(q^2)$} is specified by
\begin{equation}
\label{delk}
\Delta(q^2) = 4 \, \pi^2 \, d\,
        \left[ 4\,\pi^2\,m_t^2\,\delta^{4}(q)
                \, + \,
        \frac{1-{\rm e}^{(-q^2/[4 m_t^2])}}{q^2}\right]~,
\end{equation}
where $d=12/(33-2N_f)$, with $N_f=3$ the number of light flavors.  The first
term  provides an integrable, infrared
singularity, which generates long-range effects associated with
confinement, and the second ensures that the propagator has the correct large
spacelike-$q^2$ behavior, up to $\ln(q^2)$-corrections.  
The reason for the relationship between the
coefficients of the two terms in \Eq{delk} is that
\begin{equation}
\label{delx}
\Delta(x^2)  \equiv   \int \frac{d^4q}{(2\pi )^4} \;
{\rm e}^{i q\cdot x} \Delta(q^2) \nonumber \\
  =  d\,\left[ m_t^2
        + \case{1}{x^2}\,{\rm e}^{-x^2\,m_t^2} \right]~,
\end{equation}
shows that
the effects of $\delta^4(q)$  are completely cancelled at small $x^2$ where
\mbox{$\Delta(x^2) \rightarrow$} \mbox{$ d/x^2 + {\rm O}(x^2)$}
which is the form expected from QCD without logarithmic corrections.  Hence
$m_t$ can be interpreted as the mass scale  that marks
the transition from the perturbative to the nonperturbative regime. The
single parameter $m_t$ was varied and \mbox{$m_t = 0.69~{\rm GeV}$} provided 
a best fit to a range of calculated pion observables.  The pion ladder BSE
was solved with this model confirming that the chiral limit quark amplitude
$B_0(q^2)$ provides an excellent approximation for the $\gamma_5$ part
of the BS amplitude $\Gamma_\pi$ for a typical finite mass.  We note that the 
renormalization point used by  Frank and Roberts (1996) is 
\mbox{$\mu \sim 9.5~{\rm GeV}$} which is verified to be in the purely perturbative 
domain.   The cut-off used for the quark DSE integral is 
\mbox{$\Lambda \sim 5500\mu$}.  These values are typical for solution of the quark DSE
for momenta scales needed for hadronic physics applications.    The observables 
were verified to be independent of $\Lambda$ to about $3$\%.  Complete independence 
from $\Lambda$ would occur if the bare vertex were replaced by a dressed one 
that preserves multiplicative renormalizability. 

\sect{Light Mesons}
\label{sect_mesons}

The spectroscopy of light-quark mesons is made interesting because of the
role played by dynamical chiral symmetry breaking, the natural scale of which
is commensurate with other scales in this sector.  It also explores quark and
gluon confinement because most vector and axial-vector mesons have masses
that are more than twice as large as typical constituent-quark masses.
An important feature of the strong interaction spectrum is the exceptionally
low mass of the pion.  
Furthermore, the pion mass must vanish in the chiral limit; i.e., when the
bare or current mass of the quarks vanishes even though this does not entail 
a vanishing of the constituent-quark mass.  These observations are an 
indication that the characteristics of the pion as a manifestation of the 
Goldstone theorem is a particular and crucial feature of the strong-interaction 
spectrum.   These features can be captured in the form of relativistic potential 
models (Bernard, Brockmann and Weise, 1985; Le Yaouanc \etal, 1985) that
incorporate the consequences of the axial Ward identity from field theory.  Without
such a parameter-independent property embedded in the format, the Goldstone theorem 
requirements are not well simulated through specific parameter choices
(Sommerer \etal, 1994; Gross and Milana, 1994).   As a model field theory, the GCM
can easily respect the Goldstone theorem independent of phenomenological parameters.

\subsect{The Goldstone Boson Sector}
\label{sect_GB}

In the chiral limit, the pseudoscalar octet are massless Nambu-Goldstone bosons
associated with the dynamical (or spontaneous) breaking of chiral symmetry.  In
this situation, the quark propagator fully specifies the BS amplitude for these
pseudoscalar mesons.    At the level of tree level mesons in the GCM, this 
Goldstone 
theorem is generated exactly by the consistency between the ladder BS equation
for the pseudoscalar amplitudes and the rainbow SDE for the employed dressed
quark propagator.  
In particular, the chiral limit ladder BS equation for the amplitude
$\hat\Gamma_b(q;P)$ becomes identical to the ladder Dyson-Schwinger equation
for \mbox{$B_0(q^2)\equiv B(q^2;m=0)$}, the scalar part of the
dynamically generated quark self-energy (Delbourgo and Scadron, 1979).
Thus the chiral limit pion, in this approach, is automatically both a massless
Goldstone boson and a finite size $q\bar{q}$ bound state. 

Thus the only knowledge of the underlying effective gluon 
propagator that is needed in this limit is that which is already encoded in the
dressed quark propagator.   This suggests that, as one moves away from this 
limit to generate physical masses for the PS octet through small finite bare 
quark masses, explicit knowledge of the underlying effective gluon propagator
is a secondary consideration.  This is certainly true to first order 
perturbation in the bare masses.  For this reason, a considerable effort has
emerged in recent years to develop an efficient parameterization of the dressed
$u$, $d$ and $s$ quark propagators constrained only by pion and kaon physics.  
This effort has
been very successful and we will recount some of the major results here.

We first demonstrate that, in the chiral limit, and at the ladder-rainbow 
level, each member of the PS octet is massless and has the $\gamma_5$ part
of BS amplitude given by the quark  scalar self-energy. 
After projection onto the canonical covariant
\mbox{$\Lambda^b = i \gamma_5 \lambda^b \sqrt{2}/3$}, where we take $SU(3)_f$,
the ladder BS Eq.~(\ref{omsbs}) can be written as
\beq
\Gamma_b(q';P) = - \case{2}{9}
 \int \dq D(q'-q) 
{\rm tr} \left[ i\gamma_5 \lambda_b^\dagger S(q_+) 
i\gamma_5\lambda_b S(q_-)\right] 
\Gamma_b(q;P) ,
\label{psbs1}
\eeq
where \mbox{$P^2=-m_b^2$}.  Throughout this work the symbol tr denotes a trace 
over $SU(3)$ color, Dirac spin, and the relevant flavor indices.  
For a flavor symmetric $S(q)$, and using 
\mbox{${\rm tr}_f (\lambda_b^\dagger \lambda_b) = 2$}, we have at \mbox{$P=0$}
\beq
\Gamma_b(q';0) =  \case{4}{3} \int \dq D(q'-q) 
{\rm tr_s} \left[ S(-q) S(q)\right] \Gamma_b(q;0) .
\label{psbs2}
\eeq
This is equivalent to the DSE given in Eq.~(\ref{b}) for the scalar self-energy 
amplitude provided the chiral limit is taken; in this regard it is useful 
to note that \mbox{$S_0(-q)S_0(q)= (q^2A_0^2 + B_0^2)^{-1}$}. 
Thus there is the massless solution 
\mbox{$\Gamma_b(q;0) \propto B_0(q^2)$} for each PS meson.  The above argument
is unchanged if we consider the $SU(3)_f$ singlet with 
\mbox{$\lambda^0 = {\bf 1}_f/\sqrt{3} $}.  However this ($\eta'$) is not a
Goldstone boson and this circumstance ($U_A(1)$ anomaly) must be addressed
beyond the ladder BS and rainbow DSE level.   

The above argument covers only the pseudoscalar Dirac component of the pion 
BS amplitude.   A full analysis should take into account the general form of 
the \mbox{$J^{PC} = 0^{-+}$} BS amplitude which is (LLewellyn-Smith, 1969)
\beq
\vec{\Gamma}_\pi(q;P) = i\gamma_5 \vec{\tau} \left\{E_\pi(q;P) + 
\gamma \cdot P F_\pi(q;P) + \gamma\cdot q q\cdot P G_\pi(q;P)  +
[\gamma\cdot q ,\gamma\cdot P] H_\pi(q;P)  \right\}  ,
\label{piform}
\eeq
where the invariant amplitudes 
are functions of $P^2$, $q^2$ and $(q\cdot P)^2$.  
The last term does not arise in the present discussion because 
$\sigma_{\mu\nu}$ is not part of the set of matrices $\Lambda^\theta$ due to 
Feynman-like choice of gauge for the effective gluon $2$-point function.
The above result is that considering only the chiral limit $E_\pi$, one finds 
\mbox{$E_\pi(q;0) \propto B_0(q^2)$}. 

In general, without relying upon the explicit rainbow-ladder truncation, 
the Goldstone theorem for the pion, along with identification of the proper
normalization of the chiral limit BS amplitude, can be obtained from
consideration of the  axial vector vertex $\vec{\Gamma}_\mu^5(q;P) $.
The inhomogeneous
BS equation for $\vec{\Gamma}_\mu^5(q;P) $, driven by the bare term
$i\gamma_5\gamma_\mu \frac{ \vec{\tau} } {2} $, has a pole when the homogeneous
version has a solution.  To be specific, the ladder BSE equation for this
vertex is 
\beq
\vec{\Gamma}_\mu^5(q';P)  = i\gamma_5\gamma_\mu \case{ \vec{\tau} } {2}
- \case{4}{3} \int \dq D(q'-q) 
 \gamma_\nu S(q_+)\vec{\Gamma}_\mu^5(q;P) S(q_-)  \gamma_\nu   .
\label{axver}
\eeq
If we contract with $P_\mu$, and note that
$P_\mu \vec{\Gamma}_\mu^5(q;P) $ is a pseudoscalar, we see that \Eq{axver}
generates the inhomogeneous version of the ladder BSE for the pion BS
amplitude. As we have discussed before, one may introduce eigenfunctions of that
homogeneous kernel and its eigenmode expansion generates the pion propagator.
The mass-shell BS amplitude provides the residue at the physical pole.
Thus, there is a pion pole in  $\vec{\Gamma}_\mu^5(q;P) $ (Jackiw and Johnson, 
1973), and near the mass shell of the chiral limit pion we have
\beq
\vec{\Gamma}_\mu^5(q;P) = i\gamma_5\gamma_\mu \case{ \vec{\tau} } {2} -
\vec{\Gamma}_\pi(q;P)\; (\frac{1}{P^2}) \;(-iP_\mu f_\pi) 
+ \vec{\Gamma}_\mu^{5 Reg}(q;P) ,
\label{avpole} 
\eeq
where the last term is regular there, 
$\vec{\Gamma}_\pi(q;P)$ is the properly normalized pion BS amplitude, 
and the pion coupling to the point axial field is 
\beq
-iP_\mu f_\pi \delta_{ij} = {\rm tr} \int \dq \; [\Gamma_\pi^i(q;-P) S(q_+) 
i\gamma_5 \gamma_\mu \case{ \tau^j }{2} S(q_-)] \; .
\label{fpidef}
\eeq
These two results are true in general; they are not limited to ladder
approximation.  Meanwhile, the isovector axial WTI identity  
\beq
-P_\nu \vec{\Gamma}_\nu^5 (q;P) = S^{-1}(q_+)\gamma_5 \case{ \vec{\tau} } {2} 
+ \gamma_5  \case{ \vec{\tau} } {2} S^{-1}(q_-)
\label{avwti}
\eeq
also reveals the leading behavior of $\vec{\Gamma}_\nu^5 (q;P)$ at the pole 
(for the longitudinal component) in terms of the dressed quark propagator.   
In combination with \Eq{avpole}, this provides a means for connecting the 
chiral limit pion BS amplitude directly to the quark propagator (Delbourgo and 
Scadron, 1979).   In particular, from \Eq{avwti} we have
\beq
P_\nu \vec{\Gamma}_\nu^5 (q;P) =  \gamma_5 \vec{\tau} \left( - B_0(q^2)
+ i\gamma \cdot P \case{ A_0(q^2) } {2} + i\gamma \cdot q P \cdot q A_0'(q^2)
+ {\rm O}(P^2) \right)  ~.
\label{piexp}
\eeq
From this and \Eq{avpole}, the amplitude $E_\pi$ for the $\gamma_5$ component 
of the properly normalized chiral limit pion BS amplitude is identified as
\beq
E_\pi(q;P=0) =  \frac{B_0(q^2)}{f_\pi}
\label{pibs}
\eeq 
where the weak decay constant is given by Eq.~(\ref{fpidef}).  The other terms 
in \Eq{piexp} can arise from a combination of the pole term and the regular
term.  Further analysis is needed to be able to express the BS amplitudes 
$F_\pi$ and $G_\pi$ directly in terms of the quark amplitudes $A$ and $B$.
The result
in \Eq{pibs} is essentially the Goldberger-Treiman relation for quarks (instead
of nucleons in the original development). 
The rainbow-ladder approximation has the important property of preserving
the above manifestations of  Goldstone's theorem.  This is due to the
fact that contraction of \Eq{axver} with $P_\mu$ will produce exactly
\Eq{avwti} only if the propagators in \Eq{axver} are DSE solutions in rainbow
approximation.  

It is clear from the above that if the dynamically coupled
pion amplitudes $E_\pi$, $F_\pi$ and $G_\pi$ are expressed in a normalization
where the chiral limit of $E_\pi$ is $B_0$, then the normalization constant 
$N_\pi$ that must be divided into each of them for a physical amplitude is 
simply \mbox{$N_\pi = f_\pi$}.  This  dual 
role for $f_\pi$ (measuring the weak decay and providing the inverse
of the field normalization constant in the above sense) is also evident in
the ${\cal O}(p^2)$ chiral lagrangian where it is the only parameter.  The dual
role of $f_\pi$ is a useful device in model calculations.  With the 
asssumption that the 
pion is dominated by the $\gamma_5$ amplitude, use of \Eq{pibs} in the  
chiral limit BS normalization condition (see Eq.~(\ref{bsnorm}) )
\beq
 \left. \frac{\partial}{\partial P^2} tr \int \dq \: 
\bar{\Gamma}_\pi(q;-K) S_0(q_+)  \Gamma_\pi(q;K)  
S_0(q_-)\right| _{P^2=K^2= 0}  = 1  
\label{pinorm}
\eeq
will produce a value for $f_\pi$ without explicitly computing the weak decay
process. However, from the definition of $f_\pi$ given in Eq.~(\ref{fpidef}), 
and again assuming that the pion is dominated by the $\gamma_5$ amplitude,  one may 
also obtain in the chiral limit,
\beq
f_\pi^2 = 12 \int \dq B_0(q^2) \; [\case{q^2}{2} (\sigma_V' \sigma_S - \sigma_V 
\sigma_S') + \sigma_V \sigma_S ]  .
\label{fpiw}
\eeq
The two resulting expressions for $f_\pi$ do not generally agree and some caution 
is needed.  In practice, the difference is usually found to be at most $10\%$.  
The reason they do not agree is that  an approximate BS amplitude (just the $\gamma_5$ 
amplitude $E_\pi$) has been used; the equivalence of the two approaches to $f_\pi$
comes from analysis of \Eq{axver} and  \Eq{avwti} which are compatible only if the
full structure of the pion BS amplitude allowed by both of them is employed. 
The size of the contribution from the sub-dominant pion BS 
amplitudes $F_\pi$ and $G_\pi$ to various observables is a question that deserves
further attention.  Much work on modeling of the pion 
omits them.  Many studies of the ladder BS equation (Aoki, Kugo and Mitchard, 
1991; Jain and Munczek, 1993; Burden, Qian, Roberts, Tandy and Thomson, 1996)) 
confirm that the dominant amplitude in Eq.~(\ref{piform}) is $E_\pi(q;P)$, 
especially as far as contributions to the mass are concerned.
However $F_\pi$ and $G_\pi$ will contribute to $f_\pi$ via Eq.~(\ref{fpidef})  
even in the chiral limit and the contribution can be as large as $35\%$ 
(Burden, Qian, Roberts, Tandy and Thomson, 1996).  For the special case 
of \mbox{$A \rightarrow 1$}, Eq.~(\ref{piexp}) indicates that
the sub-dominant amplitudes of $\vec{\Gamma}_\pi$ produce at most an 
${\cal O}(P^2)$ contribution to \Eq{fpidef} and therefore will not contribute
to the chiral limit $f_\pi$. We also note that when \mbox{$A \rightarrow 1$},
Eq.~(\ref{fpiw}) reduces to the formula obtained by Pagels and Stokar (1979). 

To go beyond the above symmetry-based observations and calculate the 
finite pion mass 
generated by a finite bare quark mass, a dynamical formulation must be 
specified.  The Goldstone theorem at the nearby chiral point should be
preserved by the dynamics and the ladder BSE and rainbow DSE combination 
therefore suggests itself.  
It is difficult to formulate general higher order descriptions of the BS 
kernel and the dressed gluon-quark vertex for the DSE that also preserve the 
Goldstone theorem.  At order $g^4$ this has only recently been achieved 
(Bender, Roberts and V. Smekal, 1996)
We will limit our considerations to the rainbow-ladder level.

Consider now the PS octet modes for the physical situation of finite bare
quark masses. We shall continue to make the dominant covariant assumption
for simplicity and we shall proceed from the action given in 
Eq.~(\ref{mesonaction}).    
The free or quadratic part of this action  for the PS octet can be written
\beq
\hat{S}_2[\phi]=\case{1}{2}\int\frac{d^4P}{(2\pi)^4} \;
\phi_b^\ast(P) \; \hat{\Delta}^{-1}_b(P^2) \; \phi_b(P),
\label{freePS}
\eeq
where 
\beq
\hat{\Delta}^{-1}_b(P^2) = {\rm tr} \int \dq \;
[i\gamma_5\lambda_b^\dagger S(q_+)i\gamma_5\lambda_b S(q_-) ] \;
\hat{\Gamma}_b^2(q;P)
+\case{9}{2}\int d^4 r\frac{\hat{\Gamma}^2_b(r;P)}{D(r)}.
\label{delPS}
\eeq
Here the quark propagator $S(q)$ can be considered a diagonal matrix in flavor
space.
We can, at this stage, consider the $\hat{\Gamma}_b(q;P)$ to be ladder BS 
amplitudes that are not yet physically normed.  Evidently 
$\hat{\Delta}^{-1}_b(P^2)$ is an (unnormed) inverse propagator for the 
relevant local fields.    Here \mbox{$q_\pm=q\pm\frac{P}{2}$}
and $\lambda_b$ is the relevant flavor matrix.  For example, with $N_f=3$,
we have \mbox{$\lambda_b \rightarrow (\lambda^1 + i\lambda^2)/\sqrt{2}$} 
for  $\pi^+$, 
\mbox{$\lambda_b \rightarrow (\lambda^4 + i\lambda^5)/\sqrt{2}$} 
for  $K^+$, and $\lambda^8$ for $\eta$; 
with $N_f=2$, we have \mbox{$\lambda_b \rightarrow (\tau^1 +i\tau^2)/\sqrt{2}$} 
for $\pi^+$.

We note that $\hat{\Delta}_b^{-1}$ is a functional of $\hat{\Gamma}_b$ and that 
\mbox{$\delta \hat{\Delta}_b^{-1}[\hat\Gamma_b] / \delta\hat\Gamma_b =0$}  
reproduces Eq.~(\ref{psbs1}),  the ladder BS equation 
(projected onto  the $i\gamma_5\lambda_b$ channel) for  the amplitude 
$\hat{\Gamma}_b$.   Hence $\hat\Delta^{-1}_b(P^2)$ must have a zero at
the corresponding invariant mass and the general momentum dependence can be 
written as \mbox{$\hat\Delta^{-1}_b(P^2) = (P^2+m_b^2){\cal Z}_b^{-1}(P^2)$}. 
An equivalent observation is that the produced meson mass is 
a  dynamical function $m_b(P^2)$ due to $\bar{q}q$ substructure.  
The field renormalization constant is 
\mbox{${\cal Z}_b = {\cal Z}_b(P^2=-m_b^2)$} and 
a physical normalization (unit residue at the mass-shell pole) is produced 
by absorbing  \mbox{$1/\sqrt{{\cal Z}_b}$}  into the fields.  
The physically normalized BS amplitude is then the dimensionless quantity 
\mbox{$\Gamma_b(p;P)=$} \mbox{$\sqrt{{\cal Z}_b}\hat{\Gamma}_b(p;P)$}.    This
procedure satisfies the canonical Bethe-Salpeter normalization condition 
(Itzykson and Zuber, 1980).  

The functional $\hat{\Delta}_b^{-1}[\hat{\Gamma}_b]$ offers a
very useful strategy for approximate solutions of the ladder BS equation. 
One may use a variational trial function $\Gamma_b(p)$ to minimize the node 
position in the resulting $\hat\Delta^{-1}_b(P^2)$.  This is equivalent
to minimization of the mass functional $m_b[\hat{\Gamma}_b]$. This technique 
was 
introduced by Cahill, Roberts and Praschifka (1987) where it is also assumed
that $\hat\Delta^{-1}_b(P^2)$, in the required time-like region $P^2 < 0$,
can be extrapolated linearly from knowledge of its value and derivative at 
$P^2=0$.  This is equivalent to ignoring momentum dependence of 
${\cal Z}_b(P^2)$.
Then one has immediately \mbox{$m_b^2 \approx {\cal Z}_b\hat\Delta^{-1}_b(0)$}, 
and \mbox{${\cal Z}_b^{-1} = \hat\Delta^{-1~\prime}_b (0)$} where the prime
indicates a derivative with respect to the argument.  From \Eq{delPS} one has
\beq
m_b^2[\Gamma_b] = {\cal Z}_b \; {\rm tr}  
\int \frac{d^4q}{(2\pi)^4}
[i\gamma_5\lambda_b^\dagger S(q)i\gamma_5\lambda_b  S(q) ] \;
\hat{\Gamma}_b^2(q)
+\case{9}{2} {\cal Z}_b \; \int d^4 r\frac{\hat{\Gamma}_b^2(r)}{D(r)},
\label{mfnl}
\eeq
and
\beq
{\cal Z}_b^{-1} = \left. \frac{\partial}{\partial P^2}
 {\rm tr} \int \frac{d^4q}{(2\pi)^4}
[i\gamma_5\lambda_b^\dagger S(q_+)i\gamma_5\lambda_b  S(q_-) ] \;
\hat{\Gamma}_b^2(q) \right|_{P^2=0}  .
\label{zb}
\eeq
When \mbox{$\hat{\Gamma}_b(q) \rightarrow B_0(q^2)$} for the pion, 
Eq.~(\ref{zb}) indicates that $\sqrt{{\cal Z}_b} \hat{\Gamma}_b(q)$ is 
physically normed, and from Eq.~(\ref{pibs}) we see that  
\mbox{${\cal Z}_b^{-1} \approx f_\pi^2$}  is consistent with this present 
level of approximation.  
The assumptions underlying the above two expressions are best suited to small 
masses.  Numerical investigations for a wide range  of meson states were 
presented in Praschifka, Cahill and Roberts (1989).
 
The above variational approach 
requires a specification of the model gluon propagator.
To approximate the pion mass and normalization constant, an approach that has 
been used often is to set \mbox{$\hat{\Gamma}_\pi(q;P) \approx B(q^2)$}
from the DSE solution at finite $m$.  This has the correct chiral limit and
allows the last term of Eq.~(\ref{delPS}) involving $D$ to  be eliminated
by requiring that the GMOR relation (Gell-Mann, Oakes and Renner, 1968) 
\beq 
 m_\pi^2 f_\pi^2 \approx - 2 m \langle\bar{u}u\rangle
\label{gmor}
\eeq
be satisfied.  Since here \mbox{$\Delta_{\pi}^{-1}(0) = m_\pi^2 f_\pi^2$}, 
this approach yields
\begin{equation}
\hat{\Delta}^{-1}_{\pi}(P^2) = - 2m \langle\bar{u}u\rangle + 2N_c \; {\rm tr}_s
\int \frac{d^4q}{(2\pi)^4} [ S(-q) S(q) - S(-q_+) S(q_-)] \; B^2(q^2)  ,
\label{delpigmor}
\end{equation}
The conventional expression for the chiral condensate is
\mbox{$\langle\bar{u}u\rangle =$} 
\mbox{$ - 12 \int^{\mu^2} \frac{d^4q}{(2\pi)^4} \sigma_S^0(q)$}.  The
appropriate scale for such an approach is usually taken as
$\mu^2 = 1~{\rm GeV}^2$.  Mitchell and Tandy (1997) have used this procedure
to model the composite pion propagator in a study of $\rho-\omega$ mixing and
mass splitting. 

A more natural approach, that does not force the GMOR relation, follows from 
using the chiral limit approximation
for the BS amplitude, i.e. 
\mbox{$\hat{\Gamma}_\pi(q;P) \approx B_0(q^2)$}, while keeping the effect of 
a non-zero bare mass in the quark propagators.
This gives the pion mass to first order perturbation from the bare quark mass.
The  last term of Eq.~(\ref{delPS}) may then be recast
solely in terms of the chiral limit quark propagator.  
The chiral limit DSE (see Eqs.~(\ref{sde}) and (\ref{b})) in the form 
\mbox{$B_0(r)/D(r) = \frac{4}{3} {\rm tr}_s S_0(r)$}   leads to
\beq
\case{9}{2}\int d^4 r\frac{B_0^2(r)}{D(r)} = 2N_c \; {\rm tr}_s \int 
\frac{d^4q}{(2\pi)^4} B_0(q^2) S_0(q) . 
\eeq
Then Eq.~(\ref{delPS}) for the pion  becomes
\beq
\hat{\Delta}^{-1}_{\pi}(P^2) =   2N_c \; {\rm tr}_s \int \dq 
[S_0(q) B_0(q^2) - S(-q_+) S(q_-) \; B_0^2(q^2)  ],
\label{delnewpi}
\eeq
where a subscript $0$ indicates a chiral limit quantity.
Now  $\hat\Delta^{-1}_\pi$ is fully specified in terms of the quark propagator.
Numerical evaluation of Eq.~(\ref{delnewpi}) may be used to identify the pion 
mass through the solution of \mbox{$\Delta^{-1}_\pi(P^2=-m_\pi^2) =0$} and  
the normalization constant may be evaluated from
\beq
\label{fpival}
f_\pi^2 = \left.\frac{d}{dP^2}\,\Delta^{-1}_\pi(P^2)\right|_{P^2=-m_\pi^2}~.
\end{equation}
These expressions require that the extrapolation of 
the quark propagator into the complex quark momentum plane be well-defined. 
This is the case with the confining entire function parameterizations that have 
been available in recent years.  Before that development, a common
approximation was to use
\mbox{$\hat\Delta^{-1}_\pi(P^2) \approx f_\pi^2(P^2 + m_{\pi}^2)$} 
so that \mbox{$m_\pi^2 f_\pi^2 \approx \Delta^{-1}(0)$} and 
\mbox{$f_\pi^2 \approx \hat\Delta^{-1~\prime}_\pi(0)$}.  Because of the small 
mass of the pion this is quite accurate and the resulting explicit
pion mass formula (Cahill and Gunner, 1995b; Frank and Roberts, 1996) is
\beq
m_\pi^2 f_\pi^2 \approx 8N_c \int \dq
B_0(q^2) [ \sigma_S^0(q^2) - \frac{B_0(q^2)}{B(q^2)+m} \sigma_S(q^2)]  ,
\label{mpi1}
\eeq
with
\beq
f_\pi^2  \approx  \left. - \frac{d}{d P^2}
2N_c \; {\rm tr}_s \int \dq
S(-q_+) S(q_-)  \; B_0^2(q^2) \right|_{P^2=0}  . 
\label{fpi2}
\eeq  
More explicit expressions are  
\beq
m_\pi^2 f_\pi^2 \approx 2m \; 24 \int \dq
\frac{ B_0(q^2) [q^2 A_0(q^2)A_0'(q^2) + B_0(B_0' +1)] } { q^2 A_0^2 + B_0^2}
,\label{mpi2}
\eeq
to first order in $m$ with \mbox{$f' = \partial f / \partial m$} and subscript 
$0$ labelling chiral limit quantities;  and
\beq
f_{\pi}^2  \approx 
\frac{N_c}{8\pi^2}\int_{0}^\infty\,ds\,s\,B_0(s)^2\,
\left(  \sigma_{V}^2 -
2 \left[\sigma_S\sigma_S' + s \sigma_{V}\sigma_{V}'\right]
- s \hat{\sigma}_S - s^2 \hat{\sigma}_V \right)~,
\eeq
where \mbox{$\hat{f} = f f'' - (f')^2$}, and $s=p^2$.  

We note that in this treatment, the finite value of the pion mass is tied 
directly to the explicit chiral symmetry breaking behavior of the quark 
propagator.  It is determined by the  response of the amplitudes $B(q^2)$ 
and $\sigma_S(q^2)$ to a small current mass.  In realistic studies of the DSE,
it has been found that the response of $\sigma_S(q^2)$ at low momenta to a 
current mass is negative. That feature has been found to be an important element
in generating pion mass via Eq.~(\ref{mpi1}).
Numerical studies of the BS and DSE equations (Frank and Roberts, 1996) 
reveal that, for typical finite current quark masses, 
\mbox{$\hat{\Gamma}_\pi(q;P) \approx B_0(q^2)$} is indeed a good approximation. 
It is also found there that the  pion mass formula in \Eq{mpi1} accurately
reproduces the pion mass found from the BSE solution.

We also note that the pion mass formula in \Eq{mpi1} is different from the GMOR 
relation in \Eq{gmor} which is not exact in QCD.  As a measurement of explicit chiral 
symmetry breaking, the leading term on the RHS of \Eq{mpi1} is first-order in the 
current mass, it gives the response in the pion state (subject to the approximation
\mbox{$\hat{\Gamma}_\pi(q;P) \approx B_0(q^2)$}), and it is therefore determined by
vacuum quark properties.  This is consistent with the general arguments of 
Gell-Mann, Oakes and Renner (1968).  To the extent that one must make further
approximations  to obtain \Eq{gmor}, the relation  \Eq{mpi1} can be thought of as
containing corrections defined by the above treatment within the GCM. 
Hence, as a summary of properties 
of the composite pion mode from either the GCM or the DSE approach, both
truncated to $0^{th}$ order in meson loops, the inverse propagator 
$\hat{\Delta}^{-1}_{\pi}(P^2)$ given by \Eq{delnewpi} is quite accurate and practical 
near the mass-shell.  One qualification is that only the amplitude for the dominant
covariant $\gamma_5$ has been accounted for therein.  As remarked earlier, 
there are recent indications that the common omission of the pseudovector
pion component misrepresents some aspects.  More work on this topic is
warranted.

The  extension of \Eq{delnewpi} to other members of the PS octet is 
straightforward although one would not expect that such a reliance upon first
order effects from explicit chiral symmetry breaking would be as accurate.  
From Eq.~(\ref{delPS})  for the $K^+$ we have 
\beq
\Delta^{-1}_K(P^2)  = 2 N_c\,{\rm tr}_s \int \dq\, [
B^s_0(q^2) S_0^s(q)\, - B^s_0(q^2)^2\, S^u(-q_+) S^s(q_-) ]~.
\label{delinvK}
\eeq
where $S^f$ is the propagator for flavor $f$ and contains $m_f \neq 0$.  
The resulting mass formula is thus
\beq
m_{K}^2 f_{K}^2 \approx   2N_c \; {\rm tr}_s \int \dq 
[B_0(q^2) \; S_0(q)  - B_0^2(q^2)  \; S^u(-q) S^s(q) ]~.
\label{delK}
\eeq
Similarly for the pure flavor octet $\eta$ we have 
\beq
m_{\eta}^2 f_{\eta}^2 \approx    2N_c \; {\rm tr}_s \int \dq 
\left\{ B_0(q^2) \; S_0(q)  - \case{1}{6} B_0^2(q^2) \; 
[S^u(-q) S^u(q) + S^d(-q) S^d(q) + 4 S^s(-q) S^s(q) ] \right\}  .
\label{deleta}
\eeq
Again \mbox{$f_b^2 \approx \hat\Delta^{-1~\prime}_b(0)$} for each case. 
The results to first order in the bare masses can be expressed
\beq
m_\pi^2 f_\pi^2 = (m_u +m_d) \rho \; ; \;  
m_K^2 f_K^2 = (m_u +m_s) \rho \; ; \;
m_\eta^2 f_\eta^2 = \frac{(m_u +m_d + 4 m_s)}{3} \rho ,
\label{masses}
\eeq
where $\rho$ can be identified from  Eq.~(\ref{mpi2}) and it should be noted
that the constants $f^2$ are identical in the chiral limit.  With degenerate
$u$ and $d$ masses, the Gell-Mann--Okubo mass formula 
\mbox{$  m_\pi^2 + 3 m_\eta^2 = 4 m_K^2 $} follows immediately.

\subsect{Chiral Observables}
\label{sect_chiral}

In principle, if a systematic approach is taken to produce from the GCM an effective
action for all the Goldstone modes, and an expansion in momenta and masses is made, then
the results will be predictions for the coefficients of the effective action
employed in chiral perturbation theory (ChPT).  This is because the GCM respects 
dynamically broken chiral symmetry and hence will share with ChPT all the consequences
of this.   The same relationship is in principle shared between QCD and ChPT.  Given 
the difficulty of deriving the low energy constants of ChPT directly from QCD, they
are invariably determined as phenomenological parameters from a canonical set of data
in the form of pseudoscalar meson decays, mass relationships, etc.   Then predictions
are made for other low energy experimental quantities.   One may investigate 
whether a model field theory such as the GCM can provide some insight into the 
microscopic content of the low energy constants of ChPT.   Although some threshold
chiral observables including the $\pi-\pi$ scattering lengths and the pion charge radius
have been studied within the GCM, some caution should be exercised in comparing 
extracted low energy constants with values in use in ChPT. This is because 
constants that control the coupling of point interpolating fields with pion quantum
numbers, as defined by the form of an effective field theory such as ChPT,  imply
an ordering of contributing mechanisms that does not necessarily correspond to  the 
ordering associated with the coupling of distributed pionic $\bar q q$ correlations
produced by an  underlying theory at the quark-gluon level. 

The bosonization of the GCM to produce the meson action in
\Eq{mesaction} provides an illustration.   The fields are ladder $\bar q q$ objects,
the couplings are expressed as distributed vertex functions, and this is only the
tree-level result.  In principle, one may integrate up through the loop expansion, 
and integrate out the non-Goldstone mesons,
stopping when the effect on a given threshold observable has (hopefully) converged. 
For an observable at given order in external momenta, the relative importance of a 
particular type of coupling or a particular order in loops can be expected to be
influenced significantly by the treatment of the nonlocalities or finite size effects.
These are characteristics of the non-point nature of the states.   The meson loop 
integrals in the GCM are finite due to intrinsic substructure in the fields and hence
in the vertex functions.  To compare parameters in detail with ChPT, it matters 
whether one makes a derivative 
expansion of the GCM after quantum loop integrations or before.  In the 
former  case, the final importance of loops is a characteristic of the GCM; the latter
case is not really consistent.  Even a  localization of the tree-level couplings
means that the distributed nature of the dressed quark substructure of the bare mesons
and interactions is ignored and this can change the relative contributions to 
observables.  Unambiguous comparisons of the GCM and ChPT must be made through the
calculated observables.  Most of the investigations within the GCM have been at 
tree-level because loop effects are technically difficult due to nonlocalities.  As we
will discuss later, investigations of pion loop contributions within the GCM 
exist  for a few selected observables and indicate small contributions, typically at 
the $10$\% level.   Much more work remains to be done in that direction.
 
In two recent works, low energy chiral observables  were considered to fourth order 
in gradients and explicit chiral symmetry breaking masses.  The usual chiral power 
counting is observed (Gasser and Leutwyler, 1984, 1985; Donoghue, Golowich
and Holstein, 1992). 
The first derivation and analysis of the Goldstone modes and chiral 
coefficients to ${\cal O}(p^4)$ from the GCM was made by Roberts, Cahill and 
Praschifka (1988a).  The identification of the sums of quark loop integrals
that contribute to masses, decay constants, three ${\cal O}(p^4)$ coefficients,
and the coefficient of the anomalous Wess-Zumino (1971) five-pion term was
made in that work.  This is a complicated task and the interested reader is 
referred to that paper for details.  The implications for $\pi-\pi$ 
scattering from the terms through ${\cal O}(p^2)$ were also drawn in that work
and in particular the Weinberg (1966) result for the $S$-wave scattering
lengths was well reproduced.  
The work of Roberts, Cahill, Sevior, and Iannella (1994) extended the 
application for $\pi-\pi$ scattering to include the implications from the 
three ${\cal O}(p^4)$ coefficients considered and to employ a more realistic
dressed quark propagator model. 

To summarize the approach, we first note that in the chiral limit
of zero bare quark mass,  when 
\beq
S^{-1}(x-y;U)=\gamma \cdot \partial _xA_0(x-y) + B_0(x-y)V ~,
\label{propv}
\eeq
solves the DSE for $V=1$, it also does so for any constant unitary matrix $V$
of the form (Praschifka, Roberts and Cahill (1987a))
\beq
V = {\rm exp} \left( \frac{i \gamma_5 \lambda^b \phi^b} {f_\pi} \right) = 
P_R \; U + P_L \; U^{\dagger} ~.
\label{V}
\eeq 
Here the $\phi^b$ are constants and $P_{R,L}$ the standard 
right and left-handed projection operators \mbox{$(1 \pm \gamma_5)/2$}.
(One may include in \Eq{V} the $\lambda^0 \propto 1_f$
term for the singlet $\eta'$ but we restrict attention to the Goldstone 
flavor octet.)  With a set of constants $\phi^b$ chosen to select one out of the
degenerate vacuua, fluctuation fields will be incorporated via the 
coordinatization \mbox{$\phi^b \rightarrow \phi^b(\frac{x+y}{2})$}.  Without
loss of generality, the vacuum point about which to expand the action can
be taken to be $V=1=U$.  In terms of the chiral field $U(R)$  defined by
\beq 
U(R) = {\rm exp} \left( \frac{ i\lambda^b \phi^b(R)} {f_\pi} \right) ~,
\label{U}
\eeq
the action, given earlier by \Eq{s4}, may be written for the goldstone octet
as
\beq
{\cal S}[U] = -{\rm TrLn} S^{-1}[U] 
+  \case{1}{2} {\rm tr} \int d^{4}r \, d^{4}R \: \left\{ 
\left( \Sigma_V (-r) + B_0(-r) U^\dagger (R) U(R) \right) S(r;1) \right\} ~, 
\label{chirals}
\eeq
where $\Sigma_V$ is the Dirac vector part of the quark self-energy.
Although the second term of this action is irrelevant under the present 
circumstances where the constraint
$U^\dagger U =1$ is operative, it is relevant in general.  One example would
be the generation, along these lines, of the composite $\bar q q$ version of 
the linear sigma model where $U^\dagger U =\chi^2$ with $\chi$ the chiral 
radius field variable (Cahill and Roberts, 1985; Frank and Tandy, 1992).

The physically relevant action
\mbox{$\hat{ {\cal S} } [U] = {\cal S}[U] - {\cal S}[1]$}
has a real part
\beq
\hat{ {\cal S}}_R[U] = -\case{1}{2}\mbox{TrLn}
\left(  \left[ S^{-1}(U)\right]^\dagger S(1)^\dagger S(1)  S^{-1}(U) \right) .
\label{Sreal}
\eeq
With a perturbative introduction of the bare quark mass matrix 
\mbox{$ M_{\mu^2} = $} \mbox{${\rm diag} (m_u, m_d)_{\mu^2}$}, 
the low energy form through fourth order in 
gradients and masses for the terms relevant to $\pi-\pi$ scattering was
derived by Roberts, Cahill, Sevior, and Iannella (1994) in the form
\begin{eqnarray}
\hat{ {\cal S}}_R[U]& = & \int d^4x\, \left\{ 
\case{f_{\pi}^2}{4}{\rm tr}\left[\partial_\mu U\partial_\mu U^{\dagger}\right]
+ \case{\langle \bar qq \rangle_{\mu^2} }{4}{\rm tr}
\left[\left(2I_f - U - U^{\dagger}\right) M_{\mu^2}\right]
\right.\nonumber  \\
& & \!\!\!\!\!\!\!\!\!\!\!\!\!\!\!\!\!\!\!\!\!\!\!\!\!\!\!\!
\left. - N_c K_1 \; {\rm tr}\left[\partial^2 U\partial^2 U^{\dagger}\right]
+ N_c K_2 \; {\rm tr}\left[
                \left(\partial_\mu U\partial_\mu U^{\dagger}\right)^2\right]
- N_c K_3 \; \case{1}{2} \; {\rm tr}\left[\partial_\mu U\partial_\nu U^{\dagger}
                \partial_\mu U\partial_\nu U^{\dagger}\right]\right\} ~.
\label{spscat}
\end{eqnarray}
Here the  field equations of motion at second order, which could 
be used to eliminate the explicit appearance of the $K_1$ term, have not been
applied. The coefficients $K_1, \, K_2$, and $K_3$ are given by convergent quark
loop integrals which require only the chiral limit dressed quark propagator. 
In terms of the standard chiral coefficients $L_{1-10}$ (Gasser and 
Leutwyler, 1985), the form in \Eq{spscat} corresponds to 
\mbox{$2L_1 = L_2 = $} \mbox{$N_c K_3/2$}, \mbox{$ L_3 = N_c(K_1 - K_2 -K_3)$}, and
\mbox{$ L_8 = - N_c K_1/4 $}, with the remaining \mbox{$L_i = 0$}. 

The imaginary part of the action $\hat{ {\cal S}}[U]$ involves an odd number
of fields and corresponds to all of the anomalous interactions.  
The lowest order contribution is fifth order in chiral fields and is
\beq
i \hat{ {\cal S}}_I[U] = \case{\lambda N_c} {240 \pi^2} 
\epsilon_{\mu \nu \rho \sigma} \case{ {\rm tr} } {2} \int d^4x \,
\left\{ \left( U(x) - U^\dagger(x) \right) \partial_\mu U^\dagger(x)
\partial_\nu U(x)  \partial_\rho U^\dagger(x) \partial_\sigma U(x) \right\} ~,
\label{WZ}
\eeq
where $\lambda$ is given by a convergent integral involving self-energy 
amplitudes $A_0(p^2)$ and $B_0(p^2)$ and is given in Praschifka, Roberts 
and Cahill (1987a).  It is remarkable that this quark loop
integral gives exactly $\lambda=1$ independently of the detailed form of the
quark propagator.  Expansion of \Eq{WZ} to lowest order in
pion fields yields exactly the five-pion form of the Wess-Zumino term. This is
a characteristic feature in that the form and strength of anomalies required
by symmetries are properly embedded in the GCM. It is only necessary to 
properly maintain the symmetry constraints among the various dynamical 
quantities at the quark level.  In the above case, the $\lambda=1$ result is
critically dependent upon the symmetry result that $B_0(p^2)$ is both the 
chiral limit  $\gamma_5$ piece of the pion BS amplitude (as required by the 
axial Ward identity) and the 
quark scalar self-energy resulting from the dynamical breaking of chiral 
symmetry.  Further studies of the anomalies generated by the GCM have been
made by Roberts, Praschifka and Cahill (1989) where it is shown that 
extension of \Eq{WZ} to include vector mesons at the beginning of the 
derivation reproduces all of the anomalous interactions.

The first two terms of \Eq{spscat}, the "kinetic" terms for the chiral field
$U(x)$, generate the familiar Weinberg result (Weinberg, 1966) for the 
invariant $\pi-\pi$ scattering amplitude
\beq
A_W(s,t,u) = \frac{ s - m_\pi^2 }{ f_\pi^2}
\label{aw}
\eeq
that provides  the leading low-energy behavior.  The scattering amplitude
contributions from the ${\cal O}(p^4)$ action are linear in $K_{1-3}$ and were
calculated from the dressed quark substructure of the pions by 
Roberts, Cahill, Sevior, and Iannella (1994).  That work employed 
model  three-parameter forms of the quark self-energy amplitudes designed to 
reproduce the characteristic features that emerge from all studies of the 
quark propagator via the DSE.  It was demonstrated that the infra-red 
structure of the dressed quark propagator in the GCM
may be modeled to provide a sensible description,
not only for $f_\pi$, $m_\pi$ and $r_\pi$ which were the principal concerns
to that stage, but also for the main features of the $\pi-\pi$ scattering 
lengths $a^I_l$.   A reasonable description of the principal qualitative features
of the $\pi-\pi$ partial wave amplitudes for \mbox{$l \leq 2$} and $E \leq 4 m_\pi$
was also obtained.   In this work,
the leading asymptotic behavior of the propagator amplitudes corresponded
to the outcomes from the operator product expansion and QCD renormalization
(Politzer, 1976; 1982).  In this and subsequent studies, it is observed that the
logarithmic corrections inherent in such UV behavior, and associated with the
anomalous dimension of the propagator, do not provide significant numerical
contributions to physical quantities.
\begin{table}[ht]
\begin{center}
\parbox{130mm}{\caption{ A comparison between the low-energy $\protect\pi$ 
observables calculated at tree-level in the GCM using the parameters of 
Table~\protect\ref{tab_specfit}  
and their experimental values.   The superscripts indicate the reference that 
experimental values are taken from: 1 - Particle Data Group (1996); 
2 - Amendolia {\protect\it et al.}~(1986a); 3 - Pocani\protect\'c (1995) and 
Gasser (1995).   
The value for $\protect\langle \protect\bar q q \protect\rangle $ quoted as 
experimental is that typically used 
in QCD sum rule analysis.  (Adapted from Roberts (1997)). \label{tab_pion} } }

\begin{tabular}{|cll|} \hline
   & Calculated  & Experiment  \\ \hline
  $f_{\pi} \; $    &  ~0.0924 GeV &   ~0.0924 $\pm {\rm 0.001}^1$     \\ \hline
  $-\langle \bar q q \rangle^{\frac{1}{3}}_{1\,{\rm GeV}^2}$ & ~0.247 GeV &  
        ~0.220 $\pm$ 0.050\\ \hline
  $m^{\rm ave}_{1\,{\rm GeV}^2}$ & ~0.0045 GeV & ~0.008 $\pm {\rm 0.007}^1$
                \\ \hline
 $m_{\pi} \; $    & ~0.139 GeV & ~${\rm 0.138}^1$  \\ \hline\hline
 $r_\pi \;$ & ~0.547 fm & ~0.663$\pm {\rm 0.006}^2$  \\  \hline\hline
 $g_{\pi^0\gamma\gamma}\;$ & ~0.505 (dimensionless) & 
                                    ~0.500 $\pm {\rm 0.018}^1$ \\ \hline\hline
 $a_0^0 \;  $ & ~0.170  & ~0.26$\pm {\rm 0.05}^3$ \\ \hline
 $a_0^2 \;  $ & -0.0481 & -0.028 $\pm {\rm 0.012}^3$ \\ \hline
 $a_1^1 \;  $ & ~0.0301 & ~0.038 $\pm {\rm 0.002}^3$ \\ \hline
 $a_2^0 \; $  & ~0.00149 & ~0.0017 $\pm {\rm 0.0003}^3$ \\ \hline
 $a_2^2 \;  $ & -0.000214 &  ~0.00013 $\pm {\rm 0.0003}^3$ \\ \hline\hline
 $K_1   \;  $ & ~0.000702  &  \\ \hline
 $K_2   \;  $ & ~0.000867  &  \\ \hline
 $K_3   \;  $ & ~0.000875  &  \\ \hline
\end{tabular}
\end{center}
\end{table}

In \Table{tab_pion} we show results for low energy pion 
observables including the $\pi-\pi$ scattering lengths and associated chiral 
coefficients $K_{1-3}$ from a tree-level calculation within the GCM that uses 
the parameterized confining quark propagator model discussed 
in  Sec.~\ref{sect_prop}.  The original GCM calculation of the $\pi-\pi$ scattering
lengths (Roberts \etal, 1994) employed a quite different propagator parameterization;
we choose the more recent calculation here for consistency with other elements of this 
review.  The collection of quantities shown in 
Table~\ref{tab_pion} were fitted with the propagator model given in 
Eqs.~(\ref{ssb}) and (\ref{svb}) and with the parameters shown in 
Table~\ref{tab_specfit}.  Although the results for observables are very similar to 
those that also appear in the more complete set shown as \Table{tab_pk_res}, 
the relevant low energy constants are extracted only for this present calculation.  
The point meson limit of this GCM calculation, which corresponds
to the NJL model, produces \mbox{$K_i = 1/96\pi^2 = 0.00106 $} and the comparison
with \Table{tab_pion} indicates roughly that one can expect a dependence upon finite
size effects associated with meson substructure at the $20$\% level.

The values of the $\pi-\pi$ scattering lengths shown in \Table{tab_pion} are close
to the current algebra or soft pion theorem results (Weinberg, 1966).  The latter
produces \mbox{$a^0_0 = 0.16$} for example.   In ChPT up to one pion loop at 
${\cal O}(p^4)$, this value is corrected upwards by about $30$\% to become 
$0.20 \pm 0.01$ (Gasser and Leutwyler, 1983).   Indications from two-loop results at 
${\cal O}(p^6)$ are that the next correction is small (giving $0.217$) and that 
it presently appears to be unlikely that ChPT will reach $0.26$ (Ecker, 1996).
The pion loop contribution to $a^0_0$ has not been calculated in the GCM.  Pion loop
contributions in the GCM can in principle be obtained for each chiral observable
independently and this is clearly a topic that deserves attention in the future.   
A meaningful comparison with one-loop ChPT clearly requires that loop contributions 
in the GCM be obtained for a number of observables at least equal to the number of
independent low energy coefficients  at a given order in ChPT.  This is because a 
general observable is a linear combination of these coefficients.  For example, for 
\mbox{$N_f = 2$}, one would need as many observables as needed to determine the
\mbox{$\bar{l}_i, i=1-6$} of Gasser and Leutwyler, (1984).   The quantity $a_0^0$ is
a linear combination of \mbox{$\bar{l}_1, \bar{l}_2, \bar{l}_3$} that are determined
by fitting $D$-wave scattering lengths and $SU(3)$ mass relations.   

With \mbox{$N_f = 3$}, Frank and Meissner (1996) have studied 
the  ${\cal O}(p^4)$ chiral coefficients $L_{1-10}$ that can be produced at tree-level 
in the GCM.  They were particularly interested in the relationship between the values
of the coefficients  and the chosen infra-red behavior of the phenomenological 
$2$-point gluon function employed in the GCM.
In order to facilitate a chiral invariant derivation, they adopt the common
procedure of coupling the quarks to an external field 
$\psi (x)= (s + i\gamma_5 p)$ which finally will be set to the bare quark mass 
matrix (Gasser and Leutwyler, 1983; 1985; Donoghue, Golowich and 
Holstein, 1992).  The $x$-dependence of $\psi (x)$ helps prevent unphysical 
results for some of the  coefficients and also enters into 
the equation of motion which is obtained at second order and employed to 
implement relations among the coefficients.   They use
\begin{equation}
 S^{-1}(x,y;U)=\gamma \cdot \partial _r A_0(r)+\psi (R) \delta(r) 
+B_0(r)V(R) ,
\label{sinvch}
\end{equation}
with \mbox{$r=x-y$} and \mbox{$R=(x+y)/2$}, 
and show that the produced tree-level structure of the GCM, to ${\cal O}(p^4)$,
is
\begin{eqnarray}
{\cal S}_R = \int d^4x &\Biggl\{ & \case{f^2_{\pi }}{4}\mbox{tr}
\left[ (\partial _{\mu}U)(\partial _{\mu}U^{\dagger})\right] 
-\case{f^2_{\pi }}{4}\mbox{tr}\left[ U\chi ^{\dagger} +\chi U^{\dagger}\right] 
\nonumber \\
&-&L_1\left( \mbox{tr}\left[ (\partial _{\mu}U)
(\partial _{\mu}U^{\dagger})\right] \right) ^2 -L_2 \;
\mbox{tr}\left[ (\partial _{\mu}U)(\partial _{\nu}U^{\dagger})\right]
\cdot \mbox{tr}\left[ (\partial _{\mu}U)(\partial _{\nu}
U^{\dagger})\right] \label{chiact} \\
&-&L_3 \; \mbox{tr}\left[ (\partial _{\mu}U)(\partial _{\mu}U^{\dagger})
(\partial _{\nu}U)(\partial _{\nu}U^{\dagger})\right] +L_5 \;
\mbox{tr}\left[ (\partial _{\mu}U)(\partial _{\mu}U^{\dagger})
(U\chi ^{\dagger} +\chi U^{\dagger})\right] \nonumber \\
&-&L_8 \; \mbox{tr}\left[ \chi U^{\dagger}\chi U^{\dagger} + 
U\chi ^{\dagger }U\chi ^{\dagger }\right] \Biggr\} ,\nonumber
\end{eqnarray}
where $\chi (x)=-2\langle \bar{q}q\rangle\psi (x)/f^2_{\pi }$ is of order $2$
(in mass) and the remaining trace is over flavor.  
\begin{table}[ht]
\begin{center}
\parbox{130mm}{\caption{Parameters for the quark propagator model in 
Eqs.~(\protect\ref{ssb}) and (\protect\ref{svb}) that produce the fit shown in 
Table~\protect\ref{tab_pion} for the low energy pion quantities.
Note \protect\mbox{$\protect\Lambda= 10^{-4}$} is not varied. 
At the finite bare mass values, \protect\mbox{$C_m =0$}. 
\label{tab_specfit} } }

\begin{tabular}{|cc|}\hline
                  & $u/d$-quark      \\ \hline
  $\bar{m} \; $   &  ~0.008         \\ \hline
  $b_0     \; $   &  ~0.181            \\ \hline
  $b_1     \; $   &  ~2.89              \\ \hline 
  $b_2     \; $   &  ~0.555            \\ \hline 
  $b_3     \; $   &  ~0.184            \\ \hline
  $C_{m=0} \; $   &  ~0.119             \\ \hline 
  $D       \; $   &  ~0.159 ${\rm GeV}^2$  \\ \hline
\end{tabular}
\end{center}
\end{table}

The chiral coefficients, classified according to Gasser and Leutwyler (1985),
are produced in this work as convergent quark loop integrals. 
Of the complete set of ${\cal O}(p^4)$ coefficients labelled $L_{1-10}$,  only 
those shown in \Eq{chiact} occur at tree-level in the GCM.  
The coefficients $L_9$ and $L_{10}$ refer to external vector and axial vector 
fields and are not considered here.  Application of the 
$SU(3)$ relation (Donoghue, Golowich and Holstein, 1992)
between the terms fourth order in $U$ produces $L_2=2L_1$.  
Apart from the special case of $L_7$ which becomes finite when the $U_A(1)$ 
anomaly is accounted for (Witten, 1979), these relationships agree with 
what is expected in the large $N_c$ limit of QCD (Donoghue, Golowich and 
Holstein, 1992).  This is presumably related to the fact that
the large $N_c$ limit of QCD corresponds to a ladder description of mesons 
if the three and four gluon couplings are neglected as they are in the GCM
(Frank and Meissner, 1996).  A complicating element for considering a   
$1/N_c$ expansion for the GCM is that a phenomenological two-point gluon
function has an unknown dependence upon $N_c$.  The neglect of
higher mass states, explicitly excludes intermediate states of pure
glue which are ``$N_c$ suppressed".   Since the GCM action at the quark level is
consistent with explicit and dynamical breaking of chiral symmetry, the complete set
of coefficients must arise if higher mass states and non-Goldstone meson modes
were not simply discarded but integrated out, and if loop contributions were 
implemented.   The minimum steps of this type needed to generate the complete set
of coefficients in the GCM is not known. 

In principle, the four independent coefficients (which can be taken as \mbox{$L_1, 
L_3, L_5, L_8$} ) produced by the GCM at tree-level in \Eq{chiact} 
depend on the details of the dressed quark propagator and hence
implicitly on the form of the effective quark-quark interaction $D$.  With
\mbox{$g^2 D(q^2)=$} \mbox{$4\pi \alpha (q^2)/q^2 $},
Frank and Meissner (1996) have investigated three different 
two-parameter models for $\alpha (s)$, namely
\begin{eqnarray}
\alpha _1(s)&=&3\pi s\chi ^2\frac{e^{-s/\Delta}}{4\Delta^2}
+\frac{\pi d}{\mbox{ln}(s/\Lambda^2 +e)}\nonumber \\
\alpha _2(s)&=&\pi d\left[ \frac{s\chi^2}{s^2+\Delta }+
\frac{1}{\mbox{ln}(s/\Lambda^2 +e)}\right] \label{alfs} \\
\alpha _3(s)&=&\pi d\left[ \frac{1+\chi e^{-s/\Delta}}
{\mbox{ln}(s/\Lambda^2 +e)}\right]  ,\nonumber
\end{eqnarray}
where \mbox{$s=p^2$} and \mbox{$d=12/(33-2N_f)$}.
Each of these forms incorporates the one-loop perturbative result
for large $s$ and extrapolates differently into the
low-momentum region.  The two low-momentum parameters,
$\chi $ and $\Delta $, are varied with the pion decay constant held fixed
at $f_\pi =86$~MeV.  This value was considered to be appropriate for a chiral limit
calculation at zero-momentum rather than the physical value of $93$~MeV appropriate
to the physical mass-shell.   The results however are not very sensitive to this 
small difference.  By fixing $f_\pi$ the overall scale of D$\chi $SB is fixed.
The remaining independent parameter
is associated with the matching scale to the perturbative form.

The first model, $\alpha _1$ in Eq.~(\ref{alfs}), corresponds to that in 
\Eq{gcmD} which, as discussed earlier, has been used in many previous
investigations within the GCM (Praschifka \etal, 1989).  Here Frank and 
Meissner (1996) use slightly different parameters due to their chosen 
$f_\pi $ value.    The small $\Delta $ limit of this
model simulates  the infrared-dominant  behavior
which is known to model confinement (Burden \etal, 1992).
The infrared contribution to the second model, $\alpha _2$, generates a
$1/q^4$ singularity in the quark-quark interaction $D(q^2)$ in the limit as
$\Delta \rightarrow 0$.  Such a singularity has also been considered 
previously as a model of confinement (Marciano and Pagels, 1978).
The $1/q^4$ form falls much slower than the Gaussian in $\alpha _1$, and hence
matches the UV form at a much higher scale.
Finally the third model $\alpha _3$ was chosen here to be
structurally different from $\alpha _1$ and $\alpha _2$ in order to further
illustrate the independence of the results to the details of the low-momentum
parameterization.

In all three cases the same pattern is observed:
the coefficients $L_1$ and $L_3$, which are important for $\pi$-$\pi$
and $K$-$K$
scattering, are nearly independent of the form of $\alpha (s)$ and therefore
on the form of the quark-quark interaction, provided that the integrated
strength of $\alpha (s)$ is fixed by $f_\pi$.
On the other hand, the mass dependent coefficients, $L_5$ and $L_8$,
are moderately dependent on the actual form of the interaction.
For example with $L_5$,  in order to
reproduce the empirical value, forms of $\alpha (s)$
with a small matching scale, i.e. which are relatively strong in the infrared
region, are required.
This observation agrees with the findings by Meissner (1994),
where it is shown that infrared-dominant forms of the quark-quark interaction 
are required
to achieve convergence of the chiral series in the strange quark sector.
This also confirms the success of the
``delta-function-plus-tail"  models (obtained for example from $\alpha_1$
in the limit $\Delta \to 0$) in describing chiral
observables as most recently observed by Frank and Roberts (1996). 

Frank and Meissner (1996) also find that the results for the fourth-order 
coefficients are rather
insensitive to the asymptotic UV tale of $\alpha (s)$;
even omitting this tail completely gives no significant changes,
again provided that $f_\pi$ is fixed.   Typical results are illustrated in 
\Table{tab_meissner}.  The mixed quark-gluon condensate 
\mbox{$ -g_s \langle  {\bar q} \sigma G q \rangle_{1~{\rm GeV}^2}$}, as recently
calculated by Meissner (1997), is also shown in  \Table{tab_meissner}.
\begin{table}[ht]   
\begin{center}
\parbox{130mm}{ \caption{Low energy chiral coefficients, the quark condensate, 
and the quark-gluon mixed condensate from the GCM at meson tree-level. The 
interaction described by $\protect\alpha_1(s)$ in Eq.~(\protect\ref{alfs}) is 
employed with 
\protect\mbox{$\protect\Lambda = 0.2~{\protect\rm GeV}$}, 
\protect\mbox{$N_f = 3$}.  Adapted from Frank and Meissner 
(1996) and Meissner (1997).  Empirical values shown in the last row
are taken from Mei\protect\ss ner (1993) and Bernard, Kaiser and Mei\protect\ss ner 
(1995) for the 
chiral coefficients, and from the quenched lattice of Kremer and Schierholz (1987) 
for the condensates.   \label{tab_meissner}  } }

\begin{tabular}{|c|c||c|c||c|c|c|c|}
\hline
$\Delta $ & $\chi $ & $ -\langle \bar{q}q\rangle^{\frac{1}{3}}_{1~{\rm GeV}^2} $  &
$ -g_s \langle  {\bar q} \sigma G q \rangle^{\frac{1}{5}}_{1~{\rm GeV}^2}$ &
$L_1 $ & $L_3 $ & $L_5 $ & $L_8 $
\\ \hline
$[\mathrm{GeV}^2]$ & $ [\mathrm{GeV}]$ & $[\mathrm{MeV}]$ & $ [\mathrm{MeV}]
 $ & $*10^3$ & $*10^3$ & $*10^3$ & $*10^3$
\\ \hline \hline
$0.200$ & $1.65$ & $173$ & $458$ & $0.81$ & $-4.03$ & $1.66$ & $0.83$ \\
  \hline
$0.020$ & $1.55$ & $160$ & $448$ & $0.82$ & $-4.39$ & $1.13$ & $0.84$ \\
  \hline
$0.002$ & $1.45$ & $151$ & $432$ & $0.85$ & $-4.43$ & $0.97$ & $0.88$ \\
  \hline \hline
   &   &  $225$  &  $402-429$  & $0.7\pm 0.5$  & $-3.6\pm 1.3$ & $1.4\pm 0.5$ &
 $0.9\pm 0.3$ \\ \hline 
\end{tabular}
\end{center}
\end{table}

\subsect{Other Mesons}
\label{sect_other}

Beyond the light Goldstone bosons it is necessary to use the BSE and
solve with an approximate kernel.  Most common is the ladder approximation
and in the GCM or DSE approach one must adopt an Ansatz for the effective 
gluon $2$-point function $D(q)$.  It is not possible to use chiral symmetry 
to avoid
this in the manner of Sec.~\ref{sect_GB}.  To be consistent with what is
arrived at naturally in the GCM, and consistent with preservation of the
Goldstone theorem, the employed dressed quark propagators
should be from rainbow approximation DSE using the same gluon function $D(q)$.  
The most extensive and phenomenologically successful spectroscopic studies in
the rainbow-ladder framework are those of Jain and Munczek (1993) in which the
quark DSE is solved numerically for spacelike-$p^2$ using a model gluon
propagator.  In Landau gauge the behavior of the gluon propagator is
constrained by perturbation theory for $k^2>1-2$~GeV$^2$ 
(Brown and Pennington, 1989) and one
models the infrared behavior, which is presently unknown.  Such studies have
the ability to unify many observables via the few parameters that
characterize the behavior of the model dressed-gluon propagator in the
infrared.

In this approach, solving the meson BSEs is complicated by the fact that
these equations sample the dressed-quark propagator off the spacelike-$p^2$
axis.   This difficulty was circumvented by Jain and Munczek (1993) through
a derivative expansion of the dressed-quark propagator functions, $A(p^2)$ and
$B(p^2)$, and estimating the error introduced thereby.  This, however,
obscures the discussion and exploration of the role of quark and gluon
confinement.  The successful use of parameterized entire function quark
propagators for the light Goldstone bosons as described earlier has recently
led to several investigations, roughly within the framework of the GCM and
DSE approach, to see whether the constraints from pion and kaon
physics contained therein can also constrain the properties of a wider class
of light mesons.  We describe briefly  approaches by Cahill and Gunner (1995a),
and by Burden \etal~(1997), both of which are organized around use of a 
separable Ansatz for the ladder-rainbow BSE kernel.   The purpose in choosing
such an unusual representation is so that, when employed in the DSE it can
easily be arranged to reproduce specified dressed 
quark propagators previously fitted to pion and kaon physics. The question
is then asked whether subsequent use of this kernel Ansatz in the BSE can 
produce the dominant physics of certain other mesons such as the light vectors
and axial vectors. 

Consider the ladder-rainbow BSE  for a bound state of a quark of
flavour $f_1$ and an antiquark of flavour $\bar f_2$ in the form
\begin{equation}
\Gamma(p;P) = - \case{4}{3} \int \frac{d^4q}{(2\pi)^4}
D(p - q) \gamma_\mu S^{f_1}(q + \xi P) \Gamma(q;P)
        S^{f_2}(q - \bar\xi P) \gamma_\mu  ~, \label{bspi}
\end{equation}
where \mbox{$\bar\xi = 1-\xi$}, and for simplicity a Feynman-like gauge has again 
been used.   Here the quark and antiquark momenta are \mbox{$p_1 = q+\xi P$} and
\mbox{$p_2 = -q + \bar\xi P$} respectively while the relative momentum that has been
introduced is \mbox{$q = \bar\xi p_1 -\xi p_2$}.  The conjugate coordinates to $P,q$
are \mbox{$X = \xi x_1 + \bar\xi x_2$} and \mbox{$x = x_1 - x_2$}.   If we were 
dealing with the exact Bethe-Salpeter amplitude and equation, the definition of a 
relative momentum variable is a matter of convenience and $\xi$ is arbitrary 
(Itzykson and Zuber, 1980).   However the ladder approximation does not respect this
arbitrariness and physical observables depend on $\xi$. 
This is an unavoidable deficiency of the ladder approximation and one must impose a
physical constraint to fix $\xi$.  With equal masses  \mbox{$m_{f_1} = m_{f_2}$} as 
in the case of the pion,  charge conjugation symmetry requires \mbox{$\xi = 1/2$}. 
For unequal masses such as for the kaon where there is no charge conjugation symmetry,
one can, for example, choose $\xi$ so that the total charge of the neutral member is
zero.   This is discussed further in relation to the kaon charge form factor in 
\Sec{sect_emff}.  

One can write, without loss of generality,
\begin{equation}
\label{dfeynman}
D(p-k) = \sum_{n=0}^\infty\,D_n(p^2,k^2)\,p^n\,k^n\,\case{1}{2^n}\,
U_n(\hat p\cdot\hat k)~,
\end{equation}
where $\hat p$ is the unit-magnitude direction-vector for $p$ and
$\{U_n(x)\}_{n=0}^\infty$ are the complete set of orthonormal Tschebyshev
functions.  Translational invariance is preserved if all contributing 
Tschebyshev moments are retained.
Substitution of \Eq{dfeynman} into the corresponding DSEs, given in \Eq{a} 
and \Eq{b}, for quark propagator amplitudes
shows that, because the Tschebyshev orthogonality weight function matches
the angle integration measure that arises naturally 
in the DSE,  only the  moment $D_1(p^2,k^2)$ enters the
DSE for $A(p^2)$  and only the moment $D_0(p^2,k^2)$ enters the DSE for
$B(p^2)$.  Hence using the simplest separable representation
\begin{equation}
\label{sepD}
D(p-k) = G(p^2)\,G(k^2)\,+\, F(p^2) \,p\cdot k\, F(k^2) ~,
\end{equation}
both DSEs, \Eq{a} for $A$  and \Eq{b} for $B$, are satisfied if
\begin{equation}
\label{sepone}
\begin{array}{cc}
 G(s) = \case{1}{b} B(s) \; ~, &  \;
 F(s) = \case{1}{a} \left(A(s) -1 \right)~,
\end{array}
\end{equation}
where \mbox{$s=p^2$} and $a$ and $b$ are constants that are determined by
\begin{eqnarray}
\label{aparam}
a^2 & = & \frac{1}{24\pi^2}\int_0^\infty\,dt\,t^2\,[A(t)-1]\,\sigma_V(t)~,\\
\label{bparam}
b^2 & = & \frac{1}{3\pi^2}\int_0^\infty\,dt\,t\, B(t)\,\sigma_S(t)~.
\end{eqnarray}
This can be done for any quark flavor.  Note that the UV behavior of the
propagator amplitudes  in general require some regulation to allow $a$
and $b$ to be determined. Therefore the applications are restricted suitably to 
soft physics.   The separable Ansatz BSE kernel is now determined 
totally by the quark
propagator and it can be seen that in the chiral limit the Goldstone theorem
with a massless pseudoscalar is preserved.  With given quark propagator 
amplitudes ($\sigma_S$ and $\sigma_V$, or $A$ and $B$)
constrained by pion and kaon
physics, this BSE kernel Ansatz can be used to investigate the extent to which 
the major features of other light mesons are predicted.    
\begin{table}[ht]
\begin{center}
\parbox{130mm}{\caption{ Selected meson quantities obtained
by Cahill and Gunner (1995a) from  application of a simple form of the
separable BSE Ansatz fitted to the observables shown.  References for 
experimental values can be found in Table~\protect\ref{tab_pion}. \label{tab_reg} }  }
\begin{tabular}{|l|l|l|} \hline
 Quantity          & Theory            &  Expt.     \\ \hline
 $m_\pi~(m_{av}=$6.5~MeV)  & ~138.5~MeV (fit)   & ~138.5~MeV  \\ 
 $f_\pi$           & ~93.00~MeV (fit)   & ~93.00~MeV  \\ 
 $m_K~(m_s=$135~MeV)            & ~496~MeV   (fit)   & ~496~MeV    \\ 
 $m_{a_1}$         & ~1230~MeV  (fit)   & ~1230~MeV   \\ \hline
 $m_{\omega/\rho}$  & ~804~MeV           & ~782~MeV     \\ \hline
 $r_\pi^{ch}$      & ~0.55~fm           & ~0.663$\pm$0.006~fm     \\  \hline
  $ a^0_0$         & ~0.1634            &  ~0.26$\pm$ 0.05 \\ \hline
 $ a^2_0$         & -0.0466            & -0.028$\pm$ 0.012 \\ \hline
  $ a^1_1$         & ~0.0358            &  ~0.038$\pm$ 0.002 \\ \hline
  $ a^0_2$         & ~0.0017           &  ~0.0017$\pm$ 0.0003 \\ \hline
  $ a^2_2$        & -0.0005           &                      \\ \hline
\end{tabular}
\end{center}
\end{table}

Consider the case of the pion BSE with given $u$ and $d$ quark propagators taken
as being identical. Then use of \Eq{sepD} in \Eq{bspi} with \mbox{$\xi = 1/2$}
for the BS amplitude allows the representation 
\mbox{$\Gamma_\pi(p,P) = G(p^2)\;\lambda_\pi(P)$} where $\lambda_\pi(P)$ is a 
sum of a few amplitudes multiplied by matrix covariants.  The integration
in \Eq{bspi} now involves only functions known from the propagator model.
The distinct amplitudes in  $\lambda_\pi(P)$ can be separated by trace
techniques.  In general for mesons of other transformation character both
separable form factors $G(p^2)$ and $F(p^2)$ are involved.  The result is 
that the BSE has  been reduced to a low rank matrix
eigenvalue problem of the form \mbox{$K\;\Gamma = (\alpha(P^2)-1)\Gamma$}, 
with the bound state mass identified from \mbox{$\alpha(P^2=-M^2)=0$}.
The rank of the matrix is the number of distinct covariants that this ansatz
allows to be coupled by the BSE.  For the pion, the rank is $2$ corresponding
to both $\gamma_5$ and $\gamma_5 \gamma \cdot P$ covariants. 

In the application of this framework by Cahill and Gunner (1995a), the chiral
limit dressed quark propagator, in a certain parameterized entire function
form, was used to determine the separable form factors.  Because the pion
mass is one of the constraints that must be satisfied, it is necessary to
have propagators that correspond to finite bare mass for the quarks.  This
was obtained by use of the separable kernel in the DSE to recalculate the 
quark such propagators which were then re-parameterized back into the entire
function form.   The simplified BSE was solved under the assumption that
only the dominant or canonical matrix covariant is important.  Some of the
results obtained that way are displayed in \Table{tab_reg}.  Many other 
quantities have been calculated from this approach  by Cahill and Gunner 
(1995a) including mass estimates for several types of diquark constituents 
for baryons, a nucleon mass without meson dressing.  

\begin{table}[ht]
\begin{center}
\parbox{130mm}{\caption{ Selected meson quantities obtained
by Burden {\protect\it et al.}~(1997) from  application of a simple form of the
separable BSE Ansatz fitted to the observables shown.  Top panel pertains to
meson weak decay constants; bottom panel pertains to meson masses.  The column
labeled "Dom" refers to the results obtained using only the single dominant
covariant; for the other column all contributing covariants have been used.
References for experimental values can be 
found in Table~\protect\ref{tab_pion}. \label{tab_sep} } }
\begin{tabular}{|l|l|l|l|}\hline
 & $f_M^{\rm calc.}$~GeV & $f_M^{\rm calc.~(Dom)}$~GeV  & Expt.\\ \hline
   $\pi$
 & 0.0924~(fit)  & 0.056   & $\pi^+$(0.0924)     \\ \hline
   $K^\pm$
 & 0.113~(fit) & 0.76  &$K^+(0.113)$         \\ \hline\hline
 & $m^{\rm calc.}$~GeV & $m^{\rm calc.~(Dom)}$~GeV & Expt. \\ \hline
        $\pi$ ($0^{-+}$)
 & 0.139 (fit)  & 0.116   & $\pi^\pm(140),\pi^0(135$) \\ \hline
        $K$
 & 0.494 (fit) & 0.412  &$K^\pm(494),K^0(498)$\\ \hline
   $f_0/a_0$ ($0^{++}$)
 & 0.715         & 0.743         &  $f_0(980)$/$a_0(982)$ \\\hline
   $0^{+-}$
 & 1.082          & 1.092         & Not Seen   \\\hline
   $0^{--}$
 & 1.319          & 1.299         & Not Seen  \\\hline
   $K_0^\ast$
 & unbound      &    unbound    & $K_0^\ast(1430)$  \\\hline
   ${\eta }(\theta_P=5^{\rm 0})$
 & 0.549  &0.472 &  $\eta(547)$ \\
   $\eta(\theta_P=0^{\rm 0}) $
 & 0.513 & 0.441 &  \\\hline
   $ \omega/\rho$
 & 0.736 & 0.755 & $\omega(782)/\rho(770)$ \\ \hline
   $a_1/f_1$
        & 1.34    & 1.37       &$a_1(1260)/f_1(1285)$ \\\hline
   $K^{\ast}$
 & 0.854  & 0.866        & $K^{*}$(892)  \\  \hline
   $K_1$
 & 1.39   &   1.39      & $K_1(1270)$, $K_1(1400)$ \\  \hline
   $\phi$ ($\bar s s$ $1^-$)
 & 0.950       & 0.957        &$\phi(1020)$ \\ \hline
   $\bar s s$ $1^+$
 & 1.60        & 1.60         &$f_1(1510)$ \\ \hline
\end{tabular}
\end{center}
\end{table}

In the application of this separable Ansatz by Burden \etal~(1997), a UV
regulated form of the confining parameterized quark propagators described
in Sec.~\ref{sect_prop} was used with finite  quark bare masses to construct
the kernel form factors.  This procedure also preserves the Goldstone theorem.
With the four quantities $m_{\pi/K}$ and $f_{\pi/K}$ fit, the predictions for 
other meson masses are shown in \Table{tab_sep}.   Mass estimates
for a variety of diquark constituents of baryons were also considered in that
work.   A very good description of the ground-state, $SU_f(3)$,
isovector-pseudoscalar, vector and axial-vector meson spectrum was obtained.

The result in \Table{tab_sep} for the $\eta$ meson corresponds to the BSE
projected onto the $SU(3)$ mixed octet-singlet flavor covariant
\begin{equation}
F_\eta = \case{1}{\sqrt{2}} \left(\lambda^8 \cos\theta_P - 
\lambda^0 \sin\theta_P\right)~,
\end{equation}
with $\lambda^0 = \sqrt{2/3}\,\mbox{diag}(1,1,1)$ and $\theta_P$ the mixing 
angle.  The exact kernel of the BSE would lead to a prediction for $\theta_P$.
The $\eta^\prime$ meson corresponds to the orthogonal projection produced by
$\theta_P \rightarrow \theta_P- \pi/2$.   One expects quark annihilation
to timelike gluons, forbidden in the flavor octet channel, to be important in 
this mainly singlet channel.  The ladder kernel can only provide coupling of 
flavor components such as $\bar u u$ and $\bar s s$  through the 
symmetry-breaking masses.  This is insufficient to 
reproduce the large mass increase required by $M_{\eta'}^{\rm expt}  = 
958$~MeV compared to $M_\eta^{\rm expt} = 547$~MeV.  The ladder approximation 
is found to be adequate for the $\eta$ when the mixing angle is treated as an 
external parameter on which the mass and other properties  depend.
A small positive value for the
mixing angle is favored by the ladder approximation here while the experimental
value is $\theta_P = -10^\circ$ from Particle Data Group (1996) 
and has significant uncertainties.

Scalar mesons  were poorly 
described and this is to be expected in ladder approximation which is missing
important contributions including quark annihilation to multi-gluon 
configurations  
and open channels like $\pi\pi$ and $K\;K$  which can provide a width and
a significant change to the real mass.  The scalars, and also the charge
parity  exotics (${\cal C}= -1$), are displayed in \Table{tab_sep} to point
out these shortcomings.   Within the limited reliability and applicability of 
the rainbow-ladder approximation to the quark-DSE/meson-BSE complex, these
results imply that this crude separable Ansatz can convey to certain other
meson channels, via the dressed quark propagator, the dominant 
ladder physics constrained by the Goldstone mechanism in the psuedoscalar 
channel. 

It was found by Burden \etal~(1997) that in the isovector-pseudoscalar meson 
channel the sub-leading Dirac components of the BS amplitude provide 
quantitatively important contributions.  In particular, the pion solution was 
obtained in the form
\beq
\Gamma_\pi (p;P)  =  G_u(p^2)\,\left[i\gamma_5\; \hat\lambda_1 -
      \gamma_5\; \gamma\cdot\hat P \; \hat\lambda_2 \right]  ~, 
\label{seppi}
\eeq
where $\hat P_\mu$  is the time-like unit-vector associated with
$P_\mu$, i.e. $[\hat P^2 = -1]$. This is not the most general form of the pion 
amplitude (see \Eq{piform} given earlier), but it is the most general form 
allowed by the separable Ansatz.  The amplitudes are displayed in 
\Table{tab_sepampls}.  The sub-leading
(pseudovector) amplitude is more significant than it appears here since, as is
evident from \Table{tab_sep}, it contributes $\sim$ 15\%  to the mass  and 
$\sim$ 35\% to the  
weak decay constant $f_\pi$.  For the kaon, a similar situation is found to 
hold and there is an indication that for the  $\eta-\eta^\prime$ complex 
the sub-leading covariants are also important (along with the processes missing
from the ladder approximation as mentioned above).   As is evident from 
\Table{tab_sep} the sub-leading covariants  are found to be unimportant for 
mesons heavier than the pseudoscalar octet. 

In the vector meson case, the solutions of the BSE in \Eq{bspi} are labelled
by a Lorentz index, and satisfy the transverse condition 
\mbox{$P_\nu \Gamma_\nu(p;P) =0$}.  Due to the Feynman-like gauge employed, 
the general form is a linear combination of the  Euclidean covariants:
\begin{equation}
\label{tvcov}
\begin{array}{ccccc}
p_\nu^T, & \gamma_\nu^T, & p_\nu^T \gamma\cdot p, &
p_\nu^T \gamma\cdot P p\cdot P, &
\gamma_5 \epsilon_{\mu\nu\lambda\rho}\gamma_\mu p_\lambda P_\rho~,
\end{array}
\end{equation}
weighted by invariant amplitudes ${\cal F}_i(p^2,P^2, p\cdot P)$.   The 
superscript $T$ denotes a vector transverse to $P_\mu$.  The
separable Ansatz \Eq{sepD} introduces further simplification since it 
cannot support contributions to $\Gamma_\nu(p;P)$ that are
bilinear in $p$.  Hence, at the mass-shell, the produced vector meson
amplitude is
\begin{equation}
\Gamma_{\nu}^T(p;P) = p_{\nu}^T F_u(p^2)\hat{\lambda}_1
+ i\gamma_{\nu}^T G_u(p^2)\hat{\lambda}_2
+ i\gamma_5 \epsilon_{\mu \nu \lambda \rho}
\gamma_{\mu} p_{\lambda} \hat{P}_{\rho} F_u(p^2)\hat{\lambda}_3 ~,
\label{vamp}
\end{equation}
where the $\hat\lambda_i$ are produced as an eigenvector of a $3 \times 3$
matrix.
For the axial-vector meson the transverse Euclidean covariants are
\begin{equation}
\begin{array}{ccccc}
p_{\nu}^T\gamma_5 p \cdot P, & \gamma_5\gamma_{\nu}^T, &
p_{\nu}^T\gamma_5\gamma \cdot p, &
p_{\nu}^T\gamma_5\gamma \cdot P p \cdot P, &
\epsilon_{\mu \nu \lambda \rho} \gamma_{\mu} p_{\lambda} P_{\rho}~.
\end{array}
\label{tavcov}
\end{equation}
Again the terms bilinear in $p$ do not contribute and, using the separable
Ansatz, the axial-vector BSE reduces to a $2 \times 2$ matrix eigenvalue
problem.  The produced amplitude is
\begin{equation}
\Gamma_{5\nu}^T(p,P) =
i\gamma_5\gamma_{\nu}^T G_u(p^2)\hat{\lambda}_1
+ i\epsilon_{\mu \nu \lambda \rho} \gamma_{\mu} p_{\lambda} \hat{P}_{\rho}
F_u(p^2)\hat{\lambda}_2 ~.  \label{avamp}
\end{equation}
The solution amplitudes are displayed in \Table{tab_sepampls}.
\begin{table}[ht]
\begin{center}
\parbox{130mm}{\caption{Selected Bethe-Salpeter amplitude solutions from the
separable Ansatz of Eq.~(\protect\ref{sepD}).  The amplitudes are physically 
normalized according to Eq.~(\protect\ref{bsnorm}).  
Adapted from Burden {\protect\it et al.}~(1997).  \label{tab_sepampls} } }
\begin{tabular}{|c|c|ccc|}\hline
   & $J^{PC}$ & $\hat\lambda_1$  & $\hat\lambda_2$ & $\hat\lambda_3$ \\ \hline
 $\pi$         & $ 0^{-+}$  & 0.61   &  -0.045 &                     \\ \hline
 $\rho/\omega$ & $ 1^{--}$  & 0.075  & -0.33   & 0.049               \\ \hline
 $a_1/f_1$     & $ 1^{++}$  & 0.056  & -0.28   & 0.0                 \\ \hline
\end{tabular}
\end{center}
\end{table}

These studies also show that separable Ans\"atze have a number of 
shortcomings.  In the
pseudoscalar channel one finds that the $\gamma_5$ and $\gamma_5\gamma\cdot
P$ components of the meson BS amplitude are characterised by the
same function, $B(p^2)$, which is not true in general.  One also finds that
the dominant components in the BS amplitudes of the vector and
axial-vector mesons are characterised by the same functions that characterize
these components of the pseudoscalar mesons, $B(p^2)$.  More realistic
studies indicate that the vector meson amplitudes are narrower in momentum
space.  The separable amplitudes  should be used
with caution in the calculation of  meson interactions, and should eventually
be replaced by more realistic calculations.

The shortcomings notwithstanding, there are areas of study in hadronic physics
for which the BS amplitudes provided by a separable Ansatz have significantly
greater dynamical justification than the phenomenological amplitudes currently 
used.   For example,  as discussed in Sec.~\ref{sect_interact}, hadronic 
coupling constants such as $g_{\rho\pi\pi}$ and $g_{\gamma\pi\rho}$
have been reproduced  from the $\bar{q}q$ structure of
the mesons in terms of a single dominant Dirac covariant if the
amplitude is allowed some phenomenological freedom.  At present, a more 
realistic treatment is facilitated by use of the separable Ansatz.

Studies such as these, in combination with analysis (Bender \etal, 1996)
of processes in the BSE 
kernel that are of higher order than the bare ladder term, provide a better 
understanding of the domain of applicability of the ladder BSE. 
The ladder kernel has the defect that it is purely attractive in both
the color-singlet $\bar q$-$q$ and color-antitriplet $q$-$q$ channels.
This entails that it yields bound color-antitriplet diquarks.  This is a
peculiarity of ladder approximation.  Measuring ``order'' by the number of
dressed-gluon lines in the Bethe-Salpeter kernel, ladder approximation is the
lowest order kernel.  Repulsive terms appear at every higher order.  It has
been shown by Bender \etal~(1996) that in the isovector-pseudoscalar 
and vector meson
channels, these repulsive terms are cancelled by attractive terms of the same
order.  This explains why ladder approximation is phenomenologically
successful in these channels.  That study also showed that in the 
color-antitriplet diquark channel the
algebra of $SU_c(3)$ entails that the repulsive terms are stronger; they are
not completely cancelled and eliminate the diquark bound states.

\sect{Electromagnetic Coupling}
\label{sect_EM}

\subsect{From Quark Currents to Composite Meson Currents}
\label{sect_EMder}

To allow for electromagnetic (EM) coupling to the meson modes,
the bosonization method can be generalized (Frank and Tandy, 1994) to account 
for a background electromagnetic field minimally coupled
(\mbox{$\partial_\nu \rightarrow \partial_\nu -i\hat{Q}A_\nu$})
to the bare quarks.  Thus the action
\beq
S[\bar{q},q,A]=\int d^{4}x \bar{q}(x)
\bigl(\dslash -i \hat Q \Aslash + m  \bigr) q(x)
 + \case{1}{2} \int d^{4}xd^{4}y j_{\mu
}^{a}(x)D(x-y)j_{\mu }^{a}(y) ,\label{sgcmA}
\eeq
where $\hat{Q}$ is the quark charge operator,  is invariant under the gauge
transformation
\begin{eqnarray}
q(x)\rightarrow q'(x) &=& e^{i\hat{Q}\theta (x)}q(x) , \nonumber \\
\bar{q}(x)\rightarrow \bar{q}'(x) &=& \bar{q}(x)e^{-i\hat{Q}\theta (x)} ,
 \nonumber \\
A_{\nu }(x)\rightarrow A'_{\nu }(x) & = & A_{\nu
}(x)+\partial _{\nu }\theta (x).\label{gtrans}
\end{eqnarray}
The standard bosonization technique can be applied
in the presence of the external EM field to the expose $\bar{q}q$ meson 
fields and their electromagnetic interactions.   Gauge invariance 
of the EM coupling at the quark level can be translated to the level 
of extended meson modes (Frank and Tandy, 1994).  One obtains
$Z[A] =N\int D{\cal B} \; {\rm exp}(-{\cal S}[ {\cal B}; A])$
where the gauge invariant boson action is
\begin{equation}
{\cal S}[ {\cal B}; A ] = -{\rm TrLn} \; {\cal G}^{-1}\left[
{\cal B}; A\right] +  \int d^{4}xd^{4}y\frac{{\cal
B}^{\phi }(x,y){\cal B}^{\phi }(y,x)}{2D(x-y)} , \label{actionBA}
\end{equation}
with \mbox{${\cal G}^{-1}[{\cal B}; A] = \dslash -i\hat{Q} \Aslash + m +
\Lambda^\phi {\cal B}^\phi $}. 
The saddle point configuration of the auxilary fields ${\cal B}^\phi$,
defined in the same manner as previously,
now becomes a nonlinear functional ${\cal B}_0[A]$ of the background EM field.  
It produces a ladder dressing of the photon-quark vertices as well as the 
translationally invariant quark self-energy.   It is convenient to define
\mbox{$\tilde\Sigma [A] = \Lambda ^{\phi }{\cal B}^{\phi}_{0}[A]$} which
plays the role of the quark self-energy in the background EM field.  It 
satisfies
\beq
\tilde\Sigma (x,y)=\case{4}{3}D(x-y)\gamma _{\nu }\tilde{\cal G}(x,y)\gamma
_{\nu }.\label{SDEA}
\eeq
where 
\beq
\tilde{\cal G}^{-1}[A]= \dslash -i\hat{Q} \Aslash + m + \tilde\Sigma [A]  .  
\label{GinvA}
\eeq
The structure of this formalism is exactly the same as when \mbox{$A_\nu =0$}
except that now the generalized quantities $\tilde{\cal G}$ and 
$\tilde\Sigma[A]$ are not translationally 
invariant.  The first-order dependence of $\tilde{\cal G}^{-1}$ upon 
$A_{\nu}$  generates the EM vertex with dressed quarks via
\beqar
\Gamma_{\nu}(p,k;Q) &=& \left. \frac{\delta \tilde{\cal G}^{-1}(p,k)}
{\delta A_{\nu}(Q)} \right|_{A_{\nu}=0} \nonumber \\
&=& (2\pi)^4 \; \delta(p-k-Q) \Gamma_{\nu}\left(\case{p+k}{2};Q\right)
.\label{qver}
\eeqar
Application to \Eq{SDEA} immediately yields the ladder BSE for this vertex
\beq
\Gamma_{\nu }(q;Q) =-i\gamma_\nu \hat Q -\case{4}{3}
\int \frac{d^{4}k}{(2\pi )^{4}} \; D(q-k)\gamma _\mu
S( k_+) \Gamma _{\nu }(k;Q) S(k_-) \gamma _\mu , \label{emBSE}
\eeq
where \mbox{$S =\tilde{\cal G}[A_\nu =0]$} and \mbox{$k_\pm=k\pm \frac{Q}{2}$}.
Use of the rainbow SDE for the quark propagator $S$
also confirms that  this vertex satisfies the Ward-Takahashi 
identity (WTI) (Takahashi, 1957)
\beq
Q_{\nu }\Gamma _{\nu }(q;Q)= \hat Q \left\{S^{-1}( q_-) - S^{-1}( q_+)\right\}
, \label{wti}
\eeq
and also  the Ward identity (Ward, 1950)
\beq
\Gamma_\nu(q;Q=0) = - \; \hat Q \; \frac{\partial}{\partial q_\nu} S^{-1}(q)  . 
\label{ward}
\eeq
These are properties the ladder vertex shares with the exact vertex.  For this
reason the ladder-rainbow pair of equations for the dressed quark propagator
and EM vertex constituent a very useful approximation in that EM current
conservation can be exactly preserved.   If instead of the EM vertex, we 
the consider the ladder BSE for the Goldstone bosons, it was pointed out in 
Sec.~\ref{sect_GB} that the ladder-rainbow combination preserved chiral
symmetry and the Goldstone theorem.  It is therefore interesting and 
advantageous that, at meson tree level of the bosonized GCM, the produced
quark substructure has exactly the ladder-rainbow content. 

The action may be expanded about the saddle point configuration ${\cal B}_0[A]$
to obtain
\begin{equation}
{\cal S}[{\cal B}; A] = {\cal S}[ {\cal B}_0[A];A] + 
\hat{{\cal S}}[\hat{{\cal B}};A] , \label{SBA}
\end{equation}
where $\hat{{\cal B}}^{\phi }={\cal B}^{\phi }
-{\cal B}^{\phi }_{0}[A]$ are the new field variables for the propagating
meson modes and the corresponding action for the meson sector containing
electromagnetic interactions is
\beq
\hat{{\cal S}}[ \hat{{\cal B}}; A] = {\rm Tr} \sum
_{n=2}^{\infty }\frac{(-1)^{n}}{n}\left( \tilde{\cal G}[A] \; \Lambda^{\phi}
\hat{{\cal B}}^{\phi}\right) ^{n}
+ \case{1}{2} \left( \hat{{\cal B}}^{\phi \ast} D^{-1} \hat{{\cal B}}^\phi 
\right) . \label{ShatBA}
\eeq
The EM coupling to the $\bar q q$ meson modes is contained in the first term
of \Eq{ShatBA} and can be unfolded through use of \mbox{$\tilde{\cal G}[A] =
S -\Gamma_\nu A_\nu S + \cdots$}. 
The meson EM current is
\begin{equation}
J_{\nu}(x) =
- \left. \frac{\delta {\cal S}[{\cal B}^{\phi };A] }{\delta
A_{\nu }(x)}\right| _{A_{\nu }=0} = -\left.
\frac{\delta \hat{{\cal S}}[\hat{{\cal B}}^{\phi};A]}{\delta
A_{\nu}(x)}\right| _{A_{\nu}=0}  .  \label{2b1}
\end{equation}
The second equality in \Eq{2b1} expresses the fact that the saddle
point action ${\cal S}\left[{\cal B}_0[A];A\right]$ does not 
contribute to the current.  The functional
dependence of ${\cal S}\left[{\cal B}_0[A];A\right]$ upon $A_{\nu}$
that enters implicitly through ${\cal B}_{0}\left[A\right]$
makes no contribution since $\frac{\delta {\cal S}}{\delta {\cal B}_{0}} = 0$.
The explicit $A_{\nu}$ dependence generates a  bare
coupling of the photon to a vacuum quark loop which vanishes due
to symmetric integration.  The saddle point action can contribute
to higher order in $A_{\nu}$ beginning with a vacuum polarization
insertion $\Pi_{\mu \nu}^{\gamma \gamma} (Q)$ for the photon propagator.  
In particular, the part of ${\cal S}[{\cal B}_0[A] ;A]$ which is 
second order in  $A_\nu$ is
\beq
{\cal S}^{(2)}[{\cal B}_0[A] ;A]  = \case{1}{2} \int \dbq A_\mu(-Q) \; 
\Pi_{\mu \nu}^{\gamma \gamma}(Q) \; A_\nu (Q) , \label{SA2}
\eeq
where
\beq
\Pi_{\mu \nu}^{\gamma \gamma}(Q)
= - {\rm Tr} \; [ i \gamma_\mu \hat{Q} \; S \; \Gamma_\nu (Q) \; S ] .
\label{phpol}
\eeq
More explicitly this is
\beq
\Pi_{\mu \nu}^{\gamma \gamma}(Q) = - {\rm tr} \int \dq \; [i \gamma_\mu \hat Q 
\; S(q_+) \; \Gamma_\nu(q;Q) \; S(q_-)] \; .
\label{phpolint}
\eeq
This has the same form as the exact polarization tensor although the dressed
vertex here is in ladder approximation as is appropriate for a treatment 
at zeroth level of the meson loop expansion.  Due to the WTI in \Eq{wti}, it
follows that \mbox{$Q_\mu \Pi_{\mu \nu}^{\gamma \gamma}(Q) = 0$} producing
the transversality
required by gauge invariance.  To go further in this direction and consider
the renormalization of the photon propagator and field in conjunction with the
required counter terms, the photon needs to be treated as a dynamical field 
variable instead of an external field.  An important element is that  the
transverse amplitude of $\Pi_{\mu \nu}^{\gamma \gamma}(Q)$ has a zero at 
\mbox{$Q^2=0$}, that is \mbox{$\Pi_{\mu \nu}(Q) = $}
\mbox{$(Q^2 \delta_{\mu \nu} -Q_\mu Q_\nu) \tilde {\Pi}(Q^2) $}, 
thus allowing the physical renormalized photon to remain massless.  This 
property applies to \Eq{phpol} as long as $\Gamma_\mu$ satisfies the Ward
identity (Burden \etal, 1992a).

One of the pleasing features of the action for electromagnetic coupling
to interacting mesons that follows from \Eq{SBA} and \Eq{ShatBA} is that 
the current conservation required by gauge invariance is manifest in each term.
One can then continue on and derive photon couplings to  mesons
in a way that maintains meson EM current conservation in the face of meson 
substructure or form factors that are accountable to the conserved currents
at the quark level.  This has been done for finite size pions by Frank and 
Tandy, (1994) where pion Ward-Takahashi identities are also derived.   There
it is also shown that when the pion action is restructured to an equivalent 
form governing effective local pion fields, then EM gauge invariance in the
presence of the extended charge form factor will only be manifest if one 
recognizes the gauge transformation behavior of the effective pion inverse 
propagator that results from the substructure field content that has 
necessarily been absorbed into it.  Such observations can greatly assist in the 
task of gauging effective lagrangians for hadrons that involve form factors
related to substructure. 

An inconvenience of this approach is that there is no explicit term in 
\Eq{ShatBA} that is linear in a meson field such as the 
\mbox{$\pi^0 \rightarrow \gamma \gamma$} process, for example.  Also the
physical processes involving the direct coupling of a photon to a vector meson 
(as needed for the decay \mbox{$\rho^0 \rightarrow e^+ e^-$})
are not evident. Such processes form the basis of the empirically successful
Vector Meson Dominance (VMD)  hypothesis and it is desirable to unfold 
them from \Eq{ShatBA}.   In the point coupling limit, e.g. the NJL model, it 
is known that a point vector meson field redefinition via a simple shift 
\mbox{$\rho_\nu^\prime = \rho_\nu + c A_\nu$} reveals the VMD terms (Volkov 
and Ebert, 1982; Volkov, 1984; Ebert and Reinhardt, 1986) and in fact allows 
the vector meson fields
to be chosen so that they mediate all electromagnetic coupling to hadrons. 
This idealization entails a mass-type (i.e. constant amplitude) mixing of the 
photon with point
vector mesons, and is often taken to be the essential meaning of VMD. 
In this situation, an exact current-field identity (Sakurai, 1969) exists; 
that is, the EM current is a linear combination of the point vector meson 
fields.  

Since the EM 
current is a local object, and in reality the vector mesons (both from the 
present GCM model and in QCD) are distributed or 
nonlocal objects, the VMD current-field identity can only hold approximately.
Only the soft photon coupling to a nonperturbatively interacting $\bar{q}q$
vector correlation can be implemented through realistic vector meson degrees 
of freedom; there is a point \mbox{$\bar q \gamma q$} coupling 
(\mbox{$-i \gamma_\mu \hat Q$}) that  cannot be properly simulated
by a bound state object.  Only in the context of effective lagrangian models 
with point-like vector meson fields can such fields take over the total role of 
the photon.  For quark models (and QCD) where realistic finite size meson 
modes arise, a meson field redefinition  that attempts to absorb the point
\mbox{$\bar q \gamma q$} coupling does not make sense and leads to
dynamical problems.
However, one still expects to be able to implement resonant photon-hadron
coupling via the composite vector meson fields leaving a non-resonant direct 
photon coupling to quarks while maintaining manifest gauge invariance.
We will indicate how this can be done.
\begin{figure}[ht]
\unitlength1.pt
\begin{picture}(612,164)(-20,240)
\includegraphics{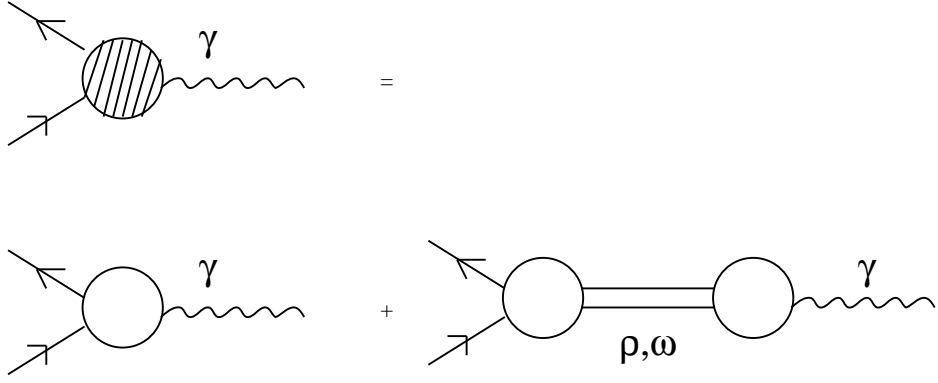}
\end{picture}

\centering{\
\parbox{130mm}{ \caption{The separated non-resonant and vector meson pole components 
of the dressed photon-quark vertex.  \label{vmdver} } } }

\end{figure}

The quantity $\tilde \Sigma[A]$ that enters \Eq{GinvA} contains ladder dynamics
for the coupling of one or more photons with dressed quarks.  For example, the 
$1$-photon part must contain $\gamma \rho$ and $\gamma \omega$ processes and
the $2$-photon part must contain a \mbox{$\pi^0 \gamma \gamma$} process.  The
former  are contained within  the  dressed quark-photon vertex $\Gamma_\nu$ 
that was produced.  This becomes evident for 
photon momenta near the vector meson mass shell where the homogeneous 
version of \Eq{emBSE} has a bound state solution.  This generates a bound state
pole in the inhomogeneous solution for the vertex $\Gamma_\nu$.  Thus, 
with respect to the invariant mass of a given vector meson, $\Gamma_\nu(q;Q)$ 
can be separated (non-uniquely) into a resonant or pole piece (which is 
transverse) and a background or non-resonant piece (which is both longitudinal
and transverse).  That is, \mbox{$\Gamma_\nu = \Gamma_\nu^{nr} + 
\Gamma_\nu^{pole} $} and we shall consider the $\rho$ pole to be the point of 
interest.   We may formally make this separation using the eigenmode analysis of 
the kernel of the ladder BSE vertex \Eq{emBSE} as discussed in \Sec{sect_eigen}.   
With the dressed part of  $\Gamma_\nu^{nr}$ representing the contributions 
from the eigenmodes other than the $\rho$, this yields
\beq
\Gamma_\nu(q;Q) = \Gamma_\nu^{nr}(q;Q) - \Gamma^\rho_\mu(q;Q) 
\; \frac{T_{\mu \sigma}(Q)}{Q^2 + m_\rho^2(Q^2)}
\; \Pi^{\rho \gamma}_{\sigma \nu}(Q)  .
\label{polevertex}
\eeq
The $\rho\gamma$ polarization tensor that is generated is given by
\beq
\Pi^{\rho \gamma}_{\sigma \nu}(Q) = - {\rm Tr} \; [ \bar 
\Gamma_\sigma^\rho(-Q) \; S \; i\gamma_\nu \hat{Q} \; S ]  .
\label{grmixing}
\eeq
This pole separation of the photon-quark vertex is illustrated in \Fig{vmdver}.
The employed $\rho$ BSE amplitude is physically normalized  and the associated
dynamical mass function $m_\rho^2(Q^2)$ is defined in \Sec{sect_eigen}. 
Although a ladder BS structure is under discussion here, the exact 1PI photon 
vertex can also be decomposed in this form.  There will in general be an imaginary
part to the mass function arising from physically accessible hadron channels (such
as \mbox{$\rho \rightarrow \pi \pi$}) in a more complete analysis.   The point of
the present discussion does not require that such aspects be made explicit.
The off-mass-shell
behavior of individual terms in separations such as in \Eq{polevertex} will differ
in different approaches and adjustments can be made to accommodate convenient
approximations for $\Gamma_\nu^{nr}$. 

With the non-resonant  $\Gamma_\nu^{nr}$ satisfying both the 
WTI and the Ward identity, then the above pole term of the vertex is zero in 
the soft photon limit.  That is, with
\mbox{$\Pi^{\rho \gamma}_{\mu \nu}(Q) = T_{\mu \nu}(Q) 
\Pi^{\rho \gamma}_T(Q^2)$}, we have $\Pi_T^{\rho \gamma}(0)=0$.  Thus, due to 
manifest EM current conservation, the photon decouples from the composite 
vector mesons at $Q^2=0$.    Similarly, with the separation in 
\Eq{polevertex}, the $\bar{q}q$ self-energy or polarization tensor for the 
photon, from \Eq{phpol}, can be written
\beq
\Pi_{\mu \nu}^{\gamma \gamma}(Q)
= - {\rm Tr} \; [ i \gamma_\mu  \hat{Q} \; S \; \Gamma_\nu^{nr} (Q) \; S ] 
- \Pi^{\gamma \rho}_{\mu \delta}(Q) \; \frac{T_{\delta \sigma}(Q)}{Q^2 + 
m_\rho^2(Q^2)} \; \Pi^{\rho \gamma}_{\sigma \nu}(Q)  .
\label{phpol2}
\eeq
Since the first term can be shown to be zero in the soft photon limit if 
$\Gamma_\nu^{nr}$ satisfies the WTI and the Ward identities 
(Burden \etal, 1992a), the result \mbox{$\Pi^{\rho \gamma}_T(0)=0$} 
is again evident and linked to the massless condition of the photon.  
In general there can be many pole
contributions to the spectral strength in the timelike domain  
that can be separated out.  The common assumption is that
selected vector mesons, such as $\rho$, $\omega$ and $\phi$, saturate the 
spectral strength.  

Although the pole term in the photon vertex admits an integral representation 
in terms of vector meson fields,  there are already explicit vector meson 
fields elsewhere in the effective meson action of \Eq{ShatBA}.  It is an 
awkward matter to consistently unfold photon-meson mixing from the previous
formalism this way.  It is more convenient to make the following field
redefinition. We expand the bosonized action \Eq{actionBA} about 
a new configuration $\tilde{{\cal B}}^\theta [A]$ of the auxiliary fields, that 
is not the saddle point ${\cal B}_0[A]$ as used before, but is chosen so that 
the resulting photon-quark vertex is  the non-resonant dressed vertex 
$\Gamma_\nu^{nr}$.   Specifically, the new auxiliary field configuration  
$\tilde{{\cal B}}^\theta [A]$ about which we expand is defined so that the 
inverse quark propagator in the absence of the meson fluctuation fields is
\beq
\tilde S^{-1}[A] = \dslash -i \hat Q \Aslash + m + 
\Lambda^\theta \tilde{{\cal B}}^\theta [A] =
\dslash + m + \Sigma +  \Gamma_\nu^{nr} A_\nu   .
\label{newexpt}
\eeq
Comparison with \Eq{GinvA} indicates that the previous quantity 
$\tilde\Sigma[A]$ is now replaced  by a quantity that is linear in
the photon field $A_\nu$, is devoid of bound state poles, but can still
allow current conservation.  A practical Ansatz for $\Gamma_\nu^{nr}$ is
considered shortly.   

To proceed with the meson field redefinition, the bosonized action 
\Eq{actionBA} is expanded about $\tilde{{\cal B}}^\theta [A]$  to produce
\beq
{\cal S}[{\cal B}; A] = {\cal S}[\tilde{\cal B}[A];A] + 
\hat{\cal S}[\hat{\cal B}; A]  ,
\label{newSBA}
\eeq
where  \mbox{$\hat{{\cal B}}^\theta = {\cal B}^\theta
-\tilde{\cal B}^\theta [A]$} are the new field variables for the propagating
meson modes (although we use the same notation $\hat{\cal B}^\phi$ as before).
The new action for the meson sector with electromagnetic interactions is
\beq
\hat{{\cal S}}[ \hat{{\cal B}}; A] = \hat {\cal S}_1[\hat{\cal B}; A]
+ {\rm Tr} \sum
_{n=2}^{\infty }\frac{(-1)^{n}}{n}\left( \tilde{S}[A] \; \Lambda^{\phi}
\hat{{\cal B}}^{\phi}\right) ^{n}
+ \hlf \left( \hat{{\cal B}}^{\phi \ast} D^{-1} \hat{{\cal B}}^\phi \right) ,
\label{newShatBA}
\eeq
where the term $\hat {\cal S}_1[\hat{\cal B}; A]$ is first order in meson 
fields and  is present because the  expansion is not about an extremum.  
The second term is at least quadratic in mesons and contains conventional
meson charge form factors processes such as \mbox{$\gamma\pi\pi$}. 
The processes involving one or more photons coupled to a single meson 
(including the so-called VMD processes) are exposed by the first term and its 
explicit form is
\beq
\hat {\cal S}_1[\hat{\cal B}; A] = - {\rm Tr} \; [\tilde{S}[A] \; \Lambda^{\phi}
\hat{{\cal B}}^{\phi}]
+  \left( {\tilde{\cal B}^\phi [A] }^\ast D^{-1} \hat{{\cal B}}^\phi \right) .
\label{Shat1BA}
\eeq
The ${\cal O}(A^0)$ terms of $\hat {\cal S}_1$ cancel out, i.e.
\mbox{$\hat {\cal S}_1[\hat{\cal B}; A=0]=0$}, because then the expansion
point is the \mbox{ $A=0$} saddle point configuration and the cancelling 
two terms in \Eq{Shat1BA} form the rainbow DSE. The ${\cal O}(A^1)$ 
terms of 
$\hat {\cal S}_1$ provide the sought-after photon-meson mixing terms.  The 
${\cal O}(A^n)$ terms of $\hat {\cal S}_1$ with \mbox{$n \geq 2$}, 
such as \mbox{$\pi^0 \gamma \gamma$},  are 
contained only in the quark loop term  and are revealed by the expansion 
\mbox{$\tilde S[A] = S - S \Gamma_\nu^{nr} A_\nu S$} 
\mbox{$+ S \Gamma_\nu^{nr} A_\nu S \Gamma_\nu^{nr} A_\nu S + \cdots$}.  
We now use  the composite meson field representation of \Sec{sect_eigen}
to introduce the relevant BS amplitudes and associated local field variables,
\mbox{$ \Lambda^{\phi} \hat{{\cal B}}^{\phi} \rightarrow 
\Gamma_n(q;Q) \phi_n(Q)$}.
The ${\cal O}(A^1)$ term of \Eq{Shat1BA} then becomes closely related to the 
ladder BSE kernel, allowing  the $D^{-1}$ term to be recast as a quark loop
by using the methods of \Sec{sect_eigen}.   Consideration of the relation between
$\Gamma_\nu^{nr}$ and the eigenmodes of the ladder BSE kernel indicates partial
cancellation between the two terms of \Eq{Shat1BA} due to orthogonality. The 
results can be illustrated by the explicit \mbox{$\gamma \rho$} mixing term which is
\beq
\hat{\cal S}[\gamma \rho]  =  \int \dbq A_\mu(-Q) \; 
\Pi_{\mu \nu}^{\gamma \rho}(Q) \; \rho_\nu (Q) , \label{Sgr}
\eeq
where the mixed self-energy tensor that is produced is
\beq
\Pi_{\mu \nu}^{\gamma \rho}(Q) =  - \; {\rm Tr} \; 
                [i \gamma_\mu \hat Q \; S \; \Gamma_\nu^\rho(Q) \; S ]  ~.
\label{grmix}
\eeq 
Here, at the $\rho$ mass-shell,  $\Gamma_\nu^\rho(q;Q)$ is the physically 
normalized BS amplitude, and for general $Q$ it is the relevant eigenvector of
the ladder BSE kernel defined in \Sec{sect_eigen}.  The result in \Eqs{Sgr} and
(\ref{grmix}) is consistent with the original analysis in \Eqs{polevertex} and
(\ref{grmixing}) and that $\rho$ pole contribution to each photon-meson coupling
would be recovered through integrating over the $\rho$ field of the reformulated
action. 

We summarize by retaining only a few low mass meson modes so that the 
reformulated  meson action of \Eq{newShatBA} can be written in the more explicit
form 
\beqar
\hat{\cal S}[\pi,\rho, \cdots] = - &&{\rm Tr} \; \left[\tilde{S}[A] \; 
( \vec{\Gamma}_\pi\cdot\vec{\pi} 
+\vec{\Gamma}_\mu^\rho \cdot \vec{\rho}_\mu  + \cdots) \right]
+\case{1}{2} \sum_{n,m} \int \dP \:
\phi_n^\ast(P) \;  \Delta_{nm}^{-1}([A];P) \; \phi_n(P)
\nonumber \\
&&+ {\rm Tr} \sum_{n=3}^\infty\frac{(-)^n}{n}
\left[\tilde S[A]\; (\vec{\Gamma}_\pi\cdot\vec{\pi}
+\vec{\Gamma}_\mu^\rho \cdot \vec{\rho}_\mu  + \cdots)\right]^n ~,
\label{emmesaction}
\eeqar
where the local fields are \mbox{$\phi_n = \pi^i,\; \omega_\mu,\; \rho_\mu^i,
\; \cdots$}.   When \mbox{$A_\nu \rightarrow 0$}, this action reduces to 
\Eq{mesaction} that was obtained earlier.  Here the generalized inverse 
propagator
$\Delta^{-1}_{nm}([A];P)$ is defined in an obvious way from the quadratic
terms of \Eq{newShatBA}.  We see in \Eq{emmesaction} that  the photon
coupling to hadronic matter consists of a resonant coupling 
mediated by single vector mesons which is part of the first term, as well as a 
direct non-resonant coupling to the conserved EM current of dressed quarks and
represented by the other terms.   A by-product of this reformulation is that 
the processes involving multiple photon coupling to a single meson are 
uncovered and are found in the first term.   The latter include the chiral 
anomalies such as \mbox{$\pi^0 \gamma \gamma$} and \mbox{$\eta \gamma \gamma$}.
This formulation is most useful when the $\Gamma_\nu^{nr}$ satisfies the 
Ward-Takahashi (Takahashi, 1957) identity and the Ward identity (Ward, 1950); 
then EM current conservation
is manifest for each term of the action in \Eq{emmesaction}.

The solution of the DSE for the dressed quark-photon vertex
is a difficult problem.   Only recently has the solution of the ladder version
given in \Eq{emBSE} begun to be addressed (M.R. Frank, 1995). 
However, much progress has been made in constraining
the form of $\Gamma_\mu(p;Q)$ and developing a realistic
Ansatz (Ball and Chiu, 1980; Curtis and Pennington, 1992; Burden and Roberts,
1993; Dong, Munczek and Roberts, 1994).  The bare vertex 
$\Gamma_\mu(p;Q) \rightarrow - i \gamma_\mu \hat Q$ is clearly
an inadequate approximation when the quark propagator has momentum dependent 
dressing because it violates the Ward-Takahashi identity  would lead to an 
EM current that is not conserved.
An Ansatz  with the desirable properties:
a) satisfies the Ward-Takahashi identity; b) is free of kinematic
singularities; c) reduces to the bare vertex in the free field limit as
prescribed by perturbation theory; and d) has the same transformation
properties as the bare vertex under charge conjugation and Lorentz
transformations, is (Curtis and Pennington, 1992)
\begin{equation}
\Gamma_\mu(p;Q) = \hat Q \left(\Gamma_\mu^{bc}(p;Q) + \Gamma_\mu^T(p;Q) \right)
\label{curtispen}
\end{equation}
where, with \mbox{$S^{-1}(p) = i\gamma\cdot p A(p^2) + B(p^2)+ m$}, the
term advocated by Ball and Chiu (1980) is
\begin{equation}
\Gamma_\mu^{bc}(p;Q)=-i\gamma_{\mu}\hlf\Bigl(A(p_+)+A(p_-)\Bigr)
+ \frac{p_{\mu}}{p \cdot Q}\Bigl[ i\gamma \cdot p
\Bigl(A(p_-)-A(p_+)\Bigr) +
\Bigl(B(p_-)-B(p_+)\Bigr) \Bigr] . \label{ballchiu}
\end{equation}
with \mbox{$p_\pm=p\pm \frac{Q}{2}$}.  Note that the undetermined piece
$\Gamma_\mu^{\rm T}$ is transverse, $Q_\mu\,\Gamma_\mu^{\rm T}(q;Q) = 0$ 
and vanishes at zero momentum $\Gamma_\mu^{\rm T}(q;0) =
0$.  It is also the case that with a bare quark propagator, which has $A=1$ 
and $B=$~constant, $\Gamma_\mu^T =0$.  In this Ansatz the $\Gamma_\mu^{bc}$ 
piece is completely determined by the dressed quark propagator and it has both 
transverse and longitudinal components.    A feature of this approach which 
follows from criterion c), is that with a quark propagator that incorporates 
asymptotic freedom the quark-photon vertex will reduce to the correct term at 
large spacelike-$Q^2$, in the manner prescribed by
perturbation theory in QCD. This Ansatz therefore provides a realistically
constrained extrapolation of the quark-photon vertex to small spacelike-$Q^2$.

The representation for  $\Gamma_\mu^{nr}$ that has been used in most investigations of 
electromagnetic interactions of hadrons within the GCM or DSE format is the above 
Ball-Chiu Ansatz, that is, \mbox{$\Gamma_\mu^{nr}(p;Q) \approx \hat{Q} 
\Gamma_\mu^{bc}(p;Q)$}.  The additional
resonant or vector meson pole processes, that we have earlier identified 
within the GCM, provide a specific dynamical model  representation for the
$\Gamma_\mu^T$ term in \Eq{curtispen}.   The resonant terms are obviously necessary 
in the timelike region sufficiently close to the poles; however in the spacelike 
region, the relative contribution of $\Gamma_\nu^{nr}$ and $\Gamma_\nu^{pole}$ to
specific hadronic processes is very much dependent upon the off-mass-shell behavior 
of the particular separation employed.  In another language, while the individual 
contributions depend upon choice of interpolating field variables, the summed 
contribution to an S-matrix element should not.   Within the GCM and DSE approaches, 
it has been found that the spacelike pion charge form factor can be described quite 
successfully (obtaining $85-90$\% of the charge radius) through use of the non-resonant
$\Gamma_\nu^{bc}$  Ansatz.  We shall comment further on this point later.
There are several other choices available to model the non-resonant EM vertex 
(Roberts and Williams, 1994).    However no inadequacy of the Ball-Chiu 
Ansatz has been found in practical calculations of form factors at spacelike
momenta. 

In a recent analysis of effective local lagrangian models  
for electromagnetic coupling to point fields for vector mesons and other 
hadrons,  O'Connell \etal,  (1995) have pointed out the merits of a form in 
which
the VMD hypothesis is implemented in precisely the "direct plus resonant"
manner arrived at above. A property of that particular local representation of 
VMD (called VMD1), is that the photon-meson 
mixing amplitude is momentum-dependent and vanishes in the soft photon limit.
A point field redefinition via, e.g. \mbox{$\rho_\nu^\prime = 
\rho_\nu + c A_\nu$} can produce the equivalent, and more common form (VMD2)
of local effective lagrangian where the photon only couples to vector mesons 
and the relevant amplitude is a constant.  This form is inconvenient 
as the zero value of the physical photon mass is only
recovered as a cancellation between finite bare and dynamically produced 
contributions; also the reproduction of the total charge of hadrons requires
fine tuning of parameters which is indicative of a formalism wherein 
current conservation is not manifest.

The vanishing at zero momentum for the transverse vector self-energy amplitudes 
we have discussed is related to, but somewhat more general than, 
the  node theorem for the self-energy amplitudes for point vector bosons 
coupled to conserved local currents that has been put forward recently 
(O'Connell \etal, 1994).  In the present analysis, it is the participation of 
the photon that precipitates the result, for it alone couples to a local 
conserved current. 
There is no aspect of the self-energies for finite size vector mesons  that 
are involved in the dressed meson propagator of \Eq{phpol2} that would
produce a zero momentum node.  In a similar fashion, with flavor asymmetric 
quark bare masses, the \mbox{$\rho-\omega$} mixed
self-energy amplitude contained in the coupled channel meson propagator that
would now appear in \Eq{phpol2} would  not  have a
zero-momentum node.  This is because the meson modes here, and in QCD, are
extended objects.  One would expect however, that a nearby node might occur
if the meson size was suitably small.  Indeed, a calculation of the 
$\rho-\omega$ amplitude at the quark loop level based on the GCM produces a
node at about $0.1~{\rm GeV}^2$ in a way related to meson size (Mitchell and 
Tandy, 1997).  
The only  model based at the quark level where the zero momentum 
node property would extend to the vector meson self-energies is evidently
the NJL model which allows an exact VMD current-field identity to be realized.
This follows from the observation that, in this point coupling approximation,
the vector meson fields are  proportional to the various flavor components of
the photon; the relevant meson currents are local and conserved and the 
analysis reduces to the electromagnetic one.

A recent study within a generalized NJL model utilizes just this
correspondence between the vector mesons of the model and the flavor components 
of the photon to analyze VMD, $\rho-\omega$ mixing and the pion charge form
factor from a quark basis organized around local current correlation functions
(Shakin and Sun, 1997).  In that work the NJL model is generalized in the sense
that a regulated linear confining potential is added to dress quark vertices
to eliminate spurious production thresholds from $\bar q q$ polarizations.

\subsect{Pion and Kaon Electromagnetic Form Factors}
\label{sect_emff}

\begin{figure}[ht]
\unitlength1.pt
\begin{picture}(612,240)(20,-330)  
\includegraphics{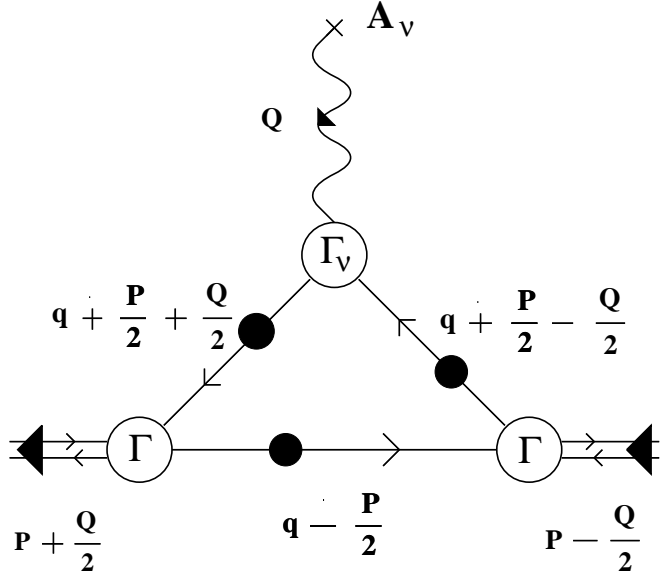} 
\end{picture}
\centering{\
\parbox{130mm}{\caption{Diagram for the generalized impulse approximation 
to the $\protect\gamma\protect\pi\protect\pi$ vertex. \label{gppdiag} } } }  
\end{figure}

The action component describing the $\gamma\pi\pi$ interaction,  
in the form $- J \cdot A$, can be written as
\beq
\hat{\cal S}\left[\gamma\pi\pi\right]= - \int 
\frac{d^4P,Q}{(2\pi)^8} \; \pi_-^\ast (P+\case{Q}{2}) \, 
A_\mu(Q) \, \Lambda_\mu(P;Q) \, \pi_- (P-\case{Q}{2})
\eeq
which emphasizes the coupling of a photon with momentum $Q$ to the 
electromagnetic current of a $\pi^+$ with 
initial momentum $P+\frac{Q}{2}$ described by the field 
\mbox{$\pi_- = (\pi_1 -i \pi_2)/\sqrt{2}$}.  An equivalent expression is
\beq
\hat{\cal S}\left[\gamma\pi\pi\right]= \case{i}{2} \int 
\frac{d^4P,Q}{(2\pi)^8}
\; A_\mu(Q) \; \hat{3} \cdot \vec{\pi}(-P-\case{Q}{2})
\times\vec{\pi}(P-\case{Q}{2})
\, \Lambda_\mu(P;Q) .
\label{sgpp}
\eeq
The vertex 
\mbox{$\Lambda_\mu = (e_u) \Lambda_\mu^u + (-e_d) \bar{\Lambda}_\mu^d$} 
consists of contributions from the photon coupling to the quark and antiquark.
With $u$ and $d$ quarks taken to be identical except for their electric charges
$e_u$ and $e_d$, it is easily shown that 
\mbox{$\Lambda_\mu^u = \Lambda_\mu^d  = \Lambda_\mu$}.  The second term
of the action produced  in \Eq{newShatBA} yields 
\beq
\hat{\cal S}\left[\gamma\pi\pi\right]=
-{\rm Tr}\left[S \;  \Gamma_\mu A_\mu \;
(S \; \vec{\pi} \cdot \vec{\Gamma}_\pi  )^2 \right] . \label{Sgpp}
\eeq
One then obtains the $\gamma\pi\pi$ vertex 
$\Lambda_\mu$ at meson tree level in the GCM as the expression
\begin{eqnarray}
\label{piemver}
\lefteqn{\Lambda_\mu(P;Q)=} \\
& &  2N_c \, {\rm tr}_s \int \dq \, 
 \overline{\Gamma}_\pi(q+\case{Q}{4}; -P_+)
        S(q_+ + \case{Q}{2}) \Gamma_\mu(q_+;Q)S(q_+ - \case{Q}{2})
        \Gamma_\pi(q- \case{Q}{4};P_-) S(q_-)~,
\nonumber
\end{eqnarray}
where we have used  \mbox{$P_\pm = P \pm \frac{Q}{2} $} 
for the final and initial momenta of the pion and also
\mbox{$q_\pm = q \pm \frac{P}{2} $}.  The momentum assignments used for this
vertex are illustrated in Fig.~\ref{gppdiag}.  The trace over color and flavor
has been carried out leaving only the trace over Dirac spin.   
The quark charge operator
$\hat{Q}$  has contibuted to the  flavor trace relations   
\mbox{$ {\rm tr}_f \tau_+ \hat{Q} \tau_- = 2 e_u$} and
\mbox{$ {\rm tr}_f \tau_- \hat{Q} \tau_+ = 2 e_d$}.
In Eq.~(\ref{piemver})  
$\Gamma_\pi(q;P)$ denotes the pion Bethe-Salpeter amplitude in Dirac spin 
space,  \mbox{$ \overline{\Gamma}_\pi(q;P) = [C^{-1}\Gamma_\pi(-q;P) C]^T$} 
is for the corresponding antiparticle and here reduces to $\Gamma_\pi(q;P)$. 
The dressed quark-photon
vertex associated with in and out quark momenta $q \pm \frac{Q}{2}$ is
denoted $\Gamma_\mu(q;Q)$.    The above result is also what is 
obtained from the DSE
approach to QCD truncated at the generalized impulse approximation.  The 
generalization refers to the propagators and vertices being 
dressed.  The dressing of each element is consistent in a way demanded by 
chiral symmetry and EM current conservation.   

In the specific study (Roberts, 1996) that we summarize here, the Ball-Chiu 
Ansatz \mbox{$\Gamma_\mu(q;Q) \rightarrow \Gamma_\mu^{bc}(q;Q)$} is used
thus making the generalised impulse approximation given in \Eq{piemver}
{\it completely determined} by the quark propagator without the need for new 
parameters.  The pion charge form factor data was used, along with
the other pion properties mentioned in Sec.~\ref{sect_GB}, to set the 
propagator parameters.  (Various predictions from that parameterization are
summarized in Sec.~\ref{sect_interact}.)  This approach thus entails that 
vector meson contributions are explicitly excluded  and must be treated as an 
additional contribution.  

Using  the charge conjugation properties: \mbox{$S(-k)^T = $} \mbox{$ C^\dagger\,S(k)\,C$};
\mbox{$\Gamma_\pi^T(-q;P)  = $} \mbox{$  C^\dagger\,\Gamma_\pi(q;P)\,C$}; and
\mbox{$\Gamma_\mu^T(-q;Q)  = $} \mbox{$ - C^\dagger\,\Gamma_\mu(q;Q)\,C$},
where \mbox{$C=\gamma_2\gamma_4$}, one finds
that the $\gamma\pi\pi$ vertex has the symmetry properties
\mbox{$\Lambda_\mu(P;Q)= $} \mbox{$ -\Lambda_\mu(-P;Q) = $} \mbox{$ \Lambda_\mu(P;-Q)$}. Hence the 
most general form is
\beq
\Lambda_\mu(P;Q) = 2P_\mu \; F_\pi (P^2,Q^2,(P\cdot Q)^2)
+ 2Q_\mu P\cdot Q \; H_\pi (P^2,Q^2,(P\cdot Q)^2).
\label{genpiver}
\eeq
For elastic scattering, 
\mbox{$(P-\frac{Q}{2})^2=(P+\frac{Q}{2})^2$}, that is $P\cdot Q=0$.  Thus the
conservation of the electromagnetic $\pi$-current 
\begin{equation}
Q_\mu \Lambda_\mu(P;Q) = 0~,
\end{equation}
follows automatically in this formulation independently of the details of
the parameterization of the dressed quark propagator.  The physical form of
the vertex is
\begin{equation}
\label{physpiver}
\Lambda_\mu(P;Q) =  2 P_\mu F_\pi(P^2,Q^2)~,
\end{equation}
where the pion mass-shell condition is
\mbox{$P^2=-m_\pi^2-\frac{Q^2}{4}$}. 

At \mbox{$Q^2 = 0$}, the condition that the correct physical charge is 
obtained is equivalent to the physical normalization condition for the pion
BS amplitude.  To see this, the Ward identity for the photon-quark vertex
\mbox{$\Gamma_\mu(q;0) = - \partial S^{-1}(q)/ \partial q_\mu$} can be used in
Eq.~(\ref{piemver}) to produce 
\begin{eqnarray}
\label{pich}
\lefteqn{\Lambda_\mu(P;0) = 2 P_\mu\,F_\pi(P^2,0) = }\\
& & \left.  \frac{\partial}{\partial P_\mu} 2N_c {\rm tr}_s \int \dq  
\overline{\Gamma}_\pi(q;-K) S(q_+) \Gamma_\pi(q;K) S(q_-) 
\right|_{P^2=K^2=-m_\pi^2} ~. \nonumber
\end{eqnarray} 
Comparison with the ladder BS normalization condition in Eq.~(\ref{bsnorm})
shows that the former is equivalent to \mbox{$F_\pi(P^2,0)=1$}.  
In other words, generalised impulse approximation combined with a
$P$-independent Bethe-Salpeter kernel provides a consistent approximation
scheme.  Also the condition of physical net electric charge may be viewed as
the normalization condition for BS amplitudes.  The expression for $f_\pi^2$
given in Eq.~(\ref{fpi2}) can be obtained directly from the chiral limit of 
Eq.~(\ref{pich}).   Since the quark propagator in 
use here is an entire function, the loop integral in \Eq{piemver} is free of
``endpoint'' and ``pinch'' singularities.  This is a particular, sufficient
manner in which to realise the requirement that Fig.~\ref{gppdiag} have no
free-quark production thresholds.  It also facilitates the numerical 
integration. 

The results obtained for $F_{\pi}(Q^2)$ are shown in Fig.~\ref{fig_piff1}
and \Fig{fig_piff2} 
taken from Roberts (1996).  The parameterization employed for the quark 
propagator in that work differs slightly from the later form that we have
summarized in Sec.~\ref{sect_prop}.  However Fig.~\ref{fig_piff1} and
\Fig{fig_piff2} are a 
faithful representation of the charge form factor resulting from the
parameterization in Table~\ref{tab_pk}.  We recall that the data for $F_{\pi}(Q^2)$
are included in the set of pion and kaon observables to which the propagator 
parameters are fit.   Other pion and kaon observables arising from
the same model are listed in Table~\ref{tab_pk_res}.   Since at zero momentum 
the photon-quark vertex used here satisfies the Ward identity,  the pion charge 
radius $r_{\pi^\pm}$ generated by the generalized impulse approximation is
totally determined by the quark propagator.  The value of $f_\pi r_{\pi^\pm}$ 
preferred by the fit was rather stable and having chosen to fit to the experimental
$f_\pi$, the  $r_{\pi^\pm}$ value shown in Table~\ref{tab_pk_res}, which is some 
$15\%$ below the experimental value, is strongly preferred.   Earlier GCM studies
(Roberts \etal, 1994), with quite different representations of the dressed quark
propagator, also find essentially a $10-15$\% low value of $r_\pi$ from the generalized 
impulse approximation.  Interestingly, the additional pion loop contribution to 
$r_\pi$ has been calculated within the GCM format (Alkofer, Bender and Roberts, 1995) 
and was found to have an upper limit of that size. 
The small role for pion loops is consistent with the lattice QCD studies of 
Leinweber and Cohen (1993).  We discuss this point in more depth in \Sec{sect_loops}.

These results imply that with photon coupling to the dressed quark core of the pion
as we have described, the $\rho$-meson contributes little to  $F_\pi(Q^2)$ in the
spacelike region, even though it is important for timelike-$Q^2$ near the 
$\rho$ pole.  This is especially clear in the NJL model limit of the
formulation described here where, for example, our dressed photon-quark vertex
\mbox{$\Gamma_\mu^{bc}(q;Q) \rightarrow - i \gamma_\mu$}.  The additional
$\rho$ pole term has been included in some NJL model studies (Blin, Hiller 
and Schaden, 1988; Bernard and Mei\ss ner, 1988).  In \Sec{sect_loops} we present
an analysis of whether such a reduced role for VMD processes is consistent with the 
GCM/DSE implementation of ladder structure for the pion in terms of dressed quarks.

The asymptotic behavior of the generalized impulse approximation to $F_\pi(Q^2)$ 
was also investigated  by Roberts (1996). 
The form of the quark propagator in Eqs.~({\ref{ssb}) and 
(\ref{svb}) leads to all elements of the quark loop being consistent with 
the UV behavior of QCD up to logarithmic corrections.  In particular, the quark 
propagator generates 
\mbox{$\Gamma_\pi(q;0) \propto B(q^2) \rightarrow \frac{1}{q^2}$}
which, up to the $\ln p^2$ corrections associated with the anomalous
dimension, reproduces the ultraviolet behavior of the BS amplitude in QCD 
(Miransky, 1990).  It also requires, via the Ball-Chiu Ansatz, that the 
dressed photon-quark vertex evolve from soft to hard behavior in accord with such 
behavior of the quark propagators that attach to it as constrained by the 
Ward-Takahashi identity.  
From the non-factorized quark loop integral in \Eq{piemver},
the asymptotic behavior of $F_\pi(Q^2)$ is found to be determined by that of 
the bound state BS amplitude in the form (Roberts, 1996)
\beq
\left. F_\pi(Q^2) \right|_{Q^2\rightarrow \infty} \propto \; 
\int \dq \; \chi_\pi(q) \, \chi_\pi(q-\case{Q}{2})~, 
\label{asff}
\eeq
where $\chi_\pi(q)$ is the chiral limit BS wavefunction 
\mbox{$\chi_\pi(q) = S(q) B(q^2) S(q)$}.   In the asymptotic $Q^2$ analysis leading to 
this, the photon vertex tends to its bare limit, but the integrand can always find 
support in a domain that leaves several of the loop integral elements evaluated at 
soft values of their momentum arguments.  Here the mass-shell pion wavefunctions
dominate.  With a constituent quark mass approximation for propagators, one has the
approximate behavior \mbox{$\chi_\pi(q) \sim $} 
\mbox{$\gamma_5 (q^2 + \Lambda^2)^{-2}$} consistent with asymptotic freedom.  
Thus the integration in \Eq{asff}, after a now standard evaluation, gives 
\begin{equation}
\left. F_\pi(Q^2) \right|_{Q^2\rightarrow \infty} \propto \; 
\frac{\ln Q^2}{Q^4}~. \label{UVFpi}
\end{equation}
This is confirmed to be the leading behavior of the numerical evaluation of the full
loop integral (Roberts, 1996). 
The $\ln p^2$ corrections that arise due to the anomalous dimension of the 
propagator and Bethe-Salpeter amplitude in QCD, are found to only lead to the 
modification of Eq.~(\ref{UVFpi}) in which 
\mbox{$\ln[Q^2] \rightarrow \ln[Q^2]^\gamma$}, where $|\gamma|$ is 
${\cal O}(1)$.   This asymptotic term is found to dominate only for 
\mbox{$Q^2 \geq 10~{\rm GeV}^2$} where it departs from the perturbative QCD
factorization form ($1/Q^2$).  The latter is based on a factorization of the loop
integral to separate out a hard factor leaving a soft integral.   This requires
assumptions about the favored loop kinematics guided by the singularity 
structure of the loop integral appropriate to non-confining quark propagators
and is not applicable here.  The result in \Eq{UVFpi} is due to full evaluation of
the loop integral without factorization and it is found that both bound state
amplitudes contribute asymptotically.  The resulting peak in $Q^2 F_\pi(Q^2)$ 
about  \mbox{$Q^2 \sim 3~{\rm GeV}^2$},  which is also found 
for the $K^\pm$ form factors, is  a signal of non-perturbative quark-antiquark
recombination into the final state meson in this exclusive elastic scattering 
process.   The same phenomena is found to occur with  the dressed loop integral for the
$\gamma ^*\pi ^0 \rightarrow \gamma $ transition form factor discussed in 
\Sec{sect_pgg}.  There the dressed vertex for the final real photon replaces the 
final pion amplitude; the real photon vertex remains soft. 

The extension of this approach to the kaon electromagnetic form factors has been
studied by Burden, Roberts and Thomson (1996).  In that work  
the generalized impulse approximation to the $\gamma K K$ vertex for the $K^+$
is written
\begin{eqnarray}
\label{giaus}
\Lambda_\mu^{K^+}(P;Q)= (e_u)\Lambda_\mu^u(P;Q)
        + (-e_s) \bar\Lambda_\mu^s(P;Q)
\end{eqnarray}
where the $u$ quark contributes
\begin{eqnarray}
\label{uloop}
\lefteqn{\Lambda_\mu^u(P;Q)=} \\
& &  2N_c \, {\rm tr}_s \int \dq \, 
 \overline{\Gamma}_K(q+ \bar \xi \case{Q}{2}; -P_+)
        S^u(q_+ + \case{Q}{2}) \Gamma_\mu^u(q_+;Q)S^u(q_+ - \case{Q}{2})
        \Gamma_K(q - \bar \xi \case{Q}{2};P_-) S^s(q_-)~,
\nonumber
\end{eqnarray}
and the $\overline s$ quark contributes
\begin{eqnarray}
\label{sloop}
\lefteqn{\bar\Lambda_\mu^s(P;Q)=} \\
& &  2N_c \, {\rm tr}_s \int \dq \, 
 \Gamma_K(q + \xi \case{Q}{2}; -P_+)
        S^s(q_+ + \case{Q}{2}) \Gamma_\mu^s(q_+;Q)S^s(q_+ - \case{Q}{2})
       \overline{\Gamma}_K(q - \xi \case{Q}{2};P_-) S^u(q_-)~.
\nonumber
\end{eqnarray}
The $K^-$ vertex is the negative of this.
The momentum assignments are the same as used for the pion in 
Fig.~\ref{gppdiag}.  We have used  \mbox{$P_\pm = P \pm \frac{Q}{2} $} 
for the final and initial momenta of the kaon.   The conventions are the same 
as for the pion vertex, \Eq{piemver}, except that
now the required flavor dependence of the quark propagators and the photon-quark
vertex are indicated as $S^f$ and $\Gamma_\mu^{f}$ respectively, and 
the parameters $\xi$ and \mbox{$\bar \xi = 1 - \xi$} reflect the choice of $\bar q q$
relative momentum variable for the kaon BS amplitude $\Gamma_{K}$.   That is, 
the kaon BSE is of the ladder form given in \Eq{bspi} and the amplitude $\Gamma_K(k;P)$
employs the choice of relative momentum variable \mbox{$k = \bar\xi p_1 -\xi p_2$}.
The notation used here for the momentum dependence of
the quark propagators is \mbox{$S^u(q_\pm) = S^u(q \pm \xi P)$} and
\mbox{$S^s(q_\pm) = S^s(q \pm \bar\xi P)$}.  As with the pion, the physical 
elastic scattering vertex can be written \mbox{$\Lambda_\mu^{K^+}(P;Q)=$} 
\mbox{$ 2P_\mu F_{K^+}(Q^2)$}.
\begin{figure}[ht]

\centering{\
\psfig{figure=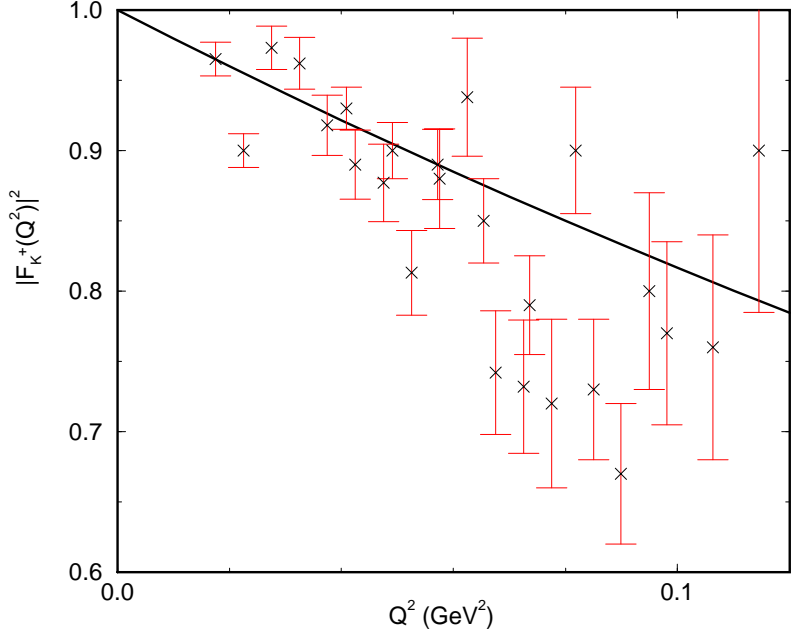,rheight=9.5cm,height=10.0cm} }

\parbox{130mm}{\caption{Kaon charge form factor from Burden, Roberts and 
Thomson (1996).  The available data is shown and is from Amendolia 
{\protect\it et al.}~(1986b), and Dally {\protect\it et al.}~ (1980). \label{kff} } } 
\end{figure}

As discussed in \Sec{sect_other}, the underlying ladder BSE that is assumed here
does not preserve the independence with respect to the partitioning parameter $\xi$ 
that would be a property of an exact treatment.  The constraint imposed by
Burden, Roberts and Thomson (1996) to fix $\xi$ is to require that the neutral kaon 
have exactly zero charge, i.e. $F_{K^0}(Q^2=0)=0$.  This works in the following
way.   The generalized impulse approximation to the $K^0$ vertex is given by
\begin{eqnarray}
\Lambda_\mu^{K^0}(P;Q)= (e_d)\Lambda_\mu^d(P;Q)
        + (-e_s)\bar\Lambda_\mu^s(P;Q)~.
\label{K0FF}
\end{eqnarray}
The $u$ and $d$ quarks are taken as identical except for the electric charge
and hence \mbox{$\Lambda_\mu^d =\Lambda_\mu^u = \bar\Lambda_\mu^u$}.  
If the $u$ and $s$ quark properties, except for the charge, are set equal in 
Eqs.~(\ref{uloop}) and (\ref{sloop}) the generalised
impulse approximation to the $\gamma\pi\pi$ vertex, \Eq{piemver}, is
recovered provided that $\Gamma_K \rightarrow \Gamma_\pi$, $f_{K}
\rightarrow f_{\pi}$ and $\xi = 1/2$, which is required by charge conjugation
symmetry. This provides a consistency check.  The same limit 
applied to \Eq{K0FF} verifies the charge conjugation 
symmetry requirement that there is no photon coupling to the (self-conjugate) 
$\pi^0$.  This formulation ensures EM current conservation because the quark-photon
vertex satisfies the Ward identity and produces \mbox{$F_{K^{\pm}}(Q^2=0)= 1$}.
This happens because at $Q^2=0$ the generalized impulse
approximation for the charged kaon vertex reduces to the normalization condition
for the BS amplitude in a similar way that led to \Eq{pich} 
for the pion.   However for the non-self-conjugate $K^0$, there is not complete 
cancellation in Eq.~(\ref{K0FF}) at $Q^2=0$ unless $\xi$ is suitably different 
from $1/2$ to counteract the different dynamics for the $d$ and $s$ quarks.    
The presence of this dependence reflects a deficiency of the ladder approximation  to
the BSE.

This formulation was applied by Burden, Roberts and Thomson (1996) using the
Ball-Chiu Ansatz for $\Gamma_\mu$.  The $u/d$ quark propagator parameters 
were fixed by pion physics and the only free parameters were the two $s$ 
quark parameters $C^s_{m_s=0}$ and $b^s_2$ described in Sec.~\ref{sect_prop}.
These parameters are determined by requiring that the model reproduce, as well 
as possible, the experimental values for the dimensionless quantities
\mbox{$f_K/f_\pi = 1.22\pm 0.02$}, 
\mbox{$r_{K^\pm}/r_{\pi^\pm}= 0.88 \pm 0.07$} and
\mbox{$m_K/f_K = 4.37 \pm 0.05$}. The ansatz
\beq
\Gamma_K(q;P^2=-m_K^2)  \approx 
        i\gamma_5\,\frac{B_{m_s=0}^s(q^2)} {f_K}~,
\label{gammaK}
\eeq
is used with the associated kaon mass being obtained by solving 
$\hat\Delta^{-1}_K(P^2) =0$, where the inverse propagator is as given in 
\Eq{delinvK}.  The normalization constant is obtained from
\mbox{$f_K^2 = \hat\Delta^{-1~\prime}_K(-m_K^2)$} 
which also ensures the correct  normalisation of the charged kaon form
factor.  

The obtained set of quark propagator parameters are those shown earlier in 
Table~\ref{tab_pk}.  The fit selects differences between the $u$ and $s$ quark
propagators. The  calculated electromagnetic form factor is 
reproduced in Fig.~\ref{kff}.  The comparison between calculated and
experimental pion and kaon quantities from tha work is presented in 
Table~\ref{tab_pk_res}.  For the neutral kaon $r^2_{K^0}<0$, and 
$F_{K^0}(Q^2)$ is found to be similar in form and
magnitude to the charge form factor of the neutron.   The other quantities
shown in \Table{tab_pk_res} are obtained consistently from the same quark
propagator model.

The difference between the calculated and measured values of the charge radii
and scattering lengths in Table~\ref{tab_pk_res} is a measure of the importance
of final-state, pseudoscalar rescattering interactions and
photon--vector-meson mixing, which are not included in generalised-impulse
approximation. The above calculation suggests that such effects
contribute less than $\sim$ 15\% and become unimportant for $Q^2 >
1$~GeV$^2$.  The fact that the calculated values of $f_K/f_\pi$ and
$r_K/r_\pi$ agree with the experimental values of these ratios suggests that
such effects are no more important for the kaon than for the pion.

The results for $F(Q^2)$ found by Burden, Roberts and Thomson (1996) for $K^+$
and $K^0$ at $Q^2< 2$~GeV$^2$ are not sensitive to details of the propagator 
parametrization.  Their qualitative features are similar to those from a recent
quark loop investigation based on a separable interaction (Buck \etal, 1995).
However specific magnitudes are different.  As is the case of the pion, the
large $Q^2$ behavior is a faster fall-off than $1/Q^2$ due to both
non-perturbative bound state amplitudes playing a role in this exclusive 
scattering process.    The
behavior of $F_{K^\pm}(Q^2)$ at $Q^2>2\sim 3$~GeV$^2$ is influenced by
details of the Ansatz for the kaon Bethe-Salpeter amplitude,
Eq.~(\ref{gammaK}), that are not presently constrained by data.  
This emphasizes that measurement of the electromagnetic form
factors is a probe of the bound state structure of the meson and
of non-perturbatively generated differences between the $u$ and
$s$ quark propagators.

\sect{Meson Interactions}
\label{sect_interact}

\subsect{The $\pi^0 \gamma \gamma$ Form Factor}
\label{sect_pgg}

The pion charge form factor for space-like
momenta is one of the simplest but non-trivial testing grounds for
applications of QCD to hadronic properties.  A closely related quantity that
has received less attention is the $\gamma ^*\pi ^0
\rightarrow \gamma $ transition form factor (Ametller, Bijnens, Bramon and 
Cornet, 1992; Jaus, 1991).
Here the photon momentum dependence maps out a particular off-shell extension
of the axial anomaly (Adler, 1969; Bell and Jackiw, 1969). Presently 
available data for this transition
form factor in the space-like region $Q^2< 2.5$ GeV$^2$ is from the CELLO 
collaboration (Behrend et al., 1991) at the PETRA storage ring where the process
$e^+e^-\rightarrow e^+e^-\pi ^0$ was measured with geometry requiring one of
the two intermediate photons to be almost real.  There is renewed interest in
this transition form factor due to the prospect of obtaining higher precision
data over a broader momentum range via virtual Compton scattering from a
proton target at CEBAF (Afanasev, 1994; Afanasev, Gomez and Nanda, 1994).
The theoretical results (Frank \etal, 1995) we summarize here can be viewed
either as an application of the GCM model at the $0^{th}$ level of the meson 
loop expansion or as an application of the DSE approach truncated to 
generalized impulse approximation.  The improvements over previous
quark loop studies (Ito \etal, 1992; Anikin \etal, 1994) are: quark confinement
(thus eliminating spurious quark
production thresholds), dressing of the photon-quark vertex, and the
dynamical relation between the pion Bethe-Salpeter amplitude and the quark
propagator.  The latter elements are crucial for producing the correct
mass-shell axial anomaly independently of model details.

Using the action given in \Eq{newShatBA}, the relevant interaction is 
generated by the first term $\hat{\cal S}_1[\pi;A]$.  Expansion of the quark
loop term shown as the first part of \Eq{Shat1BA} yields  
\beq
{\cal S}[\pi^0 \gamma \gamma] = -{\rm Tr}
[S \; \Gamma_\mu A_\mu \; S \; \Gamma_\nu A_\nu \; S \; 
\Gamma_{\pi}\pi^0].  \label{spiggtr}
\eeq
That is
\begin{equation}
{\cal S}[\pi ^0\gamma \gamma ]=\int \frac{d^4Pd^4Q}{(2\pi )^8} A_{\mu }(-P-Q)
A_{\nu }(Q)\pi ^0(P)\Lambda _{\mu \nu}(P,Q) ,
\label{spigg}
\end{equation}
where the vertex function is given by the integral
\begin{eqnarray}
\Lambda _{\mu \nu }(P,Q)=-\mbox{tr}\int \frac{d^4k}{(2\pi )^4} &&
S(k-P-Q)\Gamma _{\mu }( k-\case{P}{2}-\case{Q}{2};-P-Q)
S(k)\nonumber \\
&&\times \Gamma _{\nu }( k-\case{Q}{2};Q)S(k-Q)
\Gamma _{\pi }( k-\case{P}{2}-Q;P) .\label{int}
\end{eqnarray}
\begin{figure}

\psrotatefirst
\centering{\
\psfig{figure=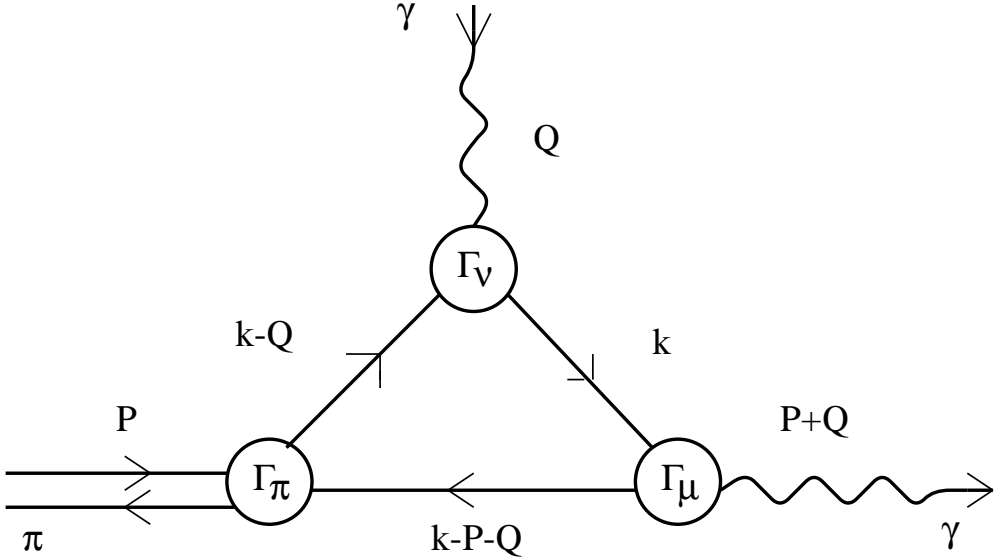,rheight=8.1cm,height=7.5cm,angle=-90} }

\parbox{130mm}{\caption{The quark triangle diagram for the generalized impulse
approximation to the $\protect\gamma^* \protect\pi^{0} \protect\gamma$ vertex. 
\label{tri} } }
\end{figure}
The momentum assignments are shown in the quark triangle diagram
of {}Fig.~\ref{tri}.  The general form of the vertex allowed by CPT symmetry is
\mbox{$\Lambda _{\mu \nu }(P,Q)=-i\frac{\alpha }{\pi f_{\pi }}\epsilon
_{\mu \nu \alpha \beta }P_{\alpha }Q_{\beta }~g(Q^2,P^2,P\cdot Q)$}
where $\epsilon_{4123}=1$, $\alpha $ is the fine-structure constant, $f_{\pi
}$ is the pion decay constant, and $g$ is the off-mass-shell invariant
amplitude.  With the one photon mass-shell condition $(P+Q)^2=0$, the
invariant amplitude, denoted by $g(Q^2,P^2)$, is the object of the 
calculation.
{}For a physical pion the shape of the $\gamma ^*\pi ^0\rightarrow
\gamma $ transition form factor is given by $g(Q^2,-m^2_{\pi })$.
The chiral limit for the physical $\pi ^0\rightarrow \gamma \gamma $ decay
amplitude is fixed at $\frac{\alpha }{\pi f_{\pi }}$ by the axial
anomaly (Itzykson and Zuber, 1980) which gives an excellent account of the
$7.7$ eV width and requires $g(0,0)=1/2$.  This follows only from
gauge invariance and chiral symmetry in quantum field theory and provides a
stringent check upon model calculations.

In the work of Frank \etal ~(1995) the dressed photon-quark vertex $\Gamma_\nu$
is taken to be the Ball-Chiu Ansatz $\Gamma_\nu^{bc}$ from \Eq{ballchiu} and
the pion BS amplitude is taken to be the $\gamma_5$ part  $B(p^2,m)/f_\pi$ as
suggested by  \Eq{pibs}.   
The $\pi^0 \gamma \gamma$ vertex function in \Eq{int} is now completely
specified in terms of the quark propagator.  An exactly parallel situation
holds for the spacelike pion charge form factor produced from the same
approach (Roberts, 1996) and discussed in Sec.~\ref{sect_emff}.  The parameters
employed there are used without adjustment for $\pi\gamma\gamma$. 
At $Q^2=0$, numerical evaluation (Frank \etal, 1995)
yields $g_{\pi^0 \gamma \gamma}= g(0,-m_\pi^2)=0.496$, in agreement with the
previous application of this model (Roberts, 1996)  and in good agreement with
the experimental value $0.504 \pm 0.019$.  The chiral limit of this approach
has been shown (Roberts, 1996) to correctly incorporate the exact result
$g(0,0)=1/2$ produced by the axial anomaly {\it independent} of the form and
details of the quark propagator.   This is only possible when all elements 
of the loop are dressed consistently as dictated by symmetries. Here this
means the chiral limit $\gamma_5$ term of the pion BS amplitude is fixed by
chiral symmetry through \Eq{pibs} in terms of the quark scalar self-energy 
amplitude, and the physical soft photon-quark vertex is fixed by EM gauge 
invariance
through the Ward identity in terms of the quark propagator.   It is
advantageous to account for the distributed nature of  pions and  dressed 
quarks, not only because it is  a realistic outcome of QCD, but also because
important symmetries are implemented this way.
\begin{figure}[ht]

\psrotatefirst
\centering{\
\psfig{figure=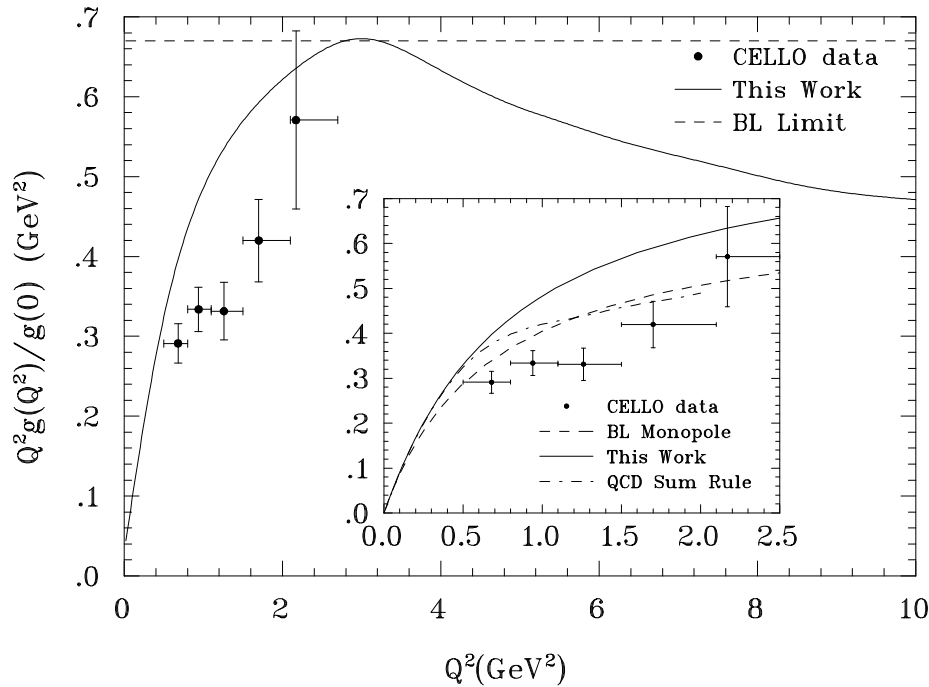,rheight=9.5cm,height=9.0cm,angle=90} }

\parbox{130mm}{\caption{The $\protect\gamma^* \protect\pi^0 \protect\gamma$ 
transition form factor
from Frank {\protect\it et al.}~(1995) along with the CELLO data  from Behrend 
{\protect\it et al.}~(1991). \label{pigg} } }
\end{figure}

The obtained form factor
$F(Q^2)=g(Q^2,-m_\pi^2)/g(0,-m_\pi^2)$ at the pion mass-shell is displayed as
$Q^2F(Q^2)$ in {}Fig.~\ref{pigg} by the solid line.   A ``radius'' or 
interaction size,  defined via
\mbox{$ r^2_{\gamma\pi^0\gamma} = $} \mbox{$ -6 \left.F'(Q^2)\right|_{Q^2=0}$},
yields \mbox{$0.47~{\rm fm}$} while a monopole fit to the data
yields (Behrend et al., 1991)  \mbox{$0.65 \pm 0.03~{\rm fm}$}.  The insert 
compares the
low $Q^2$ result with a recent result from a QCD sum rule approach
(Radyuskin, 1994), and a monopole form (Brodsky and Lepage, 1981) that 
interpolates from the
leading asymptotic behavior $F(Q^2) \rightarrow 8 \pi^2 f_\pi^2 /Q^2$ argued
from the pQCD factorization approach (Lepage and Brodsky, 1980).   In the 
latter two
approaches there is ambiguity due to: A) the unknown momentum scale at which
perturbative behavior should set in; and B) assumptions for the pion
wavefunction and how it should evolve with the momentum scale (Radyuskin, 1994).
Within the present GCM/DSE approach, both the photon coupling and the pion
wavefunction evolve with $Q^2$ in a way determined by the evolution of the
dressed quark propagators.  This produces, in a single expression, both the
ultra-violet behavior required by pQCD and the infra-red limit dictated by
the axial anomaly.  We note from the parameterization of the quark propagator
given in \Eq{ssb} and \Eq{svb} that the employed pion Bethe-Salpeter amplitude
$B(p^2,m)/f_\pi$ has the correct leading power law behavior
$m\lambda^2/p^2f_\pi$ which implements the hard gluon contribution that
dominates pQCD.

The dotted straight line in {}Fig.~\ref{pigg} is the pQCD
factorization (Lepage and Brodsky, 1980) limit \mbox{$Q^2F(Q^2) \rightarrow$} 
\mbox{$ 8 \pi^2 f_\pi^2 = 0.67~{\rm GeV}^2$}.
Although the GCM/DSE non-factorized calculation reaches this value near
\mbox{$Q^2=$} \mbox{$3~{\rm GeV}^2$}, there is a slow decrease with higher $Q^2$
consistent with a logarithmic correction.  An excellent fit to the numerical
results for \mbox{$3.3~{\rm GeV}^2 \leq Q^2 \leq 10~{\rm GeV}^2$} is provided
by \mbox{$F(Q^2) = A~[1.0+$} \mbox{$B~Q^2~\ln(C~Q^2)]^{-1}$}, where
$A=1.021$,~$B=0.461\,/m_\rho^2=0.777~{\rm GeV}^{-2}$ and
$C=1.16\,/m_\rho^2=1.45~{\rm GeV}^{-2}$.  (Neglected in this calculation is 
the anomalous dimension of the quark propagator; this would modify the
power of the $\ln$-correction but would be a numerically small effect.)

The logarithmic correction to the anticipated $1/Q^2$ asymptotic behavior can
be attributed to the persistent nonperturbative nature of the coupling to the
final state soft photon in this exclusive process (Frank \etal, 1995).
Numerically it is found that, if a bare coupling
were to be used for both photons as is implicit in the pQCD factorization
approach, $F(Q^2)$ would eventually approach $8 \pi^2 f_\pi^2/Q^2$.
The turn-over in $Q^2F(Q^2)$ near $3~{\rm GeV}^2$ predicted
in {}Fig.~\ref{pigg} is
barely within the $Q^2$ limit of $4~{\rm GeV}^2$ anticipated for measurements
at CEBAF if a $6~{\rm GeV}$ electron beam becomes available.
This turn-over and the logarithmic corrections generated by the loop integral
are features also found in the parallel approach to the pion charge form
factor (Roberts, 1996).
For a study of the behavior of the vertex function off the pion mass shell, as
needed for the anticipated CEBAF experiment, see Frank \etal~(1995).

This $\pi\gamma\gamma$ approach has been recently extended without modification
to the $\gamma \pi^\ast \rightarrow \pi\pi$ form factor (Alkofer and Roberts,
1996).  The value $F^{3\pi}(4m_\pi^2)$ of the amplitude at 
\mbox{$s=4m_\pi^2$} from that work is shown in \Table{tab_pk_res}.  
This process is also due to the chiral anomaly and it is found, again,
that the exact chiral limit current algebra result at the soft point 
\mbox{$F^{3\pi}(0) =$} \mbox{$e N_c/(12\pi^2 f_\pi^3)$} is produced 
independently of the details of consistently dressed elements of the loop.  
Together with the parallel result (Praschifka, Roberts and Cahill, 1987a)
for the anomalous five-pion Wess-Zumino term, \Eq{WZ}, these exact results
from symmetries are produced at a truncated level of the theory.  The 
correction terms, including meson loops, can apparently be absorbed into the
details of the quark propagator dressing to which the chiral anomalies
are evidently blind.

\subsect{ The $\rho\pi\pi$ Form Factor}
\label{sect_rpp}

From the second term of the action in \Eq{newShatBA},  we identify
\mbox{${\cal S}\left[\rho\pi\pi\right]=
-{\rm Tr}\left[S \; \vec{\Gamma}_\mu^\rho \cdot \vec{\rho}_\mu  \; 
(S \;  \vec{\Gamma}_\pi \cdot \vec{\pi}  )^2 \right]$}, which
yields
\beq
{\cal S}\left[\rho\pi\pi\right]= i \int \frac{d^4P,Q}{(2\pi)^8}
\vec{\rho}_\mu(Q)\cdot \vec{\pi}(-P-\case{Q}{2})\times\vec{\pi}(P-\case{Q}{2})
\Lambda_\mu(P,Q)
\label{actrpp}
\eeq
where the vertex is
\beq
\Lambda_\mu(P,Q)=\int \frac{d^4k}{(2\pi)^4}
V_\rho(k+\case{P}{2};Q) E_\pi(k+\case{Q}{4};-P-\case{Q}{2})
E_\pi(k-\case{Q}{4};P-\case{Q}{2}) T_\mu(k,P,Q)
\label{vint}
\eeq
with
\beq
T_\mu(k,P,Q)= 2 N_c {\rm tr}_s \left[
S(k+\case{P}{2}+\case{Q}{2}) i\gamma_\mu^T
S(k+\case{P}{2}-\case{Q}{2}) i\gamma_5
S(k-\case{P}{2}) i\gamma_5 \right].
\eeq
These expressions have been simplified through use of the approximations:
only the $i\gamma_5\vec{\tau}E_\pi(q;P)$ part of the pion BS amplitude 
$\vec{\Gamma}_\pi$ has been used, and the $\rho$ BS amplitude 
$\vec{\Gamma}_\mu^\rho$ has been represented by $i\gamma_\mu^T \vec{\tau} 
V_\rho(q;Q)$ where $\gamma_\mu^T$ is transverse to the $\rho$
momentum $Q$.  The $\rho \pi\pi$ vertex could have been obtained directly
from the $\gamma \pi\pi$ vertex discussed in Sec.~\ref{sect_emff}.   
Since only the isovector part ($\tau_3/2$) of $\hat{Q}$ from the photon-quark 
vertex contributes there,  it is clear that the result for $\rho^0\pi\pi$ can 
be obtained by replacing $\Gamma_\mu/2$ for the photon with 
$i\gamma_\mu^T V_\rho(q;Q)$ for the $\rho^0$. 

From symmetry properties, it is not difficult to show that
$\Lambda_\mu(P,Q)=-\Lambda_\mu(-P,Q)=\Lambda_\mu(P,-Q)$, which requires the
general form
\beq
\Lambda_\mu(P,Q)=-P_\mu \; F_{\rho\pi\pi}(P^2,Q^2,(P\cdot Q)^2)
-Q_\mu P\cdot Q \; H_{\rho\pi\pi}(P^2,Q^2,(P\cdot Q)^2).
\label{rppgen}
\eeq
With both pions on the mass-shell,
$(P-\frac{Q}{2})^2=(P+\frac{Q}{2})^2=-m_\pi^2$.
Equivalently, $P\cdot Q=0$ and $P^2=-m_\pi^2-\frac{Q^2}{4}$ so that only the
first term of \Eq{rppgen} survives. The corresponding form factor
$F_{\rho\pi\pi}(Q^2)$ contains the coupling constant as its mass-shell value,
i.e. \mbox{$F_{\rho\pi\pi}(Q^2=-m_\rho^2) = g_{\rho\pi\pi}$}. If
the form factor is held at this triple mass-shell value independent of momenta, 
then \Eq{actrpp}, when converted to coordinate representation, becomes
\beq
{\cal S}_{pt}\left[\rho\pi\pi\right]=-g_{\rho\pi\pi}\int d^4x \; 
\vec{\rho}_\mu(x)\cdot \vec{\pi}(x)\times\partial_\mu\vec{\pi}(x)~,
\label{pointaction}
\eeq
which is the point coupling limit expressed in a standard form.  There are, of course,
other forms of action for this interaction that are on-shell equivalent. The
general form of \Eq{actrpp} is just one of them (in coordinate space it is a 
complicated non-local form).   The momentum space form of \Eq{actrpp} is an efficient
way to account for the non-locality or off-shell structure of the interaction vertex
produced by the quark-gluon substructure of the tree-level quantities within the GCM. 
One should bear in mind that the off-mass-shell behavior here in the meson momenta 
depends on what has been identified as the propagator for the $\bar q q$ meson modes
as discussed in \Sec{sect_eigen}.   It is always subject to a field redefinition and
one must use a consistent definition throughout all elements used to construct an
observable or S-matrix element.
 
Previous studies of the $\rho\pi\pi$ vertex within the GCM (Praschifka \etal, 
1987b; Roberts \etal, 1988b; Hollenburg \etal, 1992) made
the approximation that $g_{\rho\pi\pi}$ could be estimated at the point 
\mbox{$P=Q=0$}.  This was necessary to avoid ambiguities associated with
numerical extrapolations of dressed quark propagators and vertex amplitudes
in the integral \Eq{vint}.  The parameterized analytic representation of the 
quark propagator described in Sec.~\ref{sect_prop} allows the correct
mass-shell constraints to be maintained for $g_{\rho\pi\pi}$ and 
like  quantities.  
Since the propagator parameters have so far been constrained only
by pion physics, there is no guarantee that continuation of $S(p)$ away from
the real $p^2$ axis by an amount proportional to $m_\rho^2$ will be correct.
The exploratory calculations involving vector mesons described here constitute 
a first investigation of this matter.
\begin{figure}[ht]

\centering{\
\psfig{figure=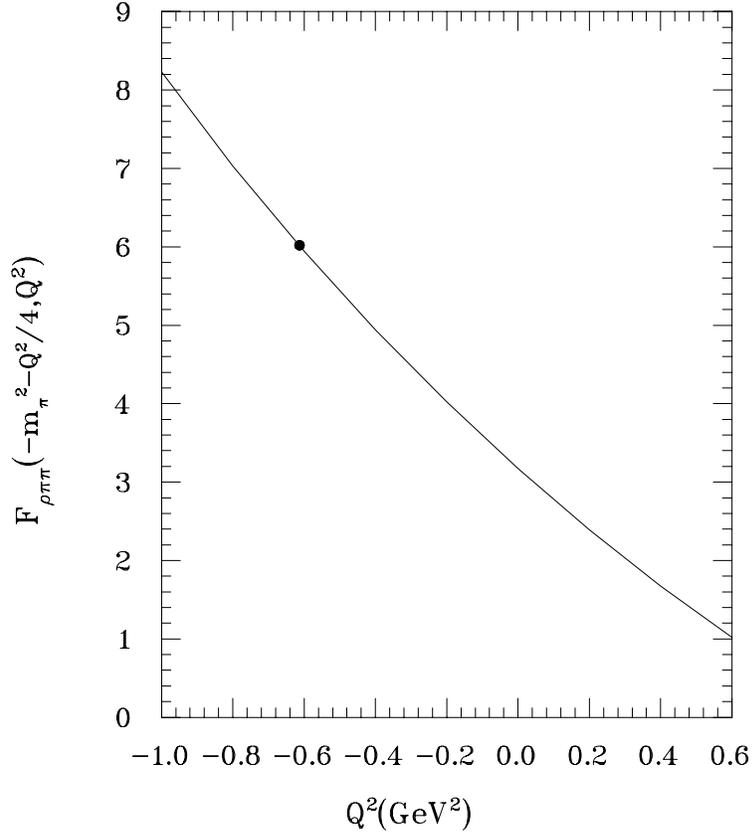,height=11.0cm} }

\parbox{130mm}{\caption{The $\rho\pi\pi$ form factor versus $\rho$ momentum
from Mitchell and Tandy, (1997). \label{rpp} } }
\end{figure}

The $\rho\pi\pi$ vertex has recently been studied within the GCM/DSE approach
by Tandy (1996) and also by Mitchell and Tandy (1997).  Those works use
\mbox{$E_\pi(q;P) = B(q^2,m)/f_\pi$} and the parameterization 
\mbox{$\Gamma^\rho(p^2)\propto e^{-p^2/a^2}$}.  The strength 
is set by the canonical normalization condition given in \Eq{bsnorm}.  
The range $a$ is then adjusted to reproduce the empirical value
$g_{\rho\pi\pi}^{\rm expt}=6.05$. This approach has proved to be 
phenomenologically successful for studies of other processes such as 
$\rho-\omega$ mixing (Mitchell \etal, 1994) and the 
\mbox{$\rho \rightarrow e^+ \; e^- $} decay (Pichowshy and Lee, 1996) from 
essentially the same theoretical framework.  The produced 
$\rho\rightarrow\pi\pi$ decay width, given by
\beq
\Gamma_{\rho\rightarrow\pi\pi} = \frac{g_{\rho\pi\pi}^2}{4\pi}
\frac{m_\rho}{12}\left[1-\frac{4m_\pi^2}{m_\rho^2}\right]^{3/2},
\label{rhowidth}
\eeq
is $151~{\rm MeV}$.  The calculated form factor
$F_{\rho\pi\pi}(Q^2)$ is shown in Fig.~\ref{rpp} for timelike and spacelike
momenta in the vicinity of the mass-shell at $-0.6~{\rm GeV}^2$ indicated by
a dot.   The main conclusion from this
calculation is that the previous approximation of using zero momentum to
extract a coupling constant (Praschifka \etal, 1987b) can underestimate the 
value by almost a factor of $2$.

These investigations are exploratory in the sense that an independent
calculation of the $\rho$ BS amplitude $\Gamma_\rho(p)$ has not been 
employed.  There are indications that an improved parameterization of 
$V_\rho(p^2)$ to include a more realistic power-law fall-off at high momenta
provides a simultaneous accounting for both  \mbox{$\rho \rightarrow \pi
\pi$} and \mbox{$\rho \rightarrow e^+ \; e^- $} decays (Roberts, 1997).
Covariants other than the canonical one $\gamma_\mu^T$ for the $\rho$ BS 
amplitude can contribute to the $\rho\pi\pi$ vertex and this has recently been 
investigated together with the effect of the sub-dominant pion covariants
(Qian and Tandy, 1997).  Truncation to the dominant $\rho$ amplitude is found
to be an accurate approximation while the effect of the pseudoscalar pion
component can be quite important.

The strong suppression displayed  in Fig.~\ref{rpp} as $Q^2$ increases from
the mass-shell point to the spacelike region indicates that the effective
coupling strength $g_{\rho\pi\pi}(Q^2)$ appropriate to the $\rho$ contribution 
to the pion charge form factor and radius is significantly smaller than
what is assumed in the standard VMD approach.  This is consistent with
findings discussed in Sec.~\ref{sect_emff}.  The strong    
dependence in Fig.~\ref{rpp} is very similar to that found recently 
(Frank and Roberts, 1996) in a study that used a $\rho$ amplitude
from solution of the BSE.  In that work the time-like region of the form factor
could not be calculated directly because the numerical solution of the $\rho$
BSE was performed only on the real $q^2 >0$ axis.  The results there suggest 
that the common procedure (e.g. see Eq.~(56) of Frank and Roberts (1996)) 
of interpolation of calculated form factors in the meson momentum from the 
Euclidean domain to the mass-shell point can be reliable if the mass is not too 
large.

\subsect{ The $\gamma\pi\rho$ Form Factor}
\label{sect_gpr}

The empirical $\rho$ BS amplitude set by $g_{\rho\pi\pi}$ as above enables
a parameter-free prediction for the $\gamma\pi\rho$ vertex. Apart from being
a consistency check in this manner, the $\gamma\pi\rho$ interaction
together with the $\gamma\pi\pi$, $\gamma\gamma\pi$, and the $\rho\pi\pi$
processes, provide important guidance for extending the present approach for
nonperturbative QCD modeling of meson physics beyond  phenomena  dictated 
by chiral symmetry.   Within nuclear physics, the associated isoscalar
$\gamma^*\pi\rho$ meson-exchange current contributes significantly
to electron scattering from light nuclei at presently available momenta.
In particular, our understanding of the deuteron EM structure functions
for $Q^2 \approx 2-4~{\rm GeV}^2$ is presently hindered by uncertainties
in the behavior of this form factor (Ito and Gross, 1993).
\begin{figure}[ht]

\centering{\
\psfig{figure=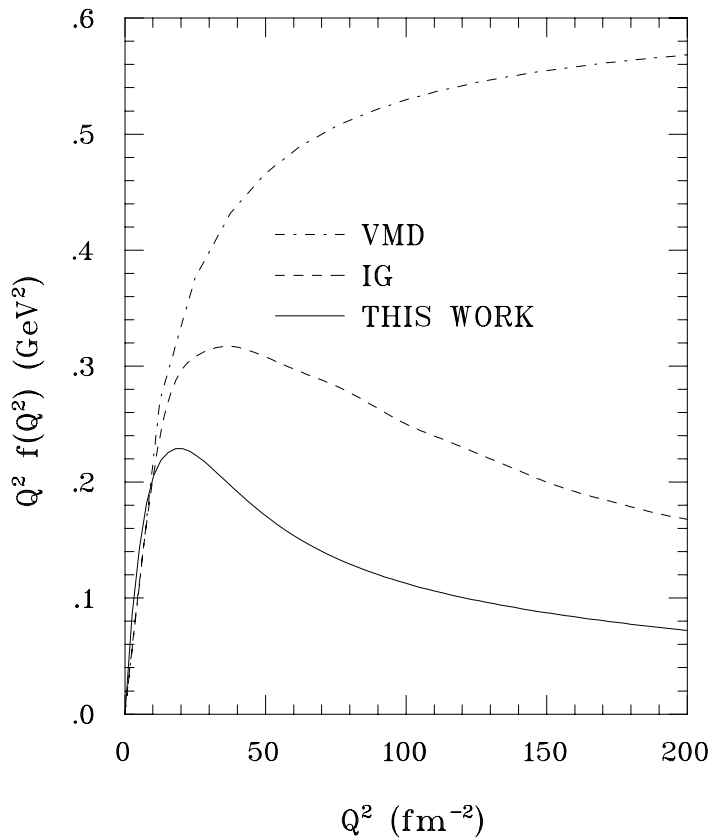,height=11.0cm} }

\parbox{130mm}{\caption{The $\gamma\pi\rho$ form factor versus photon momentum
from Mitchell (1995) and Tandy (1996). \label{rpg} } }
\end{figure}

Expansion of the second term of the action in \Eq{newShatBA} to first order in 
the EM field yields the $\gamma\pi\rho$ interaction as the pair of contributions
\beq
{\cal S}[\gamma\pi\rho]=-{\rm Tr}[S \; \Gamma_\mu A_\mu S \;
\vec{\Gamma}_\pi \cdot \vec{\pi} \;
S \; \vec{\Gamma}_\nu^\rho \cdot \vec{\rho}_\nu ] - {\rm Tr}
[S\; \Gamma_\mu A_\mu \; S\;  \vec{\Gamma}_\nu^\rho \cdot \vec{\rho}_\nu \;
S \; \vec{\Gamma}_\pi \cdot \vec{\pi} ] .
\label{gpraction}
\eeq
With a vertex function $\Lambda_{\mu\nu}(P,Q)$ defined by
\beq
{\cal S}[\gamma\pi\rho]=-\int \frac{d^4P,Q}{(2\pi)^8} \; A_{\nu}(Q) \;
\vec{\pi}(-P-\case{Q}{2})\cdot\vec{\rho}_{\mu}(P-\case{Q}{2})
\; \Lambda_{\mu\nu}(P,Q) ,
\eeq
one may combine the two terms in \Eq{gpraction} to obtain the integral
\beqar
\Lambda_{\mu\nu}(P,Q)&=&\case{e}{3}\int \frac{d^4k}{(2\pi)^4}
E_\pi(k+\case{Q}{4};-P-\case{Q}{2})
V_\rho(k-\case{Q}{4};P-\case{Q}{2}) \nonumber \\
&\times& 2 N_c {\rm tr}_s \big[S(k_+ -\case{Q}{2})\Gamma_\nu(k_+;Q)
S(k_+ +\case{Q}{2})i\gamma_5 S(k_-)i\gamma_\mu^T \big] ~.
\label{lam}
\eeqar
Here \mbox{$k_\pm = k \pm \frac{P}{2}$}.  The approximations made here for the 
$\pi$ and $\rho$ BS amplitudes are identical to those used for the $\rho\pi\pi$
interaction in Sec.~\ref{sect_rpp}. Use of the Ward-Takahashi 
identity from \Eq{wti} in \Eq{lam} shows that in this formulation the 
$\gamma\pi\rho$ EM current is conserved, that is 
$Q_\nu\;\Lambda_{\mu\nu}(P,Q)=0$.
The general form of the vertex function can be shown by symmetries to be
\beq
\Lambda _{\mu \nu }(P,Q) = -i\frac{e}{m_{\rho}}\; \epsilon
_{\mu \nu \alpha \beta } \; P_{\alpha }Q_{\beta }~g_{\rho \pi \gamma}~
f(Q^2,P^2,P\cdot Q) ~, \label{gprgen}
\eeq
as is expected for a coupling arising from the chiral
anomaly.  We have used the standard definition of the coupling constant so that,
at the triple mass-shell point, the form factor $f=1$.

The coupling constant and form factor resulting from the GCM or DSE approach 
truncated to the level shown above have been obtained
by Mitchell (1995) and by Tandy (1996).  The form of the quark propagator
parameterization is that discussed in Sec.~\ref{sect_prop} and the Ball-Chiu
Ansatz is used for $\Gamma_\nu$. Those works obtain $g_{\gamma \pi \rho}=0.5$  
The
experimental $\rho^+\rightarrow\pi^+\gamma$ partial width ($67 \pm 7~{\rm keV}$)
determines the empirical value $g_{\gamma \pi \rho}^{\rm expt}=0.54\pm 0.03$.
The $\gamma\pi\rho$ form factor obtained with on-mass-shell $\pi$ and $\rho$,
and weighted by $Q^2$, is shown by the solid curve in Figure~\ref{rpg}.
Also shown is the form factor suggested by vector meson dominance (VDM) as
initially used to treat such meson exchange effects in electron scattering
analysis (Hummel and Tjon, 1990), as well
as the result from a free constituent quark loop with bare photon
coupling (Ito and Gross, 1993; Ito \etal, 1992).   Both quark-based approaches 
produce a much softer form factor than does the VDM assumption.  
Above $50~{\rm fm}^{-2}$, which is readily accessible in
electron scattering, the differences are serious.  Since this calculation 
for $\gamma\pi\rho$ is equivalent to the
$0^{th}$ term in the meson loop expansion.  An estimate of 
meson-loop corrections to this result would be interesting.

\sect{Beyond Meson Tree Level}
\label{sect_loops}

The essence of the GCM approach is that the tree-level meson fields are ladder
$\bar q q$ bound states whose structure and interactions are governed by what we 
might call their dressed quark core.   Further quantum treatment will 
generate meson loop effects from integration over these fields. That is the 
dressing of ladder mesons by ladder mesons.  This extends the bound
state structure of the mesons to add meson cloud effects to the original dressed quark
core.  In the more general DSE approach, such processes are implemented by use of less
severe truncations of the equations of motion; also exposed that way in the DSE 
approach will be contributions from explicit gluon $n$-point functions that are not 
already accounted for by dressed quarks and mesons.   In either case,
it is clear that the approach just outlined is out of practical reach if one
has to proceed too far into the loop expansion or with higher $n$-point functions
to capture the dominant low energy physics.  A crucial question then is the size of 
meson loop contributions to the various
processes of interest, such as those observables we have described in previous 
Sections.  In the GCM/DSE approach, the composite and extended structure of the 
tree-level meson fields makes such investigations difficult.  On the other hand, the
distributed vertex functions for the coupling tend to suppress the numerical importance
of the meson dressing loop integrals, and it might be spectulated that for this 
approach the meson loops play much less of a role than for point coupling models or for
effective field theories built from point coupling of point interpolating meson fields.
We shall review the few investigations that have tackled this question in
the GCM/DSE approach.  Besides the pion loop contribution to the pion charge form 
factor, we also include here a discussion of vector meson mode contributions to this 
quantity. 

\subsect{Role of Vector Mesons in the Pion Form Factor}
\label{sect_vmdpiff} 
 
The relative unimportance of the $\rho$ for the space-like pion charge form factor 
in the present quark-gluon approach is consistent with considerations of 
vector-meson--photon mixing in QCD.   It can be understood in terms of \Eq{polevertex}
which is the dressed photon-quark vertex $\Gamma_\nu$ separated into a $\rho$ pole
term and a non-resonant or direct background term.   With both terms of that vertex 
employed for coupling to the pion in the manner discussed in \Sec{sect_emff}, one 
obtains
\beq
F_\pi(Q^2) = F_\pi^{GIA}(Q^2) + \frac{ F_{\rho\pi\pi}(Q^2) 
                     \; \Pi^{\rho \gamma}_T(Q^2) } {Q^2 + m_\rho^2(Q^2)} ~.
\label{piffrho}
\eeq
The first term is from the non-resonant photon vertex term  and is the generalized 
impulse approximation (GIA) that we have described in \Sec{sect_emff}.
The second term is the $\rho$ resonant term that arises in the GCM or
DSE approaches.   The relative importance of the two terms away from the $\rho$ 
mass-shell depends strongly upon their momentum dependence.  The off-mass-shell momentum
behavior of individual terms in photon vertex separations such as 
\mbox{$\Gamma_\nu = \Gamma_\nu^{nr} + \Gamma_\nu^{pole}$}, given in \Eq{polevertex},
and that underly \Eq{piffrho},  will differ in different approaches.  The total result 
should not.

In the naive limit of a point coupling model with structureless hadrons, the elements
of \Eq{piffrho} become \mbox{$F_\pi^{GIA}(Q^2) \rightarrow 1$}, 
\mbox{$F_{\rho \pi\pi}(Q^2) \rightarrow g_{\rho \pi\pi}$}, and 
\mbox{$ \Pi^{\rho \gamma}_T(Q^2) \rightarrow -Q^2/g_V$}.  The latter $Q^2$ factor
arises from the description of the mixing as proportional to 
\mbox{$ \int d^4x \rho_{\mu\nu}(x)\; F_{\mu\nu}(x)$}.
This produces the VMD empirical form
\beq
F_\pi(Q^2) = 1 - \frac{ g_{\rho\pi\pi} \; Q^2} { g_V \, (Q^2 + m_\rho^2) } ~.
\label{piffvmd}
\eeq
where the first (contact) term accounts for the pion charge and the charge radius
is required to come totally from the second ($\rho$ propagator) term.  This implies 
\mbox{$r_\pi^2 \sim 6  g_{\rho \pi\pi}/(m_\rho^2 g_V) $}, which under the universality
assumption of vector coupling \mbox{$g_V \approx g_{\rho \pi\pi}$}, produces 
\mbox{$r_\pi^2 \sim 0.4~{\rm fm}^2$} which compares very well with the experimental
value $0.44~{\rm fm}^2$.  In this sense it is often considered that the spacelike
pion charge form factor is the tail of the $\rho$ resonance.  A recent re-examination 
of the VMD phenomenology for this and related processes with two choices of point 
hadron lagrangian can be found in the work of O'Connell \etal, (1995).
We note that the phenomenological form (VMD1) found preferable there obeys the 
massless photon constraint \mbox{$ \Pi^{\rho \gamma}_T(0) = 0$} and produces just the 
point limit in \Eq{piffvmd}.  We also note that the more standard form of VMD, called 
VMD2 by O'Connell \etal, (1995), uses universal vector coupling and also the 
assumption that \mbox{$ \Pi^{\rho \gamma}_T(Q^2) \approx m_\rho^2/g_V$} for all 
$Q^2$; then the equivalence of \Eq{piffvmd} to \mbox{$m_\rho^2/(m_\rho^2 + Q^2)$} is 
often used to invoke the extreme form of VMD, namely that the photon couples 
{\it only} through vector mesons. 

From the perspective of the GCM or DSE approaches, the reality 
of a space-time distributed quark-gluon substructure of the pions, for example, 
demands the distributed form for the form factor shown in \Eq{piffrho} rather than
the point form in \Eq{piffvmd}.   These considerations certainly apply also
to QCD.  The question of the relative importance of the two terms in \Eq{piffrho} then
arises.  This  will receive different answers in different formulations for it depends 
on how the separation is defined off-mass-shell.  This corresponds to the observation 
that there is no unique way to separate a pole contribution from a background 
contribution except precisely at the pole.   The sum of the two terms is the 
meaningfull quantity.   With adoption of the Ball-Chiu Ansatz, this representation of 
the non-resonant term of the photon-quark vertex shares with the BSE vertex
those aspects of momentum dependence that are constrained by symmetries including the 
Ward-Takahashi identity.  Since the vector mesons are described by the same kernel, it
should not be surprizing that at least some extrapolation of  vector $\bar q q$ 
correlations into the spacelike region is accounted for through the Ball-Chiu vertex.
A similarity in strength 
between the Ball-Chiu vertex and the spacelike extension of low mass vector meson 
poles  has recently been found (Frank, 1995) through a numerical study of the ladder 
BSE for the photon vertex.   It is quite a robust result that about \mbox{$85-90$}\% 
of the experimental
value of $r_\pi$ is produced by the resulting non-resonant or generalized impulse 
term in the work of Roberts (1996); the value of 
$f_\pi r_\pi$ is quite insensitive to variation in the representation of the dressed 
quark propagator if other constaints such as $m_\pi$ and $\langle \bar qq \rangle$ are 
maintained.  

This leads to considerations of why the $\rho$ contribution should be small in such
an approach.  There is a constraint at $Q^2 =0$.
Electromagnetic gauge invariance requires the $\rho$-$\gamma$ mixing amplitude 
\mbox{$\Pi^{\rho \gamma}_T(Q^2)$}, and hence the pole term of $F_\pi(Q^2)$ in 
\Eq{piffrho}, to vanish at $Q^2=0$.   Asymptotic freedom ensures 
\mbox{$\Pi^{\rho \gamma}_T(Q^2)$} vanishes at large spacelike-$Q^2$; hence a 
small contribution to $F_\pi(Q^2)$ from the $\gamma \rho$ mixing term over a large 
range of space-like $Q^2$ can be consistent.  A more detailed investigation requires
a consistent definition of the momentum dependence of the separation in \Eq{polevertex}
as related to the $\rho$ BS amplitude and the defined non-resonant photon vertex. 
An initial investigation has utilized a mass-shell BS amplitude for
the $\rho$-meson which is well described as a $\bar{q}q$
bound state within the ladder BS approach; such a  dressed-quark core gives
$\sim $ 90\% of its mass (Hollenburg, Roberts and McKellar, 1992; Burden, Qian,
Roberts, Tandy and Thomson, 1996).   A calculation (Roberts, 1996) based on 
this confirms that \mbox{$\Pi^{\rho \gamma}_{\mu\nu}(Q^2)$} is a slowly 
varying function for spacelike $Q^2$.   The momentum suppression of the pole term of
\Eq{piffrho} due to $F_{\rho \pi\pi}(Q^2)$ is quite strong as discussed in 
Sec.~\ref{sect_rpp} and displayed in \Fig{rpp}.   We may use the representations
\mbox{$F_{\rho \pi\pi}(Q^2) =  g_{\rho \pi\pi} f_{\rho \pi\pi}(Q^2)$} and
\mbox{$\Pi^{\rho \gamma}_T(Q^2) = -Q^2 \hat{f}_{\rho \gamma}(Q^2)/g_V $}, where the two 
new functions $f$ and $\hat{f}$ depend on the details of the dynamics.   The $\rho$
pole contribution to the pion charge radius from \Eq{piffrho} can then be written in 
the form
\beq
(r_\pi^{pole})^2 = r_\pi^2 - (r_\pi^{GIA})^2 = 1.2 \;
 f_{\rho \pi\pi}(0)  \hat{f}_{\rho \gamma}(0) \; \frac{6}{m_\rho^2}  ~,
\label{rpipole}
\eeq
where we have used the empirical result \mbox{$ g_{\rho \pi\pi}/g_V \sim 1.2$}. 
Since from \Fig{rpp} \mbox{$f_{\rho \pi\pi}(0) \approx 0.5$}, one estimates that a value
\mbox{$ \hat{f}_{\rho \gamma}(0) \approx 0.4$} would account for all of the
difference between the non-resonant $(r_\pi^{GIA})^2$  in \Table{tab_pk_res} 
($0.31~{\rm fm}^2$) and the experimental value ($0.44~{\rm fm}^2$).   A reasonable 
value \mbox{$ \hat{f}_{\rho \gamma}(0) \approx 0.3$} would leave room for a contribution
as well from the pion loop process of the size estimated by Alkofer \etal (1995) and
described below.

\subsect{Pion Loop Shift of the $\rho$ Mass}
\label{sect_rhopiloop}

At the tree level of the GCM, the ladder description of the $\rho$ and $\omega$ is
degenerate, \mbox{$m_\rho = m_\omega$}.  The pion loop correction to the $\rho$
self-energy produces a mass shift and also a width due to the open $2\pi$ channel.
Within the GCM, this mass shift was originally studied by Hollenburg \etal (1992). 
We review the later work on the shift and width by Mitchell and Tandy (1997) and
highlight some issues that generally  arise in the GCM or DSE approaches.
 
Mitchell and Tandy (1997) treat the transverse $\rho$ and
$\omega$ modes and use fields that contain the transverse projector
\mbox{$T_{\mu \nu}(P)= \delta_{\mu \nu} -P_{\mu}P_{\nu}/P^2$}.   That is, 
$\vec{\rho}_\mu(P) \equiv T_{\mu\nu}(P)\vec{\rho}_\nu(P)$ and 
$\omega_\mu(P) \equiv T_{\mu\nu}(P)\omega_\nu(P)$.  
In a matrix notation where $V_{\mu}$ denotes  $(\vec{\rho}_{\mu}, 
\omega_{\mu})$,  the tree-level action from \Eq{mesonaction}, 
up to second order in the fields, can be written as 
\beq
\hat{S}_2[\rho,\omega] = \case{1}{2} \int \frac{d^4P}{(2\pi)^4}
V^T_{\mu}(-P) \Bigl[ \hat{\Delta}^{-1}_{\mu \nu}(P) 
+ \hat{\Pi}_{\mu \nu}^q(P) \Bigr] V_{\nu}(P) . 
\label{sromix}
\eeq
Here, in meson channel space, $\hat{\Delta}^{-1}$ is diagonal 
and the only non-zero elements of $\hat{\Pi}^q$ are the off-diagonal ones 
that provide $\rho^0 -\omega$ coupling via the $u-d$ current mass difference.
The superscript $q$ identifies the quark-loop contribution to distinguish from 
subsequent pion loop quantities.   The composite $\bar{q}q$ nature of the mesons is 
reflected in the fact that the diagonal inverse propagators contain a dynamical
mass function.  The general form 
\beq
\hat{\Delta}^{-1}(P^2)\, = \,(P^2+m_V^2){\cal Z}_1^{-1}(P^2)
\, = \, (P^2+ m_V^2(P^2)) 
\eeq
as calculated from the formalism in \Sec{sect_eigen} identifies the degenerate mass 
$m_V$ of the tree-level $\rho$ and $\omega$ states.   The momentum dependence of 
${\cal Z}_1(P^2)$ is due to the $\bar q q$ substructure.  Without $\rho^0-\omega$ 
mixing,  a physical normalization (unit residue at the mass-shell pole) would be 
produced by absorbing at least the on-mass-shell value of
${\cal Z}_1^{-\frac{1}{2}}$ into the fields.   The function 
${\cal Z}_1^{-\frac{1}{2}}(P^2)$ may be absorbed into the fields producing
unmixed propagators have the standard point meson form.   As in \Sec{sect_eigen} the 
associated BS amplitudes become correspondingly rescaled as
\mbox{$\Gamma_V(q;P)= \hat{\Gamma}_V(q;P)~{\cal Z}_1^{\frac{1}{2}}(P^2)$}, 
so that at the mass-shell the standard normalization condition (Itzykson and Zuber, 
1980) is satisfied. 

The  relevant low-order terms in the effective action for $\pi,\rho,\omega$ 
from the bosonization result in \Eq{mesaction} are  
\beq
\hat{S}\left[\pi,\rho,\omega\right] = \hat{S}_2\left[\rho,\omega\right]
+\hat{S}_2\left[\pi\right] + \hat{S}[\rho\pi\pi] +\hat{S}[\omega\pi\pi] 
+ \cdots  .
\label{sql}
\eeq
Here the quadratic terms $\hat{S}_2\left[\rho,\omega\right]$  and
$\hat{S}_2[\pi]$ represent respectively the $\rho-\omega$ sector containing the 
quark-loop mixing mechanism and the free pion sector. 
To integrate out the pion fields, it is convenient to first combine the
last three terms of Eq.~(\ref{sql}) so that it becomes
\beq
\hat{S}\left[\pi,\rho,\omega\right]=\hat{S}\left[\rho,\omega\right]
+\case{1}{2}\int 
\frac{d^4P,P'}{(2\pi)^8} \, \pi_i(P')D^{-1}_{ij}(P',P)\pi_j(P) .
\eeq
The three terms in
\beq
D_{ij}^{-1}(P',P)=\Delta^{-1}_{ij}(P',P)+V_{ij}(P',P)+W_{ij}(P',P)
\eeq
correspond to the tree-level pion inverse propagator given by
\beq
\Delta^{-1}_{ij}(P',P)=(2\pi)^4\delta_{ij}\delta^4(P'+P)\Delta^{-1}_\pi(P^2) ,
\eeq
the $\rho\pi\pi$ interaction term given by
\beq
V_{ij}(P',P)=2i\epsilon_{ijk} \, \rho_\mu^k(-P'-P)  \;
\Lambda_\mu^\rho(-\frac{P'+P}{2};-P'-P) ,
\eeq
and the $\omega\pi\pi$ interaction term given by
\beq
W_{ij}(P',P)=2i\epsilon_{ij3} \, \omega_\mu(-P'-P)  \;
\Lambda_\mu^\omega(-\frac{P'+P}{2};-P'-P).
\eeq
In this work the pion inverse propagator of the GCM is reduced to the standard point
form by absorbing the calculated ${\cal Z}_\pi^{-\frac{1}{2}}(P^2)$ into the pion field.
Thus the pion momentum dependence of the $\rho\pi\pi$  and $\omega\pi\pi$ vertex
functions contain the compensating aspects of the $\bar q q$ substructure of the pion.
The $\rho\pi\pi$ vertex function is described in \Sec{sect_rpp}. The corresponding
$\omega\pi\pi$ vertex function, which is generated by a $u-d$ current quark mass
difference, is calculated and included by Mitchell and Tandy 
(1997)  to extract the pion loop contribution to $\rho-\omega$ mixing.
Integration over the pion fields allows a new effective action 
$\hat{{\cal S}}[\rho,\omega]$ to be identified from
\beq
Z=N \int D\rho D\omega  D\pi \, {\rm exp}\left( -\hat{S}[\pi,\rho,\omega] \right)
= N' \int D\rho D\omega \, {\rm exp}\left( -\hat{{\cal S}}[\rho,\omega] \right) 
\eeq
by using the functional integral result
\beq
\int D\pi \, {\rm exp}\left( -\case{1}{2}\int \frac{d^4P,P'}{(2\pi)^8} \,
\pi_i(P')D^{-1}_{ij}(P',P)\pi_j(P) \right)
={\rm exp}\left( -\case{1}{2} {\rm TrLn} \; D^{-1}\right)  .
\eeq
The field-independent term \mbox{${\rm exp}(-\frac{1}{2} 
{\rm TrLn} \; \Delta^{-1} )$} from the result may be absorbed into the normalization 
constant to arrive at
\beq
\hat{{\cal S}}\left[\rho,\omega\right]=\hat{S}_2\left[\rho,\omega\right]
+\case{1}{2} {\rm TrLn} (1+\Delta(V+W)).
\label{spiloop}
\eeq
The second term here gives the vector meson coupling to all orders
produced by a single pion loop.  

\begin{figure}[ht]

\centering{\
\psfig{figure=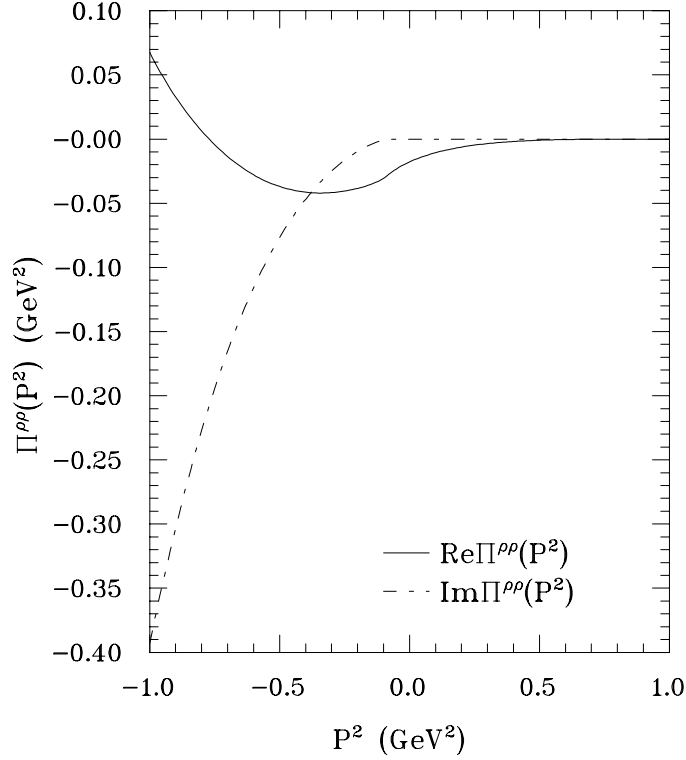,height=10.0cm} }

\parbox{130mm}{\caption{The real and imaginary parts of the pion loop contribution 
to the $\rho$ self-energy.  Taken from Mitchell and Tandy (1997). \label{rho_pl} } }
\end{figure}

The quadratic part of the second term of Eq.~(\ref{spiloop}) adds a contribution to 
the tree-level $\rho^0-\omega$ matrix  inverse propagator.  The net transverse 
result is
\beq
{\cal D}^{-1}(P^2) = \left( \begin{array}{cc} 
P^2+m_V^2 +\Pi^{\rho\rho}(P^2)    & \Pi^{\rho\omega}(P^2) \\
\Pi^{\rho\omega}(P^2) & P^2+m_V^2 +\Pi^{\omega\omega}(P^2) \end{array} \right) ,
\label{totinvpr}
\eeq
where the net mixing amplitude is \mbox{$\Pi^{\rho\omega}(P^2)= 
\Pi^q(P^2)+\Pi^\pi(P^2)$}.
The additional terms produced by the pion loop are the diagonal self-energies
$\Pi^{\rho\rho}(P^2)$ and  $\Pi^{\omega\omega}(P^2)$, and the mixing term 
$\Pi^\pi (P^2)$.   From  Eq.~(\ref{spiloop}),
the self-energy contribution $\Pi^{\rho\rho}(P^2)$ is given by the transverse  
component of
\beq
\Pi^{\rho \rho}_{\mu \nu}(P)= -4 \int \frac{d^4 q}{(2 \pi)^4} \Delta_\pi(q_+)
\Delta_\pi(q_-) \Lambda^{\rho}_{\mu}(q;-P) \Lambda^{\rho}_{\nu}(q;P)  ,
\label{rhose}
\eeq
where \mbox{$q_\pm = q \pm P/2$}, and the summation over pion isospin labels
has been carried out.   This contribution is the same for each isospin 
component of the $\rho$.   The $\omega$ self-energy contribution is
obtained from Eq.~(\ref{rhose}) by the replacement  of $\Lambda^\rho$ by
$\Lambda^\omega$.  It is quadratic in the small symmetry breaking mechanism 
and is ignored.

In the absence of mixing,  the real part of $~\Pi^{\rho\rho }(P^2)$ generates  
a mass shift for the isospin eigenstate $\rho_I$.   This mass shift now represents 
the $\rho-\omega$ mass splitting.   
For timelike momenta such that $P^2 \leq -4 m_\pi^2$, both
$~\Pi^{\rho \rho }(P^2)$ and $~\Pi^{\pi }(P^2)$ have imaginary parts associated 
with the decay $\rho \rightarrow 2 \pi$.   This mechanism should produce most of 
the $\rho$  width.
The mass modifications due to mixing are second  order in a very small quantity and are 
ignored.  The shifted $m_\rho$ is therefore identified from the position of the 
zero in the real part of the relevant inverse propagator in \Eq{totinvpr}.  That is, 
\beq
m_\rho^2 =m_\omega^2+ {\rm Re} \, \Pi^{\rho\rho}(-m_\rho^2) ,
\label{rhomass}
\eeq
where now $m_\omega = m_V$.   With the definition 
\mbox{$\hat{\Gamma}(P^2) = -{\rm Im}\, \Pi^{\rho\rho}(P^2)/m_\rho$}, 
the inverse propagator in the $\rho$ channel can be written as
\beq
{\cal D}^{-1}_\rho(P^2) = (P^2 + m_\rho^2 -i m_\rho \Gamma(P^2)) \, 
{\cal Z}_2^{-1}(P^2)  ,
\label{rinvpr}   
\eeq
where \mbox{$\Gamma(P^2)=\hat{\Gamma}(P^2)~{\cal Z}_2(P^2)$}.  The function 
${\cal Z}_2(P^2)$ arises from the momentum dependence of the real part of 
the pion loop self-energy.  The on-mass-shell value, which is the 
$\rho$ field renormalization constant due to the $2\pi$ content, is given by
\beq
{\cal Z}_2^{-1}(-m_\rho^2) = 1 + {\rm Re}\, {\Pi^{\rho\rho}}^{\prime} 
(-m_\rho^2)  ,
\label{z2pi}
\eeq
with the prime superscript denoting differentiation with respect to the 
argument.  One absorbs ${\cal Z}_2^{-\frac{1}{2}}(P^2)$ 
into the field $\rho_\mu(P)$ so that the resulting propagator has the 
conventional Breit-Wigner form.  The  physical width is given by
\beq
\Gamma_\rho = - m_\rho^{-1} \, {\cal Z}_2(-m_\rho^2) \, 
{\rm Im}\, \Pi^{\rho\rho}(-m_\rho^2) .
\label{rwid}
\eeq
\begin{figure}[ht]

\centering{\
\psfig{figure=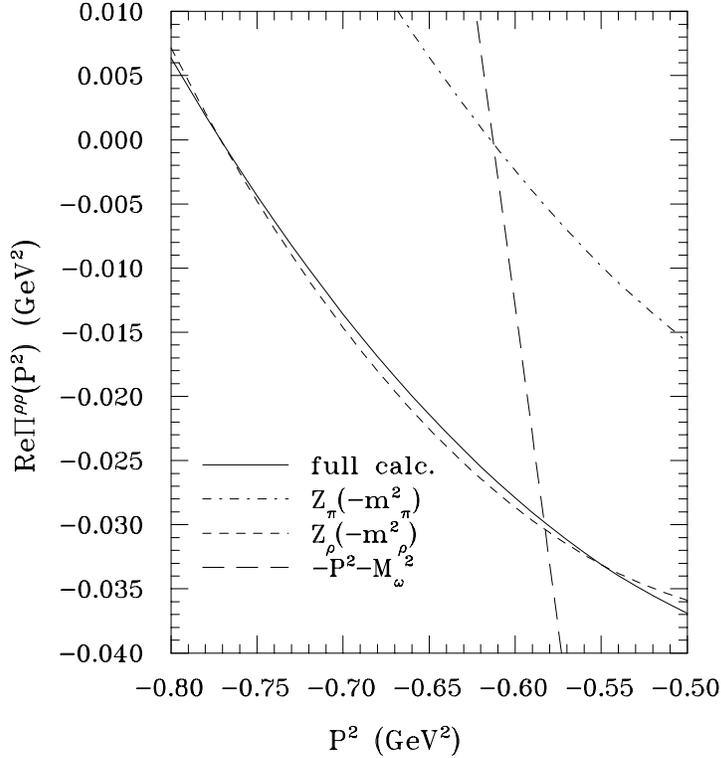,height=10.0cm} }

\parbox{130mm}{\caption{Graphical determination of the $\rho-\omega$ mass difference
due to the pion loop.  The influence of the $\protect\bar{q}q$ composite nature of 
the $\pi$ 
and $\rho$ propagators is also shown. Taken from Mitchell and Tandy (1997). 
\label{mcorr_comp} } }
\end{figure}
Numerical results for the $\rho$ mass shift and width are obtained from 
Eq.~(\ref{rhose}) in the specific form (Mitchell and Tandy, 1997)
\beq
\Pi^{\rho\rho}(P^2)=-\frac{4}{3}f_{\rho\pi\pi}^2(P^2) \int 
\frac{d^4q}{(2\pi)^4}
\frac{( q^2- (q \cdot P)^2/P^2 ) e^{-2 q^2/\lambda_{\rho}^2(P^2)}}
{\left[(q+\frac{P}{2})^2+m_\pi^2-i\epsilon\right]
\left[(q-\frac{P}{2})^2+m_\pi^2-i\epsilon\right]} ,
\eeq
where a parameterized  form of the calculated $\rho\pi\pi$ vertex function 
is used for convenience.
The numerical results for ${\rm Re}\, \Pi^{\rho\rho}(P^2)$ and ${\rm Im}\, 
\Pi^{\rho\rho}(P^2)$ are shown in Fig.~\ref{rho_pl}.  
The self-consistent solution for $m_\rho$ produced by Eq.~(\ref{rhomass}) 
is displayed in Fig.~\ref{mcorr_comp} in the following way.
The quantity $-(P^2+m_\omega^2)$ is plotted as a long dash line and its 
intercept with the solid line representing ${\rm Re}\, \Pi^{\rho\rho}(P^2)$ 
identifies the mass shell point.  With \mbox{$m_\omega=m_V=782~{\rm MeV}$}  one
finds \mbox{$m_\rho=761~{\rm MeV}$} giving
\mbox{$m_\rho - m_\omega = -21~{\rm MeV}$}.   The experimental value is 
$-12.0\pm 0.8$ MeV.  The calculated width is \mbox{$\Gamma_\rho = 156~{\rm MeV}$} 
while the experimental value is $151~{\rm MeV}$.   
Fig.~\ref{mcorr_comp} also illustrates the contribution to the mass splitting
result made by the momentum dependence of the meson self-energies  
generated by the $\bar{q}q$ substructure.  This is most marked for the pion
and enters the pion loop integral in Eq.~(\ref{rhose}) through the factors 
${\cal Z}_\pi (q_\pm^2)$
which are contributed by each pion propagator.  (In the calculational procedure
these factors have been moved into the vertex functions for convenience.)  If 
each function ${\cal Z}_\pi$ is held constant at its mass-shell value 
(as would be the case for a structureless pion), the full calculation 
in Fig.~\ref{mcorr_comp} (solid line) becomes the dot-dashed line.  The 
intercept then indicates that the mass shift would be essentially zero.  In
contrast to this, the short dashed line indicates that the influence of
the corresponding quantity ${\cal Z}_1(P^2)$, from the $\bar q q$ substructure of the 
$\rho$, is quite negligible here.  This is because the range of variation in 
$P^2$ is very small.
\begin{figure}[ht]
\unitlength1.pt
\begin{picture}(612,164)(-10,240)
\includegraphics{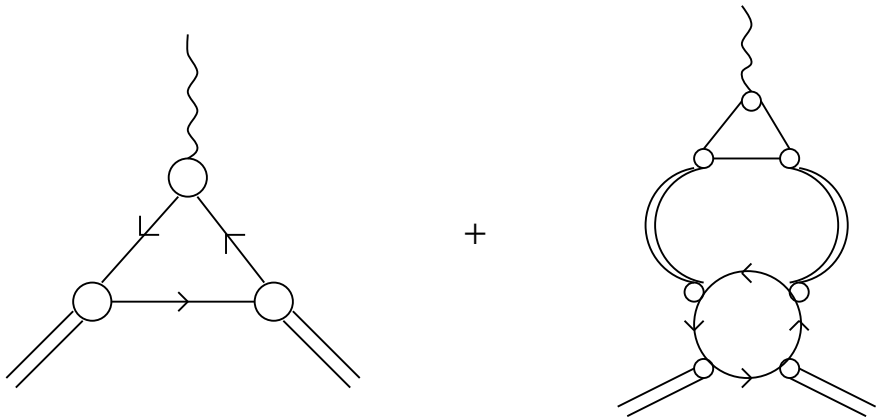}
\end{picture}

\centering{\
\parbox{130mm}{ \caption{The pion charge form factor.  The first term is the 
generalized impulse approximation, and the second term is the one pion loop 
correction.  Double lines represent the ladder $\protect\bar q q$ pions of the 
GCM/DSE 
approach. Quark lines are dressed propagators and circles are dressed vertices.  
\label{piloopff}  }  }  }
\end{figure}

\subsect{Pion Loop Part of the Pion Charge Radius}
\label{sect_rpiloop}

A somewhat more complicated task is presented by the pion loop contribution to
the pion charge radius.  This is illustrated in \Fig{piloopff}.   The analysis  has 
been carried out within the GCM framework by Alkofer, Bender and Roberts (1995)
in the following way.   Starting from the tree level GCM meson action of 
\Eq{emmesaction}, which includes coupling to a background EM field, the first-order
contributions in the pion loop expansion may be developed for various vertex
functions.   For the $\gamma \pi\pi$ vertex function, Alkofer \etal, (1995) obtain
\beq
\Lambda_\mu(P;Q) =  \Lambda_\mu^{GIA}(P;Q) + \Lambda_\mu^{loop}(P;Q)~,
\label{pivertot}
\eeq
where the first term is the generalized impulse approximation result (i.e., at meson
tree level) given by \Eq{piemver} and discussed in \Sec{sect_emff}.   The pion loop
contribution is given by
\beq
\epsilon^{3ij} \; \Lambda_\mu^{loop}(P;Q) = - \case{1}{2} \int \dK
T^{ijkl}(Q,P-K,K+P) \; \Delta_\pi(K_+) \Delta_\pi(K_-) \; \epsilon^{3kl} \;
\Lambda_\mu^{GIA}(K;Q) ~,
\label{piverlp}
\eeq
where \mbox{$K_\pm = K \pm Q/2$} and $T^{ijkl}((p_1,p_2;p_3,p_4)$ is the $\pi-\pi$ 
scattering kernel which may be written in terms of a single invariant amplitude
$A(p_1,p_2;p_3,p_4)$.   A threshold expansion for this amplitude in low powers of
momentum is not adequate here since the integral in \Eq{piverlp} samples a large
domain of spacelike momenta.  The contribution to the required $A(p_1,p_2;p_3,p_4)$
that yields both the leading order chiral limit result at low $s$, the term 
(Weinberg, 1966) given in \Eq{aw}, and the leading contribution within the GCM
at tree level for large $s$, is obtained from the action in \Eq{Sreal} in the form
\beq 
\hat{ {\cal S}}_R[U] = 
\int d^4x d^4y \;  \case{f_\pi^2}{2} {\rm tr} \left[ \partial_\mu U(x) 
\partial_\mu U(y)^\dagger \right] \hat{f}(x-y) ~,
\label{4piaction}
\eeq
which avoids the derivative expansion by  retaining the nonlocality.   Here 
$\hat{f}(x-y)$ is the fourier transform of the momentum space quark loop integral 
with two chiral insertions.   It is of the same form that contributes
to the pion inverse propagator as discussed in \Sec{sect_GB}.  

For the integral in \Eq{piverlp}, the onshell approximation
\mbox{$\Lambda_\mu^{GIA}(K;Q) \approx 2K_\mu F_\pi^{GIA}(Q^2)$} is made. The result
is expressed in the form
\beq
 F_\pi(Q^2) = F_\pi^{GIA}(Q^2) \left[ 1 + L(Q^2;m_\pi)\right] ~,
\label{totff}
\eeq
where $L$ is a result from the pion loop integral.   Since the condition
\mbox{$ F_\pi(0) = 1$} is equivalent to the proper normalization condition for the 
pion BS amplitude, the above form emphasises that the pion loop mechanism generates
additional internal structure for the pion.  A renormalization
of the modified BS amplitude is necessary.   The form \mbox{$i\gamma_5 B(q^2)/f_\pi$}
used at each $\pi \bar q q$ vertex is now effectively supplemented by an additional 
pion cloud term.  The normalization constant $f_\pi$,  
previously determined at the quark loop level, must be renormalized to adjust for
the pion cloud now being included in the pion. 
From these considerations, Alkofer \etal, (1995) report results in the form
\beq
r_\pi^2 = (r_\pi^{GIA})^2 + (r_\pi^{loop})^2
\label{rpiloop}
\eeq
where the first-order pion loop contribution $r_\pi^{loop}(m_\pi)$ is at most 
\mbox{$10-15$}\% of 
the generalized impulse approximation  (meson tree-level) result.  The loop term 
has the expected divergence as a chiral logarithm in the
\mbox{$m_\pi \rightarrow 0$} limit, but it has the stated small effect at the physical
point and is seen to not dominate until \mbox{$m_\pi \sim 10~{\rm MeV}$}. 
The pion loop correction to $f_\pi$ is found to be at most  $2$\% of the impulse or 
tree-level result.  

In ChPT with \mbox{$N_f = 3$}, the result given 
by Gasser and Leutwyler (1985)  up to one pseudoscalar loop is
\beq
r_\pi^2 = \frac{12\, L^r_9}{f_\pi^2} - \frac{1}{32\pi^2 \, f_\pi^2}
\left[ 2 \; {\rm ln}\left( \frac{m_\pi^2}{\mu^2}\right)
+ {\rm ln}\left( \frac{m_K^2}{\mu^2}\right) + 3 \right]~,
\label{rpichpt}
\eeq
where $L^r_9$ is the $9^{th}$ low energy constant at ${\cal O}(p^4)$ after the 
divergent part from  the loop  has been absorbed and $\mu^2$ is the loop regularization
scale.   Any reasonable choice of $\mu^2$ such as $m_\rho^2$,  leads to the ``chiral
logarithms'' in \Eq{rpichpt} contributing  typically \mbox{$15$}\% of 
the experimental value $0.44~{\rm fm}^2$.   The dominant $L^r_9$ term is the 
counterpart in ChPT of $(r_\pi^{GIA})^2$ obtained from the 
distributed quark-gluon substructure of the $\gamma\pi\pi$ vertex in the GCM/DSE
approaches.  However the correspondence is  only approximate; the distinction between
tree-level and loops in the two approaches is not the same due to the very different
character of the pseudoscalar fields.  This has been discussed in the early part of
\Sec{sect_chiral}.

\sect{Concluding Remarks}
\label{sect_conclude}

We have summarized the Global Color Model as it applies to a variety of meson
physics observables.  A constant theme that emerges is that the tree level
interactions among $\bar q q$ Goldstone  mesons and photons are well-described
in terms of only the dressed quark propagators.   This efficiency is due to 
the connection between the pseudoscalar Bethe-Salpeter amplitudes
and the quark propagator brought about by chiral symmetry and also by the 
connection between the photon-quark vertex and the propagator brought about by
electromagnetic gauge invariance.   Only tentative steps have been taken so far
in the task of relating the phenomenological aspects of the confining quark 
propagators used at this stage to an underlying gluon $2$-point function.  This
step is required if the GCM is to fulfil the initially designed  role of a
field theory model based upon an effective gluon $2$-point function.
This need is emphasized by investigations of mesons other than the Goldstone
bosons where more approximations must necessarily be made.  However, even with
an increased degree of phenomenological input, valuable information is
being obtained.

We have seen that various coupling constants obtained at meson tree-level  
are within $10$\% of experiment.  At that stage the meson modes from the
GCM have ladder Bethe-Salpeter content and the meson coupling is via a dressed
quark loop.  An interesting question is whether this will be a persistent
theme that leaves meson loop corrections relatively unimportant in this 
approach.  This is possible because the subsequent meson loop dressing would 
have to overcome distributed coupling caused by the finite size effects of the
tree-level mesons in this approach.  There are only several isolated pieces
of evidence.  Pion-loop corrections to the pion EM form factor were seen to 
contribute at the level of $<15$\% to the charge radius (Alkofer \etal, 1995).
Similarly, the pion-loop contribution to the downward shift in the
$\rho$ mass relative to $m_\omega$ in the GCM has been 
found to be $< 5$\% (Hollenburg \etal, 1992) and in a more recent study $2$\%
(Mitchell and Tandy, 1997); this latter result is in good agreement with 
experiment.  It would certainly be a convenience
if the low-mass mesons and their soft interactions were amenable to a 
representation in which their dressed quark content captures the dominant 
quantum loop effects.  However this remains to be determined and in this regard
future work within the GCM is called for to obtain the pion loop contributions to 
the $\pi-\pi$ scattering lengths especially $a_0^0$.   In ChPT at ${\cal O}(p^4)$,
the only distributed substructure enters through loops of the (auxiliary) point
pion fields, while at tree level in the GCM there is significant substructure
through all orders in momenta.  A systematic comparison through a complete set
of threshold observables at ${\cal O}(p^4)$ would be quite useful.   One of the
features of the GCM is that constraints from asymptotic freedom at large momenta
are automatically included and the complete momentum dependence of form factors
and scattering processes can be examined.   From this perspective, one may speculate 
that with a model field theory with a few parameters such as the GCM, high 
accuracy in threshold behavior might be incompatible with quality descriptions of 
many processes over a large momentum domain.  More work is needed to explore a 
large variety of observables.

There has been significant effort in the past devoted to investigations of
baryons (particularly the nucleon) within a covariant Faddeev
formalism truncated to a quark-diquark approximation derived within the GCM.  
Probably the first covariant Faddeev 
calculation of the nucleon (Burden, Cahill and Praschifka, 1989) was carried
out within the GCM framework.  Other works (Cahill, Roberts and 
Praschifka, 1987; 1989a; Praschifka \etal, 1989) have made important 
contributions to understanding  the properties and role of diquarks as 
constituents of baryons.  Progress in this endeavour is complicated by the 
rich singularity structure of the covariant Faddeev equations.  A more 
accessible, but necessarily more approximate approach to the nucleon has been
via chiral meson-quark mean field models of the non-topological soliton 
variety.  The GCM has provided a definitive underpinning to such methods by
producing the soliton equations of motion for these degrees of freedom as
dressed composite modes based at the quark level (Cahill and Roberts, 1985;
Frank, Tandy and Fai, 1991).   As a confining quark condensate model of the
nucleon it has provided  very efficient and sensible numerical results 
(Frank and Tandy, 1992).   Space has not permitted an analysis of such baryon
work to be included here.

\section*{\sc Acknowledgments}

This work was supported in part by the National Science Foundation under
Grant Nos. PHY91-13117 and PHY94-14291.  We wish to thank C. D. Roberts,
R. T. Cahill and M. R. Frank  for many very helpful discussions.  This review
has drawn heavily upon previous work done in conjunction with the above as 
well as K. Mitchell, S. Banerjee, C. J. Burden and Lu Qian.


\section*{\sc References}\vspace*{-\parskip}
\addcontentsline{toc}{section}{\protect\numberline{ }{\sc References}}
\label{references}
\begin{description}
\itemsep=0pt
\parskip=0pt
\item Afanasev, A. (1994). Form factors of the transitions $\gamma^\ast \pi^0
\rightarrow \gamma$ and $\gamma^\ast \pi^0 \rightarrow \gamma$,  {\it Proc. of
the Workshop on CEBAF at Higher Energies, CEBAF, Newport News}, Eds. N. Isgur
and P. Stoler,  P. 185; Afanasev, A., Gomez, J. and Nanda, S. (1994). CEBAF
Letter of Intent \# LOI-94/005, unpublished.
\item Adler, S. (1969). Axial-vector in spinor electrodynamics, 
{\it Phys. Rev.} {\bf 177}, 2426.
\item Alkofer, R., Bender, A. and Roberts, C.D. (1995). Pion loop
contribution to the electromagnetic pion charge radius, {\it Int. J. Mod. Phys.
A} {\bf 10}, 3319.
\item Alkofer, R. and Roberts, C.D. (1996). Calculation of the anomalous
$\gamma \pi^\ast \rightarrow \pi \pi$ form factor. {\it Phys. Lett. B}
{\bf 369}, 101.
\item Amendolia, S.R., et al. (1986a). A measurement of the spacelike pion
electromagnetic form-factor, {\it Nucl. Phys. B} {\bf 277}, 168.
\item Amendolia, S.R., et al. (1986b). A measurement of the kaon charge radius,
{\it Phys. Lett. B} {\bf 178}, 435.
\item  Ametller, L., Bijnens, L.,  Bramon, A. and  Cornet, F. (1992). 
Transition form-factors in $\pi^0$, $\eta$ and $\eta^\prime$ couplings to
$\gamma \gamma$, {\it Phys. Rev. D} {\bf 45}, 986.
\item Anikin, I. V., Ivanov, M. A., Kulimanova N. B. and  Lyubovitskij, V. E.
(1994). Sensitivity to form-factors in the extended Nambu--Jona-Lasinio
model with separable interactions, {\it Sov. J. Nucl. Phys.} {\bf 57}, 1082.
\item Aoki, Ken-Ichi, Kugo, Taichiro and Mitchard, Mark K. (1991).
      Meson properties from the ladder Bethe-Salpeter equation,
      {\it Phys. Lett. B} {\bf 266}, 467.
\item Ball, J.S. and Chiu, T.-W. (1980).  Analytic properties of the vertex
function in gauge theories. I and II, {\it Phys. Rev. D},
{\bf 22}, 2542.
\item Barducci, A., Casalbuoni, R., De Curtis, S., Dominici, D and Gatto, R.
(1988).  Dynamical chiral-symmetry breaking and determination of the quark
masses, {\it Phys. Rev. D} {\bf 38}, 238.
\item Bar-Gadda, U. (1980). Infrared behavior of the effective coupling in
quantum chromodynamics, {\it Nucl. Phys. B} {\bf 163}, 312.
\item Bebek, C. J., \etal  (1976).  Determination of the pion form factor up to
\mbox{$Q^2 = 4~{\rm GeV}^2$} from single charged-pion electroproduction, 
{\it Phys. Rev. D} {\bf 13}, 25.   
\item Bebek, C. J., \etal  (1978).  Electroproduction of single pions at low $\epsilon$
and a measurement of the pion form factor up to \mbox{$Q^2= 10~{\rm GeV}^2$}, 
{\it Phys. Rev. D} {\bf 17}, 1693. 
\item Behrend, H.J., {\it et al.} (CELLO Collab.), (1991). A measurement of the
$\pi^0$, $\eta$ and $\eta^\prime$ electromagnetic form factors,
{\it Z. Phys. C } {\bf 49}, 401.
\item Bell, J. and Jackiw, R. (1969). A PCAC puzzle: $\pi^0$ to $\gamma \;
\gamma$ in the sigma-model, {\it Nuovo Cimento} {\bf A60}, 47.
\item Bender, A., Roberts, C.D. and v. Smekal, L. (1996). Goldstone theorem
and diquark confinement beyond rainbow-ladder approximation, {\it Phys. Lett.
B} {\bf 380}, 7.
\item Bernard, C., Parrinello, C. and Soni, A. (1994). A lattice study of
the gluon propagator in momentum space, {\it Phys. Rev. D} {\bf 49}, 1585.
\item  Bernard, V., Brockmann, R. and Weise, W. (1985). The Goldstone pion and
the quark anti-quark pion. II. Pion size and decay, {\it Nucl.Phys. A} {\bf 440}, 
605.  
\item  Bernard, V. and  Mei\ss ner, U.-G. (1988). Electromagnetic structure of
the pion and the kaon,  {\it Phys. Rev. Lett.} {\bf 61}, 2296.
\item Bernard, V., Kaiser, N. and Mei\ss ner, U.-G. (1995). Chiral dynamics in 
nucleons and nuclei, {\it Int. J. Mod. Phys. E} {\bf 4}, 193.
\item  Blin, A.H., Hiller, B. and  Schaden, M. (1988). Electromagnetic
form-factors in the Nambu--Jona-Lasinio model, {\it Z. Phys. A.} {\bf 331}, 75.
\item Brodsky, S.J. and Lepage, G.P. (1981). Large-angle two-photon exclusive
channels in quantum chromodynamics, {\it Phys. Rev. D} {\bf 24}, 1808.
\item Brown, C. N., \etal, (1973).  Coincidence electroproduction of charged pions
and the pion form factor, {\it Phys. Rev. D} {\bf 8}, 92. 
\item Brown, N. and Pennington M.R. (1988a), Preludes to confinement: Infrared
properties of the gluon propagator in the Landau gauge, {\it Phys. Lett. B}
{\bf 202}, 257.
\item Brown, N. and Pennington, M. (1988b). Studies of confinement: How
quarks and gluons propagate, {\it Phys. Rev. D} {\bf 38}, 2266.
\item Brown, N. and Pennington, M.R. (1989). Studies of confinement: How the
gluon propagates, {\it Phys. Rev. D} {\bf 39}, 2723.
\item Buck, A., Alkofer, R. and Reinhardt, H. (1992). Baryons as bound states
of diquarks and quarks in the Nambu--Jona-Lasinio model, {\it Phys. Lett. B}
{\bf 286}, 29.
\item Buck, W.W., Williams, R.A. and Ito, H. (1995). Elastic charge form
factors for $K$ mesons, {\it Phys. Lett. B} {\bf 351}, 24.
\item Burden, C.J., Cahill, R.T. and Praschifka, J. (1989). Baryon
structure and QCD: Nucleon calculations, {\it Aust. J. Phys.} {\bf 42}, 147.
\item Burden, C.J. and Roberts, C.D. (1991). Light-cone regular vertex in
three-dimensional quenched QED, {\it Phys. Rev. D} {\bf 44}, 540.
\item Burden, C.J., Praschifka, J. and Roberts, C.D. (1992a). Photon
polarization tensor and gauge dependence in three-dimensional quantum
electrodynamics, {\it Phys. Rev. D} {\bf 46}, 2695.
\item Burden, C.J. and Roberts, C.D. (1993).  Gauge covariance and the
fermion-photon vertex in three-dimensional and four-dimensional, massless
quantum electrodynamics, {\it Phys. Rev. D} {\bf 47}, 5581.
\item Burden, C.J., Roberts, C.D. and Williams, A.G. (1992). Singularity
structure of a model quark propagator, {\it Phys. Lett. B} {\bf 285}, 347.
\item Burden, C.J., Roberts, C.D. and  Thomson, M. J. (1996).
Electromagnetic Form Factors of Charged and Neutral Kaons, {\it Phys. Lett. B}
{\bf 371}, 163.
\item Burden, C.J., Qian, Lu, Roberts, C.D., Tandy, P.C. and Thomson, M.J. (1997).
Ground state spectrum of light-quark mesons, {\it Phys. Rev. C} {\bf 55}, 2649.
\item Burkardt, M., Frank, M.R. and Mitchell, K.L. (1997). Calculation
of hadron form factors from Euclidean Dyson-Schwinger equations, {\it Phys. Rev. 
Lett.} {\bf 78}, 3059.
\item Cahill, R.T. and Roberts, C.D. (1985).  Soliton bag models of hadrons
from QCD, {\it Phys. Rev. D} {\bf 32}, 2419.
\item Cahill, R.T., Roberts, C.D. and Praschifka, J. (1987). Calculation of
diquark masses in QCD, {\it  Phys. Rev. D} {\bf 36}, 2804.
\item Cahill, R.T., Roberts, C.D. and Praschifka, J. (1989a). Baryon
structure and QCD, {\it  Aust. J. Phys.} {\bf 42}, 129.
\item Cahill, R.T., Praschifka, J. and Burden, C.J.  (1989b). Diquarks and the
bosonization of QCD, {\it  Aust. J. Phys.} {\bf 42}, 161.
\item Cahill, R.T. (1989). Hadronization of QCD, {\it Aust. J. Phys.} {\bf 42},
 171.
\item Cahill, R.T. (1992). Hadronic laws from QCD, {\it Nucl. Phys. A} {\bf
543}, 63.
\item Cahill, R.T. (1993). Private communication.
\item Cahill, R.T. and Gunner, S. (1995a). Quark and gluon propagators from 
meson data, {\it Phys. Lett. B} {\bf 359}, 281.
\item Cahill, R.T. and Gunner, S. (1995b). A new mass formula for NG bosons
in QCD, {\it Mod. Phys. Lett. B} {\bf 10}, 3051.
\item Cahill, R.T. and Gunner, S. (1997). The global color model of QCD 
and its relationship to the NJL model, chiral perturbation theory and other
models, {\it Aust. J. Phys.} {\bf 50}, 103.
\item Celenza, L.S., Shakin, C.M., Sun, W.-D., Szweda, J. and Zhu, X. (1995).
Quark model calculations of current correlators in the nonperturbative domain,
{\it Ann. Phys. (N.Y.)} {\bf 241}, 1.
\item Celenza, L.S., Shakin, C.M. and  Sun, W.-D. (1996). Calculation of the
nucleon-nucleon interaction due to vector-meson exchange, {\it Phys. Rev. C}
{\bf 54}, 487.
\item Curtis, D.C. and Pennington, M.R. (1992). Generating fermion mass in 
quenched QED in four-dimensions, {\it Phys. Rev. D} {\bf 46}, 2663.
\item Dally, E.B., \etal~(1980). Direct measurement of the negative kaon form
factor, {\it Phys. Rev. Lett.} {\bf 45}, 232.
\item Delbourgo, R. and Scadron M.D. (1979).  Proof of the Nambu-Goldstone
realisation for vector-gluon--quark theories, {\it J. Phys. G} {\bf 5}, 1621.
\item Dong, Z., Munczek, H.J. and Roberts, C.D. (1994). Gauge covariant
fermion propagator in quenched, chirally-symmetric quantum electrodynamics,
{\it Phys. Lett. B} {\bf 333}, 536.  
\item Donoghue, J.F., Golowich, E. and Holstein, B.R. (1992). {\it Dynamics of
the Standard Model}, Cambridge University Press. 
\item Ebert, D. and Reinhardt, H. (1986).  Effective chiral hadron Lagrangian
with anomalies and Skyrme terms from quark flavor dynamics, {\it Nucl. Phys. B}
{\bf 271}, 188.
\item Ecker, G. (1996). Low-energy QCD, {\it Prog. Part. Nucl. Phys.} {\bf 36}, 71.
\item Efimov, G.V and Ivanov, M.A. (1993). {\it The Quark Confinement Model of
Hadrons}, IOP Publishing, Bristol. 
\item Eguchi, T. and Sugawara H. (1974).  Extended model of elementary particles
based on an analogy with superconductivity, {\it Phys. Rev. D} 
{\bf 10}, 4257.
\item Eguchi, T. (1976). A new approach to collective phenomena in 
superconductivity models, {\it Phys. Rev. D} {\bf 14}, 2755.
\item Frank, M.R. (1995). Nonperturbative aspects of the quark-photon vertex,
{\it Phys. Rev. C} {\bf 51}, 987.  
\item Frank, M.R. and Meissner, T.  (1996). Low-energy QCD: chiral 
coefficients and the quark-quark interaction, {\it Phys. Rev. C} {\bf 53}, 2410.
\item  Frank, M. R.,  Mitchell, K. L., Roberts, C. D. and Tandy, P. C. (1995).
The off-shell axial anomaly via the $\gamma^\ast \pi^0 \rightarrow \gamma$
transition, {\it Phys. Lett. B} {\bf 359}, 17.
\item Frank, M.R. and Tandy, P.C. (1992). Confining quark condensate model
of the nucleon, {\it Phys. Rev. C} {\bf 46}, 338.
\item Frank, M.R., Tandy, P.C. and Fai, G. (1991).  Chiral solitons with
quarks and composite mesons, {\it Phys. Rev. C} {\bf 43}, 2808.
\item Frank, M.R. and Roberts, C.D. (1996). Model gluon propagator and pion and
rho meson observables, {\it Phys. Rev. C} {\bf 53}, 390.
\item Frank, M.R. and Tandy, P.C. (1994). Gauge invariance and the 
electromagnetic current of composite pions, {\it Phys. Rev. C} {\bf 49}, 478.
\item  Gasser, J. and Leutwyler, H. (1982).  Quark masses, {\it Phys. Rep.}
{\bf 87}, 77. 
\item  Gasser, J. and Leutwyler, H. (1983). Low energy theorems as precision tests of
QCD, {\it Phys. Lett. B} {\bf 125}, 325.
\item  Gasser, J. and Leutwyler, H. (1984).  Chiral perturbation theory to one 
loop, {\it Ann. Phys. (N.Y.)} {\bf 158}, 142.
\item  Gasser, J. and Leutwyler, H. (1985).  Chiral perturbation theory:  
Expansions in the mass of the strange quark, {\it Nucl. Phys. B}
{\bf 250}, 465; 517; 539.
\item  Gasser, J. (1995). The $\pi\pi$ scattering amplitude in chiral perturbation
theory, {\it $DA \Phi NE$ Physics Handbook}, $2^{nd}$ Ed., Eds. L. Maiani, G. 
Pancheri and N. Paver, P. 215. 
\item Gell-Mann, M., Oakes, R. and Renner, B. (1968). Behavior of current 
divergences under SU(3) $\otimes$ SU(3), {\it Phys. Rev.} {\bf 175}, 2195. 
\item Gross, F. and Milana, J. (1994). Goldstone pion and other mesons using
a scalar confining interaction, {\it Phys. Rev. D} {\bf 50}, 3332.
\item Hawes, F. T. and Williams, A. G. (1995).  Chiral symmetry breaking in
quenched massive strong-coupling four-dimensional QED, {\it Phys. Rev. D}
{\bf 51}, 3081.
\item Hollenberg, L.C.L., Roberts, C.D. and McKellar, B.H.J. (1992). Two loop
calculation of the $ \omega$-$\rho$ mass splitting, {\it Phys. Rev. C} {\bf
46}, 2057.
\item Hubbard, J. (1959). Calculation of partition functions, {\it Phys. 
Rev. Lett.} {\bf 3}, 77.
\item Hummel, E. and Tjon, T.J. (1990). Relativistic analysis of meson exchange
currents in elastic electron deuteron scattering, {\it Phys. Rev. C} 
{\bf 42}, 423. 
\item Ito, H., Buck, W. W. and  Gross, F. (1992). The axial anomaly and the 
dynamical breaking of chiral symmetry in the $\gamma^\ast \pi^0 \rightarrow
\gamma$ reaction, Phys. Lett. {\bf B287}, 23 
\item Ito, H. and Gross, F. (1993). Isoscalar meson exchange currents and the
deuteron form-factors, {\it Phys. Rev. Lett.} {\bf 71}, 2555.
\item  Itzykson C. and Zuber, J.-B. (1980). {\it Quantum Field Theory}.
McGraw-Hill, New York.
\item Jackiw, R. and Johnson, K. (1973). Dynamical model of spontaneously 
broken gauge symmetries, {\it Phys. Rev. D} {\bf 8}, 2386.
\item Jain, P. and Munczek, H.J. (1993).  $q\bar q$ bound states in the
Bethe-Salpeter formalism, {\it Phys. Rev. D} {\bf 48}, 5403. 
\item Jaus, W. (1991). Relativistic constituent quark model of electroweak
properties of light mesons, {\it Phys. Rev. D} {\bf 44}, 2851. 
\item Kikkawa, K. (1976). Quantum corrections in superconductor models,  
{\it Prog. Theor. Phys.} {\bf 56}, 947.
\item Kleinert, H. (1976a). Quark masses, {\it Phys. Lett. B} {\bf 62}, 77.
\item Kleinert, H. (1976b). Hadronization of quark theories and a bilocal 
form of QED, {\it Phys. Lett. B} {\bf 62}, 429.
\item Kleinert, H. (1978). Hadronization of Quark Theories, {\it Proceedings
of the 1976 School of Subnuclear Physics, Erice}, Ed. A. Zuihuchi, (Plenum).
\item Klevansky, S.P. (1992). The Nambu--Jona-Lasinio model of quantum
chromodynamics, {\it Rev. Mod. Phys. } {\bf 64}, 649.
\item Kugo, T. (1978). Dynamical instability of the vacuum in the Lagrangian 
formalism of the Bethe-Salpeter bound states, {\it Phys. Lett. B} 
{\bf 76}, 625.
\item Kusaka, K. and Williams, A. G. (1995). Solving the Bethe-Salpeter equation for
scalat theories in Minkowski space, {\it Phys. Rev. D} {\bf 51}, 7026.
\item Langfeld, K., Kettner, C. and Reinhardt, H. (1996). A renormalizable
extension of the NJL-model, {\it Nucl. Phys. A} {\bf 608}, 331.
\item Leinweber, D. B. and  Cohen, T. D. (1993). Chiral corrections to lattice
calculations of charge radii, {\it Phys. Rev. D} {\bf 47}, 2147.
\item Lepage, G.P. and Brodsky, S.J. (1980). Exclusive processes in 
perturbative quantum chromodynamics, {\it Phys. Rev. D} {\bf 22}, 2157.
\item Le Yaouanc, A., Oliver, L., Ono, S., Pene, O., and Raynal, J. C. (1985). 
A quark model of light mesons with dynamically broken chiral symmetry, {\it Phys.
Rev. D} {\bf 31}, 137.
\item Llewellyn-Smith, C. H. (1969). A relativistic formulation of the quark 
model for mesons, {\it Ann. Phys. (N.Y.)} {\bf 53}, 521.
\item  Marciano, W. and Pagels, H. (1978). Quantum chromodynamics. {\it Phys.
Rep. C} {\bf 36} 137.
\item Maris, P. and Holties H. (1992). Determination of the singularities of
the Dyson-Schwinger equation for the quark propagator, {\it Intern. J. Mod. Phys
A} {\bf 7}, 5369.
\item McKay, D. and Munczek, H.J. (1985).  Anomalous chiral Lagrangians of
pseudoscalar, vector, and axial-vector mesons generated from quark loops,
{\it Phys. Rev. D} {\bf 32}, 266.
\item McKay, D., Munczek, H.J. and Young, B.-L., (1989).  From QCD to the
low-energy effective action through composite fields: Goldstone's theorem and
$f_\pi$, {\it Phys. Rev. D} {\bf 37}, 195.
\item Meissner, T. (1994). The convergence radius of the chiral expansion in 
the Dyson-Schwinger approach, {\it Phys. Lett. B} {\bf 340}, 226.
\item Meissner, T. (1997). The mixed quark-gluon condensate from an effective
quark-quark interaction, preprint, hep-ph/9702293.
\item Mei\ss ner, U.-G. (1993). Recent developments in chiral perturbation theory, 
{\it Rep. Prog. Phys.} {\bf 56}, 903.
\item Miransky, V.A. (1990). On the wave function of light pseudoscalar
mesons, {\it Mod. Phys. Lett. A} {\bf 5}, 1979. 
\item Mitchell, K.L. (1995). {\it Meson dynamics from a QCD model field 
theory},  Ph.D. Dissertation, Kent State University, unpublished.
\item Mitchell, K.L. and Tandy, P.C. (1997). Pion loop contribution to  
$\rho-\omega$ mixing and mass splitting, {\it Phys. Rev. C} {\bf 55}, 1477.
\item Mitchell, K.L., Tandy, P.C., Roberts, C.D. and Cahill, R.T. (1994).
Charge symmetry breaking via $\rho-\omega$ mixing from model quark-gluon
dynamics, {\it Phys. Lett. B} {\bf 335}, 282.
\item Molzon, W.R., et al. (1978). K(S) regeneration of electrons from 30-GeV/c
to 100-GeV/c:  A measurement of the K0 radius, {\it Phys. Rev. Lett.}
{\bf 41}, 1213.
\item Munczek, H.J. and Nemirovsky, A.M. (1983). Ground-state $q\bar q$
mass spectrum in quantum chromodynamics, {\it Phys. Rev. D}
{\bf 28}, 181.
\item Munczek, H.J. and Jain. P. (1992).  Relativistic pseudoscalar
$q\bar q$ bound states: Results on Bethe-Salpeter wavefunctions and
decay constants, {\it Phys. Rev. D} {\bf 46}, 438.
\item Nambu, Y. and Jona-Lasinio, G. (1961). Dynamical model of elementary
particles based on an analogy with superconductivity. I., {\it Phys. Rev.}
{\bf 122}, 345.
\item Narison, S. (1989). {\it QCD Spectral Sum Rules}, (World Scientific, 
Singapore).
\item O'Connell, H.B., Pearce, B.C., Thomas, A.W. and Williams, A.G. (1994). 
Constraints on the momentum dependence of rho-omega mixing, {\it Phys. Lett. B}
{\bf 336}, 1.
\item O'Connell, H.B., Pearce, B.C., Thomas, A.W. and Williams, A.G. (1995). 
Rho-Omega mixing and the pion electromagnetic form factor, {\it Phys. Lett. B}
{\bf 354}, 14.
\item Pagels, H. and Stokar, S. (1979).  Pion decay constant, electromagnetic
form factors, and quark electromagnetic self-energy in quantum
chromodynamics, {\it Phys. Rev. D} {\bf 20}, 2947.
\item Particle Data Group (1996). Review of Particle Physics, {\it Phys.
Rev. D} {\bf 54}, 1. 
\item Pichowsky, M.A. and Lee, T.-S.H. (1996). Pomeron exchange and 
exclusive electroproduction of $\rho$-mesons in QCD, {\it Phys. Lett. B}
{\bf 379}, 1.
\item Pocan\'c, D. {\it Chiral Dynamics: Theory and Experiment}, Eds. A.M.
Bernstein and B.R. Holstein, Lecture Notes in Physics, {\bf 452}, p. 95
(Springer, Berlin, 1995)
\item Politzer, H. D. (1982). Effective quark masses in the chiral limit,  
{\it Nucl. Phys. B} {\bf 117}, 397.
\item Praschifka, J., Roberts, C.D. and Cahill, R.T. (1987a).
QCD bosonization and the meson effective action, {\it Phys. Rev. D} {\bf 36},
209.
\item Praschifka, J., Roberts, C.D. and Cahill, R.T. (1987b).
A Study of $\rho\rightarrow\pi\pi$ Decay in a Global Colour Model for QCD,
{\it Intern. J. Mod. Phys. A} {\bf 2}, 1797.
\item Praschifka, J., Cahill, R.T. and Roberts, C.D. (1989).
Mesons and diquarks in chiral QCD: Generation of constituent quark masses,
{\it Intern. J. Mod. Phys. A} {\bf 4}, 4929.
\item Qian, Lu. and Tandy, P.C. (1997). In preparation.
\item Radyushkin, A.V. (1994). Nonperturbative QCD and elastic processes at
CEBAF energies,  {\it Proc. of the Workshop on CEBAF at Higher Energies, 
CEBAF, Newport News}, Eds. N. Isgur and P. Stoler,  P.273.
\item Reinders, L.J., Rubinstein, H. and Yazaki, S. (1985).  Hadron
properties from QCD sum rules, {\it Phys. Rep.} {\bf 127}, 1.
\item Reinhardt, H. (1990). Hadronization of quark flavor dynamics, 
{\it Phys. Lett. B} {\bf 244}, 316.
\item Roberts, C.D. (1996). Electromagnetic pion form factor and neutral
pion decay width, {\it Nucl. Phys. A} {\bf 605}, 475.
\item Roberts, C.D. (1997). Private communication.
\item Roberts, C.D. and Cahill, R.T. (1987).  A Bosonisation of QCD and
Realisations of Chiral Symmetry, {\it Aust. J. Phys.} {\bf 40}, 499.
\item Roberts, C.D., Cahill, R.T. and Praschifka, J. (1988).  The
effective action for the Goldstone modes in a global colour symmetry model of
QCD, {\it Ann. Phys. (N.Y.)} {\bf 188}, 20.
\item Roberts, C.D., Cahill, R.T. and Praschifka, J. (1988b).   QCD and a 
calculation of the omega-rho mass splitting,
{\it Int.\ J.\ Mod.\ Phys.\ A} {\bf 4}, 719.
\item Roberts, C.D., Praschifka, J.\ and Cahill, R.T. (1989). A chirally
symmetric effective action for vector and axial vector fields in a global
colour symmetry model of QCD, {\it Intern.\ J.\ Mod.\ Phys.\ A} {\bf 4},
1681.
\item Roberts, C.D. and McKellar, B.H.J. (1990). Critical coupling for
dynamical chiral symmetry breaking, {\it Phys. Rev. D} {\bf 41}, 672. 
\item Roberts, C.D and Williams, A.G. (1994). Dyson-Schwinger equations and 
their application to hadronic Physics,  {\it
Prog.  Part. Nucl. Phys.} {\bf 33}, 477, Ed. 
A. F\"a\ss ler (Pergamon Press, Oxford, 1994).
\item Roberts, C.D., Williams, A.G. and Krein, G. (1992).  On the
implications of confinement, {\it Intern. Journal Mod. Phys. A} {\bf 7}, 5607.
\item Roberts, C.D., Cahill, R.T., Sevior, M.E., Iannella, N. (1994).
$\pi$-$\pi$ scattering in a QCD based model field theory, {\it Phys. Rev. D}
{\bf 49}, 125. 
\item  Rothe, H.J. (1992). {\it Lattice Gauge Theories: An Introduction},
Lecture Notes in Physics,  Vol.~{\bf 43}. World Scientific, Singapore.
\item Sakurai, J.J. (1969). {\it Currents and Mesons}, University of Chicago
Press.
\item Shakin, C.M. and Sun, W.-D. (1997). Microscopic foundations of the vector
meson dominance model and the analysis of rho-omega mixing, {\it Phys. Rev. D}
{\bf 55}, 2874.
\item Shrauner, E. (1977). Bilocal fields for the study of quantum corrections 
to dynamically broken symmetries, {\it Phys. Rev. D}  {\bf 16}, 1887.
\item  Sommerer, A.J., Spence, J.R. and  Vary, J. P.  (1994). Relativistic 
momentum space wave equations and meson spectroscopy,
{\it Phys. Rev. C} {\bf 49}, 513. 
\item Stainsby, S.J. and Cahill, R.T. (1992). The analytic structure of
quark propagators, {\it Int. J. Mod. Phys. A}, {\bf 7}, 7541. 
\item Stratonovich, R.L. (1957). On a method of calculating quantum 
distribution functions, {\it Sov. Phys. Dokl.} {\bf 2}, 416.
\item Tandy, P.C. (1996). Meson transition form factors from a QCD model field
theory, {Prog. Part. Nucl. Phys.} {\bf 36} 97.
\item Tandy, P.C. (1997). in preparation. 
\item Takahashi, Y. (1957). On the generalised Ward identity. {\it Nuovo
Cimento} {\bf 6}, 370.
\item Van Orden, J.W., Devine, N. and Gross, F. (1995). Elastic electron
scattering from the deuteron using the Gross equation. {\it Phys. Rev. Lett.}
{\bf 75}, 4369.
\item Vogl, U. and Weise, W. (1991). The Nambu and Jona-Lasinio model: its
implications for hadrons and nuclei, {\it Prog. Part. Nucl. Phys. } {\bf 27},
195.
\item Volkov, M. K. (1984). Meson lagrangians in a superconductor quark model.
{\it Ann. Phys. (N.Y.)} {\bf 157}, 282.
\item  Volkov, M. K. and Ebert, D. (1982). Four-quark interactions as a common 
dynamical basis of the $\sigma$ model and the vector dominance model. 
{\it Sov. J. Nucl. Phys.} {\bf 36}, 736.
\item Ward, J.C. (1950). An identity in quantum electrodynamics. {\it Phys.
Rev.} {\bf 78}, 182.
\item Weinberg, S. (1966). Pion scattering lengths, {\it Phys. Rev. Lett.} 
{\bf 17}, 616.
\item Wess, J. and Zumino, B. (1971). Consequences of anomalous ward
identities, {\it Phys. Lett. B} {\bf 37}, 95.
\item Wick, G.C. (1954). Properties of the Bethe-Salpeter wavefunctions,
{\it Phys. Rev.} {\bf 96}, 1954.
\item Williams, A.G., Krein, G. and Roberts, C.D. (1991). Modelling
the quark propagator, {\it Ann. Phys.} (N.Y.) {\bf 210}, 464.
\item Witten, E. (1979). Current algebra theorems for the U(1) Goldstone Boson,
{\it Nucl. Phys. B} {\bf 156}, 269.
\end{description}
\end{document}